\documentclass[a4paper,fleqn,usenatbib]{mnras}

\usepackage{newtxtext,newtxmath}
\usepackage[T1]{fontenc}
\usepackage{ae,aecompl}

\usepackage{graphicx}	% Including figure files
\usepackage{amsmath}	% Advanced maths commands
\usepackage{amssymb}	% Extra maths symbols
\usepackage{latexsym, url, lscape}

\hypersetup{draft}
\title{A Study of High-redshift AGN Feedback in SZ Cluster Samples}
\author[L.~B\^{\i}rzan et al.]{L.~B\^{\i}rzan,$^{1}$ D.~A.~Rafferty,$^{1}$ M. Br\"{u}ggen$^{1}$ and H. T. Intema$^{2}$ \\
$^{1}$ Hamburger Sternwarte, Universit\"{a}t Hamburg, Gojenbergsweg 112, 21029, Hamburg, Germany\\
$^{2}$ Leiden Observatory, Leiden University, Oort Gebouw, P.O. Box 9513, 2300 RA Leiden, The Netherlands}

\date{Accepted XXX. Received YYY; in original form ZZZ}

\pubyear{2017}

\begin{document}
\label{firstpage}
\pagerange{\pageref{firstpage}--\pageref{lastpage}}
\maketitle

% Abstract of the paper
\begin{abstract}

We present a study of AGN feedback at higher
redshifts ($0.3<z<1.2$) using Sunyaev-Zel'dovich (SZ) selected samples of
clusters from the South-Pole Telescope and Atacama Cosmology Telescope surveys.
In contrast to studies of nearby systems, we do not find a separation between
cooling flow clusters and non-cooling flow clusters based on the
radio luminosity of the central radio source. This lack may be  due to the increased
incidence of galaxy-galaxy mergers at higher redshift that
triggers AGN activity. In support of this scenario, we find evidence for evolution in the
radio luminosity function of the central radio source: while the
lower-luminosity sources do not evolve much, the higher-luminosity sources show
a strong increase in the frequency of their occurrence at higher redshifts. We
interpret this evolution as an increase in high-excitation radio galaxies (HERGs)
in massive clusters at $z>0.6$, implying a transition from HERG-mode accretion
to lower-power low-excitation radio galaxy (LERG)-mode accretion at intermediate
redshifts. Additionally, we use local radio-to-jet power scaling relations to estimate
feedback power and find that half of the cooling flow systems in our sample probably
have enough heating to balance cooling. However, we postulate that the local
relations are likely not well suited to predict feedback power in
high-luminosity HERGs, as they are derived from samples composed mainly of
lower-luminosity LERGs.

\end{abstract}


\section{Introduction}\label{intro}

The co-evolution of galaxies and their supermassive black holes (SMBH) is a key
ingredient in our understanding of how the present-day universe came to be.
SMBHs interact with (or \emph{feedback} on) their host galaxies via the
energetic emission from an active galactic nucleus (AGN), powered by accretion
of matter onto the SMBH. A breakthrough in this topic occurred with the
discovery that the masses of SMBHs correlate with the properties of their host
galaxies, specifically those of the bulge \citep[see the review of][]{korm13}.
Luminous AGN feedback, in which AGN quench the star formation (SF) in
post-starburst galaxies through energetic winds, and the averaging of SMBH
masses (inherited in galaxy and SMBH mergers) are natural candidates for the
underlying cause of these correlations \citep{silk98,fabi06b,fabi12,korm13}. It
is thought that luminous AGN feedback might help to establish these correlations during
the wet major-merger stage in galaxy evolution, which makes the classical bulges
in low to moderate luminosity elliptical galaxies. However, another mode of AGN
feedback, in which the AGN output is dominated by mechanical energy, is likely
one of the processes that maintains these correlations \citep{fabi12}.

In the nearby universe, direct observational evidence for mechanical AGN
feedback comes from observations of giant elliptical galaxies, groups and
clusters which contain large amounts of hot gas. This discovery was made
possible by a new generation of X-ray instruments with high sensitivity and
spatial and spectral resolution, namely the \emph{Chandra} and \emph{XMM Newton}
Observatories. Data from these instruments showed X-ray cavities or bubbles
filled with radio emission, which include notable examples as Perseus
\citep{fabi00}, Hydra A \citep{mcna00}, A2052 \citep{blan01}, M87
\citep{form05}, MS0735+74 \citep{mcna05} among many others
\citep{birz04,dunn04,raff06,birz12}. Additionally, these data show a lack of
spectroscopic evidence for gas cooling below 2 keV \citep{pete01}.

The existence of such feedback alleviates a
long-standing problem in cluster studies known as the cooling flow problem
\citep{fabi94}. In the initially proposed scenario by \citet{fabi94}, the gas is
heated up to 10$^{7}$ K by the initial gravitational collapse when clusters
form. In the center of clusters, where the density is high and the temperature
is low, the cooling time can be less than the Hubble time
($t_{\rm{cool}}<t_{\rm{H}}$). In this case, it was postulated that a `cooling
flow' should form, in which gas cools and flows inward in quasi-hydrostatic
equlibrium. However, searches for the high mass deposition rates predicted by
this model have found that cooling seems to be proceeding at much lower
rates than predicted \citep[e.g.,][]{pete01}. Instead of steady cooling throughout the core,
it appears that cooling occurs primarily as a result
of local cooling instabilities, regulated by feedback \citep[e.g.,][]{voit16}.

Studies of samples of systems with X-ray cavities found a strong correlation between the mechanical power
injected into the hot gas by the AGN through cavities (i.e., the $4pV$ enthalpy
of the buoyantly rising cavities) and the cooling rates of this gas
\citep{birz04,dunn04,raff06}. From this evidence, it was postulated that the AGN
are heating the gas and regulating cooling through a feedback loop. This process
is known as the \emph{jet-mode}, \emph{maintenance-mode}, or \emph{radio-mode}
AGN feedback and is commonly observed in the nearby universe, where no
significant growth of the brightest cluster galaxy (BCG) typically occurs. In jet-mode AGN feedback, the
accretion rate is thought to be well below the Eddington limit, with the bulk of
the energy released in kinetic form by two-sided radio-bright jets. Galaxies
with such AGN tend to be low-excitation radio galaxies (LERGs), based on the
presence of weak, narrow, low-ionisation emission lines
\citep{hine79,hard06,hard07,best12}. LERGs may be powered by hot-mode accretion,
when the material falls directly onto the SMBH through accretion of clumps of
gas  \citep[known as the \emph{cold feedback
mechanism},][]{pizz05,soke06,pizz10,gasp12,gasp15,gasp17,mcco12,shar12,li15,pras15,
voit15b,voit16}, with no accretion disc to ionize \citep[see also observational
support from][]{raff08,cava11,fara12,mcna14}.

Star formation is an important ingredient in the feedback process because, as
the gas cools down, some of it should to go to fuel SF (through residual
cooling). There are several pieces of evidence for this residual cooling,
such as the observed correlation between the observed star formation rates and
the residual cooling rates
\citep{mcna89,koek99,mcna04,raff08,odea08,dona10,hick10,mcdo11,oonk11,dona15,
foga15,mitt15,trem15,trem16}. Cooling should occur when the central cooling time (or
entropy, or the ratio of cooling to dynamical time) becomes low enough that
thermally unstable condensing clouds form from the hot atmosphere
\citep{raff08,voit08,shar12,gasp12,voit15,voit15b,guo14,brig15,vale15,pras15}.
X-ray cavities likely have an important role in uplifting some of the cooling
gas from the cluster centre \citep{simi08,wern10,wern11,kirk11,vant16,russ16},
and there is evidence that the soft X-ray gas cools and forms H$\alpha$ emission
and cold molecular clouds in a filamentary structure
\citep{mcdo10,wern14,trem15,mcna14,mcna16,fabi16}.

The feedback process is expected to operate in cooling flows.
The cooling time in the core is a basic selection criterion for the cooling
systems: if the central cooling time of a system is smaller than its age, which is
typically a significant fraction of the Hubble time, then one expects that this
system needs heating to prevent cooling \citep[see][]{birz12,pana14a,pana14b}.
Additionally, as mentioned above, the ratio of cooling time to free fall time
has proved to be a sensitive indicator of the presence of cooling, both in
observations and simulations \citep[e.g.,][]{voit08,birz12,shar12,li15}.

However, many of the
details of AGN feedback are missing from this picture or are poorly understood
\citep[e.g., how the energy is transported to cluster scales, either through
weak shocks, turbulence, mixing, sound
waves or cosmic rays;][]{zhur14,zhur16,bane14,wagh14,hill14,hill16,yang16,fabi17,tang17,
pfro13,rusz17}.
Additionally, there are few direct studies of feedback in clusters at high
redshifts ($z>0.5$).

\subsection{AGN Feedback at Higher Redshifts}

The question of how much heating is produced by AGN at higher redshifts is
important since it is at these redshifts that the bulk of galaxy and cluster
formation occurred and, consequently, that the effects of AGN feedback were
likely most instrumental in shaping them. On the cooling side, there is evidence
of evolution in the cuspiness of the density profile \citep{dona92,vikh07,sant10,samu11,mcdo11,mcdo13c}
since $z \sim 1$. However, recent results
from the SPT-SZ survey \citep{carl11,blee15} do not find evidence of evolution in the cooling
properties of the intracluster medium (ICM) in the redshift range $0.3<z<1.2$, only in the cuspiness \citep{mcdo17}.

On the heating side, there is little evidence for evolution in jet-mode AGN
feedback in the general population of radio loud (RL) AGN using deep-field
surveys up to $z \sim 1.3$, suggesting that jet-mode feedback starts to operate
as early as 7 Gyr after Big Bang and does not change since
\citep[e.g.,][]{simp13}, thus maintaining the same approximate balance between
AGN heating and radiative cooling as in the local universe \citep{best06}. The
only direct study of jet-mode feedback in higher-redshift systems was done by
\citet{hlav12} using the MAssive Cluster Survey (MACS) sample and reached
only to redshifts of $z \sim 0.5$. Other studies of AGN feedback at high
redshift \citep{lehm07,smol09,dani12,ma13,best14} rely on indirect methods of
inferring AGN feedback powers, such as scaling relations between the jet
(mechanical) power and the radio luminosity
\citep{birz04,birz08,cava10,osul11,daly12,anto12,godf13}.

Until recently, the majority of complete cluster samples were X-ray flux-limited
samples, e.g., the B55 \citep{edge90}, HIFLUGCS \citep{reip02}, and REFLEX
\citep{bohr04} samples. However, recently a number of SZ surveys have been
undertaken, such as the ATACAMA \citep[ACT;][]{fowl07,marr11,hass13}, South Pole
Telescope \citep[SPT;][]{carl11,reic13,blee15}, and Planck \citep{plan13}
surveys. The main advantages of SZ surveys is that the SZ signal is independent
of redshift \citep{song12,reic13} and is closely related to the
cluster mass with very little scatter \citep{motl05}. Consequently, these
surveys have identified the most massive clusters up to and beyond redshifts of
$\sim 1$, and thus are important for understanding the high redshift universe.

For AGN feedback studies, such a sample allows for
comparisons of feedback properties among samples with different selection
criteria (i.e., mass versus X-ray flux), allowing us to identify potential
biases. For example, it is known that at higher redshift (above
0.5) $\sim$ 50  per cent of  radio-loud quasars  (RLQs), and other powerful radio galaxies, are located in
rich clusters of galaxies \citep{yee87,yate89,hall98,hill91}. These powerful
sources may obscure the thermal signature of the ICM and result in incomplete
X-ray flux-limited cluster samples.

In this paper, we use the SPT and ACT SZ cluster samples to study AGN feedback
at $z>0.3$. These samples are well studied at a variety of wavelengths and
have extensive archival \emph{Chandra} data, making them ideally suited to our
purposes. In addition to archival \emph{Chandra} data, we use radio data from
SUMSS and NVSS (plus targeted GMRT observations for a small subsample) and
star formation rates from \citet{mcdo16}.  We assume $H_{0}=70$ km
s$^{-1}$Mpc$^{-1}$, $\Omega_{\Lambda}=0.7$, and $\Omega_{\rm{M}}=0.3$ throughout.

\section{Sample}

Our sample consists of 99 systems with archival \emph{Chandra} data from the SZ
surveys of the southern and equatorial sky \citep[SPT and ACT;][]{carl11,fowl07}.
The SPT survey covers an area of 2500 deg$^2$, with 677 cluster candidates above
a signal-to-noise threshold of 4.5 \citep{ruel14,blee15}, which represents a
mass-limited sample ($\sim 80$ per cent complete at $M > 5 \times 10^{14}$ M$_{\odot}$) to arbitrarily large distances. From this sample, the
80  cluster candidates with the highest SZ-effect detection significance have been observed with \emph{Chandra},
through a \emph{Chandra} X-ray Visionary Project (PI: Benson) or other GO/GTO
programs (e.g., PI: Mohr, Romer), resulting in $\sim$ 2000 counts per system. To
this sample of 80 clusters, we added a number of other SPT systems which have
archival \emph{Chandra} data (e.g, RDCS J0542-4100, PI: Ebeling; RXC
J0232.2-4420, PI: B\"{o}hringer). We did not include SPT-CL J0330-5228 (z=0.44), since the clusters
A3125/A3128 (z=0.06) are in the foreground, and SPT-CL J0037-5047 (z=1.026)
because of insufficient counts in the X-ray data.

The ACT SZ survey is a sample of $91$ systems ($\sim 90$ per cent complete at $M > 5 \times 10^{14}$ M$_{\odot}$) from within a nearly 1000 deg$^2$
area \citep{hass13}, identified during the 2008 \citep[southern
survey;][]{marr11}, 2009 and 2010 \citep[equatorial survey;][]{hass13}
campaigns. Of these, 18 clusters with the most significant SZ-effect detections were observed with \emph{Chandra}
(PI: Hughes). As with the SPT clusters, the exposure times were
such as to obtain $\sim$ 2000 counts per system (or a minimum of 20-30 ks). In
addition to these systems we added four extra ACT clusters that had
\emph{Chandra} observations: ACT-CL J0326-0043 (MACS J0326-0043; PI: Ebeling),
ACT-CL J0152-0100 (A267; PI: Vanspeybroeck), ACT-CL J2337-0016 (A2631; PI:
Bonamente), ACT-CL J2129-0005 (RXC J2129.6+0005; PI: Allen).

We note that recent simulations \citep{lin15} have shown that SZ observations
can be biased by the presence of a cool core and a
radio-loud (RL) AGN, in the sense that a cool core increases the SZ signal and a
RL AGN decreases it. However, such biases are expected to be small
overall in SZ samples \citep{lin15,gupt16}.

In summary, we have constructed a sample of 99 massive southern and
equatorial clusters with $\sim$ 2000 X-ray counts per system. These data allow
us to obtain reliable temperature and pressure profiles \citep[see
also][]{mcdo13c} to achieve our goal of understanding the state of the system,
such as its cooling time (see Section \ref{analysis}), and, along with
complementary radio data, the impact of AGN feedback.

\section{Data Analysis}\label{analysis}

\subsection{X-ray Analysis}\label{xray_analysis}
All systems were observed with the \textit{Chandra} ACIS detector in imaging
mode, and the X-ray data were obtained from the \textit{Chandra} Data Archive.
Details of the observations are given in Table \ref{Xray_table}.

The \textit{Chandra} data were reprocessed with CIAO 4.8 using CALDB 4.7.2 and
were corrected for known time-dependent gain and charge transfer inefficiency
problems.  Blank-sky background files, normalised to the count rate of the
source image in the $10-12$ keV band, were used for background
subtraction.\footnote{See \url{http://asc.harvard.edu/contrib/maxim/acisbg/}.}

Analysis of the X-ray data closely followed that of \citet{raff08} and \citet{birz12}.
However, in contrast to these works, where 2000 counts per spectrum were commonly used,
in this sample we have only 2000 counts in total for a majority of the systems.
As a result, to obtain spectra at at least two radii, the X-ray spectra were
extracted in circular annuli with as low as $\sim 750$ counts centred on the
centroid of the cluster emission. The lower number of counts results in larger
errors on the derived quantities, but not so large that they are not useful in
assessing the state of the system.  A majority of the systems have at least
three radial bins, except for 10 systems, marked in Table \ref{Xray_table}
(in the $kT$ column), which have two bins only.  In some systems, due to the diffuse
nature of the cluster emission or to substructure, the centroid of the X-ray
emission was difficult to identify precisely. These systems are noted in Table
\ref{Xray_table} (in the X-ray-core column). Spectra and their associated weighted responses
were made for the annuli using CIAO and were fit in XSPEC version 12.5.1.

Gas temperatures and densities (listed in Table \ref{Xray_table}) were found by deprojecting the spectra with a
single-temperature plasma model (MEKAL) with a foreground absorption model
(WABS) using the PROJCT mixing model. In this fit,
we fixed the redshift to those listed in Table \ref{lum_table}, the hydrogen column density
to the value of \citet{dick90} at the cluster position, and the abundance of the MEKAL component to be at least 0.3
times the solar abundance \citep[see][]{mern17}.
Our central values are typically within a factor of two of the
central values reported by \citet{mcdo13c}, with larger discrepancies
attributable to differences in the annuli and deprojection techniques between
our study and theirs.

We derived the cooling times using the deprojected densities and temperatures
found above and the cooling curves of \citet{smit01}. The pressure in each
annulus was calculated as $p=nkT,$ where we have assumed an ideal gas and $n
= 2n_{\rm{e}}$. To derive densities as close to the core as possible, we
used the onion-peel deprojection method described in \citet{raff08}. This method
assumes that changes in the surface brightness within this region are dominated
by changes in the density. Therefore, the temperature and
abundance of the gas are assumed to be constant in the inner region used in spectral
deprojection. We then extrapolated the density profile inward using the
surface-brightness profile, accounting for projection effects under the
assumption of spherical symmetry. The surface-brightness profiles were derived
in annuli with a width of 10 pixels ($\approx 4.9$ arcsec), with typical annulus containing $\sim
100$ counts.

Within the cooling radius, radiative energy losses must be replaced to prevent
the deposition of large quantities of cool gas. Therefore, to assess whether a
system has enough energy to balance cooling, ones needs the luminosity of the
cooling gas inside the cooling radius.  To be consistent with previous works
\citep[e.g.,][]{raff08,birz12}, we define the cooling radius as the radius
within which the gas has a cooling time less than 7.7 $\times$ 10$^{9}$ yr. To
find the total luminosity inside the cooling radius, we performed the
deprojection using a single-temperature model, extracting the spectra in annuli
matched to this cooling radius (i.e., the outer radius of one annulus falls on
the cooling radius). Table \ref{Xray_table} gives the values of $t_{\rm cool}$
and Table \ref{lum_table} gives the values of $r_{\rm cool}$ and $L_{\rm
X}(<r_{\rm cool})$. However, in some cases it was not possible to measure a
cooling region and an X-ray luminosity (e.g., for faint, diffuse clusters that
are likely non-cooling flow clusters).

Furthermore, we also fit the bolometric luminosity inside the $R_{500}$ region,
$L_{\rm X}(<R_{500})$, and these values are listed in Table \ref{lum_table}.
$R_{500}$ is defined as the region at which the mean mass density is 500 times
the critical density at that cluster redshift \citep[see][]{prat09}. We
calculated $R_{500}$ using the masses, $M_{500}$, derived from the SZ signal
$Y_{\rm{SZ}}$ \citep{reic13,hass13,hilt13,blee15}.
\footnote{$R_{500}=(\frac{M_{500}}{500 \rho_c(z)  4 \pi/3})^{1/3}$, with
$\rho(z)=\frac{h(z)^{2} 3 H_{0}^{2}}{8 \pi G}$ and
$h(z)^{2}=\Omega_{\rm{M}}(1+z)^{3}+\Omega_{\Lambda}$}

\subsection{Cooling Flow Clusters}\label{Sec:cf}
We investigated three different diagnostics to identify the cooling flow systems in
our sample: central cooling time, the minimum thermal instability and the ratio of
cooling time to free-fall time. Each of these diagnostics should be sensitive to
the presence of gas that is unstable to cooling.

To calculate the central cooling times, we used the deprojection technique
described in Section \ref{xray_analysis}. The minimum radius at which we could derive
reliable cooling times depends on the central surface brightness (see Table
\ref{Xray_table} for the radius of the inner annulus used in the deprojection).
Since we want to compare the cooling times for all the systems at a single
physical radius, as close as possible to the nucleus, we computed the cooling
time at 10 kpc using the surface brightness profiles to extrapolate the densities inward (see
Section \ref{xray_analysis}). For some systems the extrapolation did not work well,
as the surface brightness (SB) profile is too noisy or drops towards the centre or
there is significant substructure that is inconsistent with the deprojection
method (i.e., systems with large SB errors, see Section \ref{morph}). For these
systems we did not calculate a cooling time at 10 kpc, and these are the ones
with missing values for $t_{\rm{cool}}$(10kpc) in Table \ref{Xray_table}.

Additionally, we calculated the central temperature drop for each system, since
the temperature is expected to drop towards the centre in a cooling flow
cluster. We calculated the drop as the ratio between the highest temperature in
the profile and the temperature of the innermost annulus. Table \ref{Xray_table}
lists the temperature drop values for all systems. Some systems have no entry
since the temperature profile increases towards the centre, or the profile was
too noisy and the temperature drop value was insignificant within errors.
Generally, the calculated temperature drop depends on the size of the innermost
annulus. However, the temperature typically varies slowly with radius, so
variations in the size of the annuli should not affect our estimates
significantly. We note that the temperature drop is not used in this paper as a
criterion to separate the cooling flow systems from non-cooling flow systems.

An alternative way to select cooling flows is based on the thermal stability of
the gas \citep{voit08,voit16,shar12}.\citet{voit08} found that star formation and H-$\alpha$ emission
(and hence cooling) seem to occur only if, at some location in the cluster, the
following condition is met: \begin{equation} {\eta_{\rm min}=\rm
min}\left(\frac{\kappa T}{\Lambda (T) n_{e} n_{H} r^{2}}\right) \sim
\frac{1}{f_{c}} \lesssim 5, \end{equation} where $\Lambda(T)$ is the cooling
function calculated using the APEC spectral model \citep{smit01},  and $f_{c}$ is
the factor by which the magnetic field suppresses the conductivity below the
Spitzer value. Assuming that the effective thermal conductivity can be expressed
as a multiple, $f_{c}$, of the Spitzer value, this parameter provides a measure
of the stability of the gas to local cooling. For large values of $\eta_{\rm
min}$, thermal conduction overwhelms radiative cooling, preventing local cooling
throughout the ICM. For small values of this parameter, local cooling can run
away, so that some regions of the ICM may cool to low temperatures, resulting in
a multiphase medium and the deposition of the cooled gas.

If the AGN are
fuelled by the cooled ICM, a process known in literature as the cold feedback mechanism
\citep{pizz05,pizz10}, chaotic feedback \citep{gasp13,gasp15} or precipitation
\citep{voit15,voit15b,voit16}, then $\eta_{\rm min}$ determines the systems where
cooling should occur. \citet{voit08} found that values of $\eta_{\rm min} \lesssim 5$
correspond approximately to an inner cooling time of $5\times 10^{8}$ yr.

Recently, the multiphase threshold has been interpreted as resulting from the
coupling between conduction and thermal instability \citep[for a review,
see][]{voit16}, since simulations have shown that thermal instability can
produce a multiphase medium when the ratio of cooling time to free-fall time is
$\lesssim$ 10 \citep{mcco12,shar12,gasp12}. There are 22 systems in our sample
which meet this multiphase threshold.
Additionally, we find that a multiphase
threshold of $t_{\rm cool}/t_{\rm ff} \lesssim~10$ corresponds to a central
cooling time of $t_{\rm{cool}}$(10kpc) $\lesssim~ 2 \times  10^{9}$~yr and
$\eta_{\rm min} \lesssim~10$, with 2 exceptions: SPT-CL J2248-4431 and ACT-CL
J0438-5419, where $t_{\rm{cool}}$(10kpc) $> 2 \times  10^{9}$~yr. These 2 systems show elongated X-ray morphologies, with higher
central temperatures, and therefore may have gone through a merger recently. The
20 systems with $t_{\rm cool}/t_{\rm ff} \lesssim~10$, $t_{\rm{cool}}$(10kpc)
$\lesssim~ 2 \times 10^{9}$~yr and $\eta_{\rm min} \lesssim 10$ are our sample
of cooling flow (CF) systems (see Section \ref{disc:tcool_lrad}). Furthermore,
there are eight intermediate systems with
$t_{\rm{cool}}$(10kpc)~$\lesssim~ 2 \times~10^{9}$~yr, but
$t_{\rm cool}/t_{\rm ff} >10$, and four systems with
$t_{\rm{cool}}$(10kpc)~$>~ 2 \times~10^{9}$~yr, $t_{\rm cool}/t_{\rm ff}>10$, but
$\eta_{\rm min} \sim 10$.

Our sample of 20 CF systems is smaller than that of \citet{mcdo13c}, who
identified 29 cooling flow systems through their short central cooling times (a
cooling time of the inner bin $\lesssim~10^{9}$~yr) from a similar parent
sample. Our CF sample and the \citet{mcdo13c} CF sample have 13 systems in
common. Among the seven systems that are only in our CF sample, four systems
were not analysed in the \citet{mcdo13c} sample (e.g., some ACT systems); for
the remaining three, \citet{mcdo13c} did not find a short cooling time  (e.g.,
El Gordo). For eight of the 16 CF systems that appear in the \citet{mcdo13c} CF
sample but not in ours, we obtain higher inner temperatures than those from the
extrapolation used in \citet{mcdo13c} The remaining eight systems are either
borderline CFs or have large errors in their X-ray profiles.

\subsection{Radio Properties}\label{radio_props}

The SPT and ACT southern sources are covered at 843 MHZ by the Sydney University
Molonglo Sky Survey \citep[SUMSS,][]{bock99}. The ACT galactic sources are
covered by the NRAO VLA Sky Survey at 1.4 GHz \citep[NVSS,][]{cond98}. For
ACT-CL J0326-0043, we use the flux density at 1.4 GHz from the Faint Images of
the Radio Sky at Twenty-cm (FIRST) survey \citep{helf15}, as the central source
and an unrelated source at $\sim 80$ kpc separation are blended together in the
NVSS image (see Figure \ref{images_cav}). For ACT-CL J0102-4915 (El Gordo) and ACT-CL J0152-0100 we use the
flux density from deep GMRT images at 610 MHz \citep{lind14,kale13}.
Additionally, for the
following systems, we have obtained targeted GMRT observations at 325 MHz with integration times of 5.5~h per target
(Intema et al. 2017, in preparation):
SPT-CL J0123-4821,
SPT-CL J0142-5032, SPT-CL J0212-4657, SPT-CL J0304-4401, ACT-CL J0304-4514,
SPT-CL J0307-5042, SPT-CL J0348-4514,  SPT-CL J0411-4819,
SPT-CL J2031-4037, SPT-CL J2258-4044, and SPT-CL J2301-4023. These observations were made between May and November 2014,
and were reducted with the SPAM package \citep{inte09,inte14}.

We searched the radio images for evidence of a central radio source (cRS). We consider
a source to be a central radio source if the peak of the emission lies within a
radius of 2 arcsec of the BCG optical core (when more than one BCG is present,
we use the BCG that lies closest to the cluster X-ray core). The 2 arcsec radius corresponds to
the approximate positional accuracy of NVSS and SUMMS for sources with flux
densities typical of those in our sample \citep[$\gtrsim 15$ mJy at the
frequency of the survey][]{cond98,bock99}, and equates to uncertainties of $\sim
8$--20 kpc for our sample, depending on the redshift of the source. Since this
radius is typically within the envelope of the BCG and chance superposition of
an unassociated source within this radius is unlikely \citep[$<
0.001$][]{cava08}, we do not expect significant contamination by non-central
radio sources. In total, 46 sources in our sample have a detected cRS (two of
which were found in our targeted GMRT observations: SPT-CL J2301-4023 and ACT-CL J0304-4921; see Figure \ref{images_cav}
and Figure \ref{images_no_cav}).

Table \ref{lum_table} lists the rest-frame 843 MHz monochromatic radio luminosities for
the central radio sources, calculated as follows: \begin{equation}
L_{843\rm{MHz}}=4 \pi D_{\rm{L}}^{2} S_{843\rm{GHz}} (1+z)^{\alpha-1},
\end{equation} where $\alpha$  is the spectral index assuming $S_{\nu} \sim
\nu^{-\alpha}$ and $S_{843\rm{MHz}}$ is the (observed-frame) flux density at 843 MHz.  Since no
spectral index information was available for the sources in our sample, a value of 1.0 was
adopted.

For the systems with no detected central radio source, Table
\ref{lum_table} lists the upper limits from the SUMSS, NVSS and GMRT images.
The 5-$\sigma$  sensitivity limits of the SUMSS images are 6--10 mJy beam$^{-1}$, depending on the
declination \citep{mauc03}. For the NVSS catalog, the 5-$\sigma$ sensitivity limit is 2.5 mJy beam$^{-1}$ at 1400 MHz
\citep{cond98}, which implies a limit of 4.5 mJy beam$^{-1}$ at our reference
frequency of 843 MHz for a source with $\alpha = 1$. For our GMRT images at 325 MHz, we obtained
5-$\sigma$ sensitivities of 1--40 mJy beam$^{-1}$ (Intema et al., in preparation),
giving limits as low as 0.4 mJy beam$^{-1}$ at 843 MHz.

We note that some sources may have a radio mini-halo in addition to the central
radio source \citep[e.g.,][]{mitt09}, and recently it was found that radio mini-haloes are common in massive CF clusters
up to $z<0.35$
\citep{giac17}.
For example, in SPT-CL J2344-4242 (the Phoenix cluster),
\citet{vanw14} found a probable radio mini-halo. Other systems also have diffuse
relic emission at the periphery of the cluster \citep[e.g., ACT-CL J0102-4915,
also known as El Gordo, has a double radio relic,][]{lind14,bott16}, and based
on the SUMSS images, there might be other sources with possible relic emission
(see Section \ref{merging}).

\subsection{Systems with Possible Cavities}\label{cav}

\begin{figure*}
\begin{tabular}{@{}cc}
\includegraphics[width=150mm]{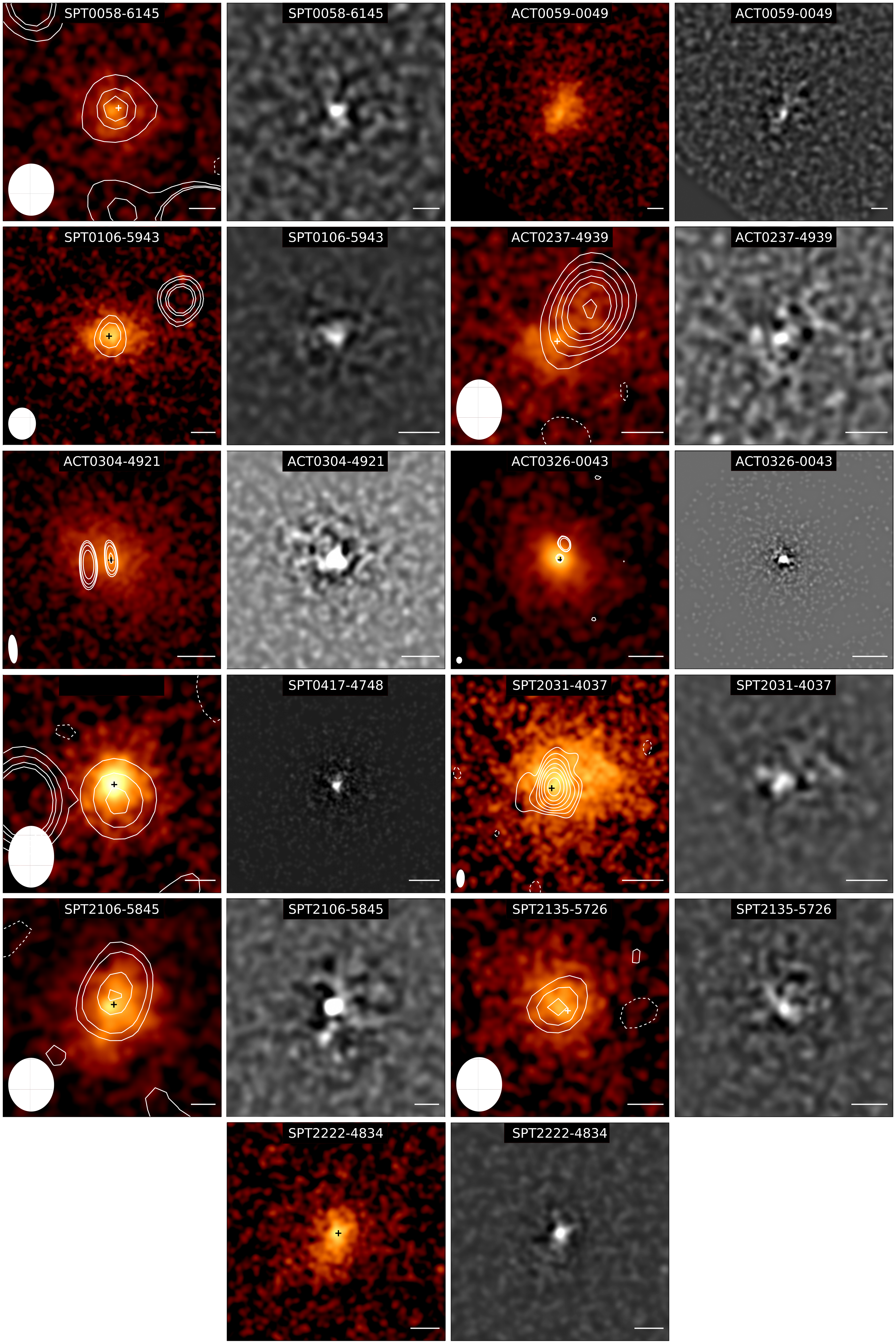}
\end{tabular}
\caption{Smoothed and unsharp-masked images for ten systems which
show visible structure in \emph{Chandra} images (see Section \ref{cav}). The
radio contours from FIRST (ACT-CL J0326-0043), GMRT (ACT-CL J0304-4921 and SPT-CL J2031-4037), or SUMSS (all others)
images are overlaid, and the restoring beam is indicated by
the white ellipse in the lower left corner. The BCG location is marked with a cross,
and the line in the lower right corner denotes a scale of 200 kpc.}
\label{images_cav}
\end{figure*}

\citet{mcdo13b} have shown that there are cavities in the Phoenix cluster, and
\citet{hlav15} used unsharp-masking techniques to identify possible cavities
in seven other SPT clusters. Through visual inspection of the X-ray images, we
found evidence for significant structure in eleven systems
(see Figure \ref{images_cav}).
For these systems, we made unsharp-masked images to
make any such structure more evident. Among the CF sample there are possible
cavities in SPT-CL J2106-5845 at $z=1.132$,\footnote{SPT-CL J2106-5845 is one of
the most X-ray luminous systems in the sample, the fifth most luminous after the
Phoenix cluster ($z=0.595$), El Gordo ($z=0.87$), AS1063 ($z=0.351$) and ACT-CL
J0438-5444 ($z=0.421$), and the most massive one above $z>0.6$.} which has an
S-like enhancement of X-ray emission, possibly due to X-ray cavities which
lie along an axis with a small angle to the line of sight  \citep[see
also NGC4636,][]{bald09}. Alternatively, such an arm-like structure could arise from turbulence
driven by core sloshing \citep{ahor16}. The BCG and X-ray centre are displaced by about 55 kpc
\citep[see Figure \ref{images_cav} and][]{mcdo16}, and an interesting
question is how AGN feedback operates in this case (see Section
\ref{agn_feedback}).\footnote{In the case of SPT-CL J2106-5845, there may also be
a separation between the BCG location and radio source position; however,
high-resolution radio observations are needed to verify this.} Additionally,
there is some evidence for structure in ACT-CL J0304-4921 and SPT-CL J0417-4748.
In the case of ACT-CL J0304-4921 we found a lower-power cRS in our GMRT image
(below the SUMSS detection limit; see Figure
\ref{images_cav}). Both ACT-CL J0304-4921 and SPT-CL J0417-4748 have a radio source displaced
from the BCG. However, in these cases the unsharp-masked images are not consistent
with cavities, and as a result the structure we see in the X-ray images might be
due to merger activity.

There are also two cavity candidates among the eight intermediate CF systems,
SPT-CL J2222-4834 at $z=0.652$ and SPT-CL J0058-6145 at $z=0.83$ (see Figure
\ref{images_cav}). Additionally, as with SPT-CL J2106-5844, SPT-CL J0058-6145
shows a separation between the X-ray core and the BCG (of $\approx$ 70 kpc). For
both SPT-CL J2222-4834 and SPT-CL J0058-6145, the unsharp-masked images support
the presence of cavities to the east and west. In the case of
SPT-CL J0058-6145, a central radio source is also present.
There are also cases of possible cavity systems among the NCF systems, e.g.,
SPT-CL J2135-5722\footnote{In the case of SPT-CL J2135-5726, the radio source is
displaced from the BCG location, and there is also a separation between the
X-ray core and BCG position of 50 kpc.} at $z=0.427$, ACT-CL J0237-4939 at
$z=0.334$, SPT-CL J0106-5943 at $z=0.348$ and SPT-CL J2031-4037 at
$z=0.342$.

The best cavity system candidate from our sample is SPT-CL J2031-4037,
which is one of the NCF systems. In this case, there is also
evidence in the GMRT image that the radio emission extends towards the cavity (see Figure
\ref{images_cav}). We measure a cavity in this system as a ellipsoid with
semi-major and semi-minor axes of $9.3 \times 4.9$~arcsec, situated at a
projected distance of 18.3 arcsec from the cluster centre. By assuming that
the cavity rose buoyantly from the cluster centre to the current location, we
estimate an age of 1.6 $\times$ 10$^{8}$ yr and a mechanical power (considering
only $pV$ work) of 2.9 $\times$ 10$^{44}$ erg s$^{-1}$ \citep{birz04}, enough
to balance the X-ray luminosity inside the cooling region (see
Table \ref{lum_table}). However, the cooling region in the system is small, as
only in the very centre does the cooling time drop below 7.7 Gyr.

In summary, there are possible cavities in both CF and NCF systems in our
sample. Although we do not expect cavities in NCFs, many of the NCFs might
harbour small cool cores \citep[e.g.,][]{sun07} that would only be visible in deeper \emph{Chandra}
observations (see Section \ref{morph} for more possible CF candidates).

\subsection{Merging Activity}\label{merging}

Signs of merging activity or interactions are often apparent in X-rays images
(e.g., a distorted morphology), radio images (e.g., relic emission) and optical
images (e.g., a separation between X-ray centre and BCG centre). We briefly
outline below such evidence in our sample.

\begin{figure*} \begin{tabular}{@{}cc}
\includegraphics[width=150mm]{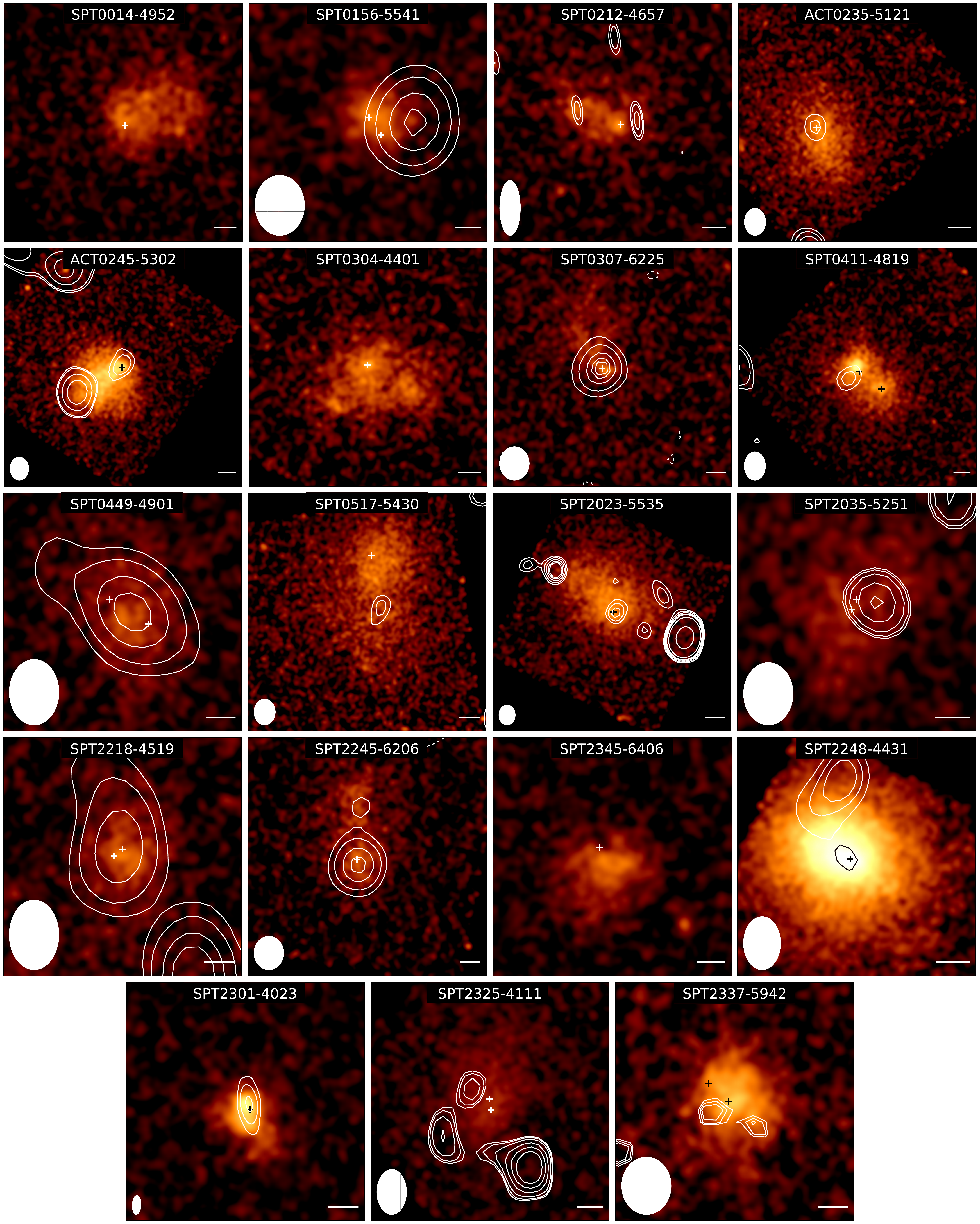} \end{tabular} \caption{Smoothed
\emph{Chandra} images of systems with signs of interactions (see Section
\ref{merging}). The radio contours from GMRT (SPT-CL J0212-4657 and SPT-CL J2301-4023) and SUMSS (all others) images are overlaid,
  and the symbols are the same as in
Figure \ref{images_cav}. } \label{images_no_cav}
\end{figure*}

In some systems in our sample, the X-ray images show direct evidence of
interactions, such as a tail-like structure (e.g., SPT-CL J0307-6225) or the
presence of multiple subclusters (see Figure \ref{images_no_cav}). For example,
SPT-CL J0304-4401 shows at least three interacting systems; SPT-CL J0411-4819
shows two interacting systems; SPT-CL J0212-4657 shows an excess of diffuse
X-ray emission at the end of an X-ray tail, perhaps a subgroup, which has its
own central radio source (there is also  radio emission just a little ahead
of the cluster core).

Another indicator of merging activity is the relic radio emission, which is
thought to be due to cluster-cluster mergers \citep[see the review
of][]{brun14}. Based on the SUMSS images of our sample, besides the already known relics in El
Gordo \citep{lind14,bott16}, there are possible relics in some other systems
(e.g.; SPT-CL J2023-5535, which shows evidence
of a subcluster on the cluster periphery, see Figure \ref{images_no_cav}).
However, one cannot exclude the possibility that in some of these systems the radio
emission may be associated with AGN activity (even if there are no apparent
optical counterparts to the radio emission). Deeper optical images and radio images at different
frequencies are needed to confirm the putative radio relic emission.

Further evidence of cluster-scale merging activity is the presence of two or more
cD galaxies \citep{mcdo16}. This is the case for SPT-CL J0156-5541 and SPT-CL
J0411-4819 (with displaced radio  emission, see Figure
\ref{images_no_cav}), among others. Additionally, a large offset between the
X-ray core and the BCG location, as seen in El Gordo (for images see Figure
\ref{images_cav} and Figure \ref{images_no_cav}), is evidence of significant
sloshing, thought to be often triggered by a merger. Evidence of other (e.g.,
galaxy-galaxy) merging activity is the presence of a nearby companion galaxy to
the BCG \citep[e.g., SPT-CL J0000-5748]{mcdo16} or of asymmetric emission at UV wavelengths
with a minimum of two peaks.

\subsection{X-ray Morphology and the Central Radio Source}\label{morph}

\begin{figure*} \begin{tabular}{@{}cc}
\includegraphics[width=84mm]{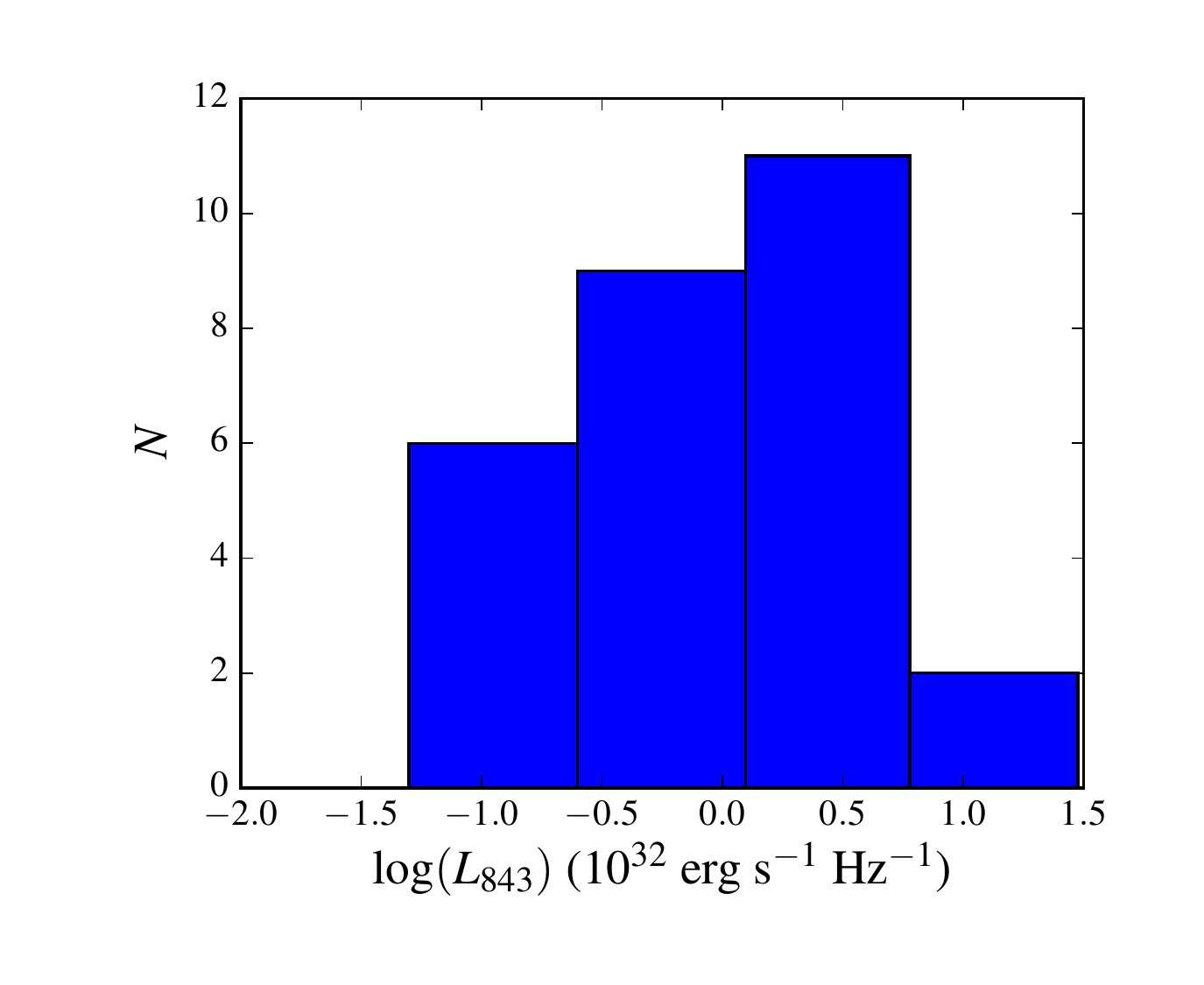} &
\includegraphics[width=84mm]{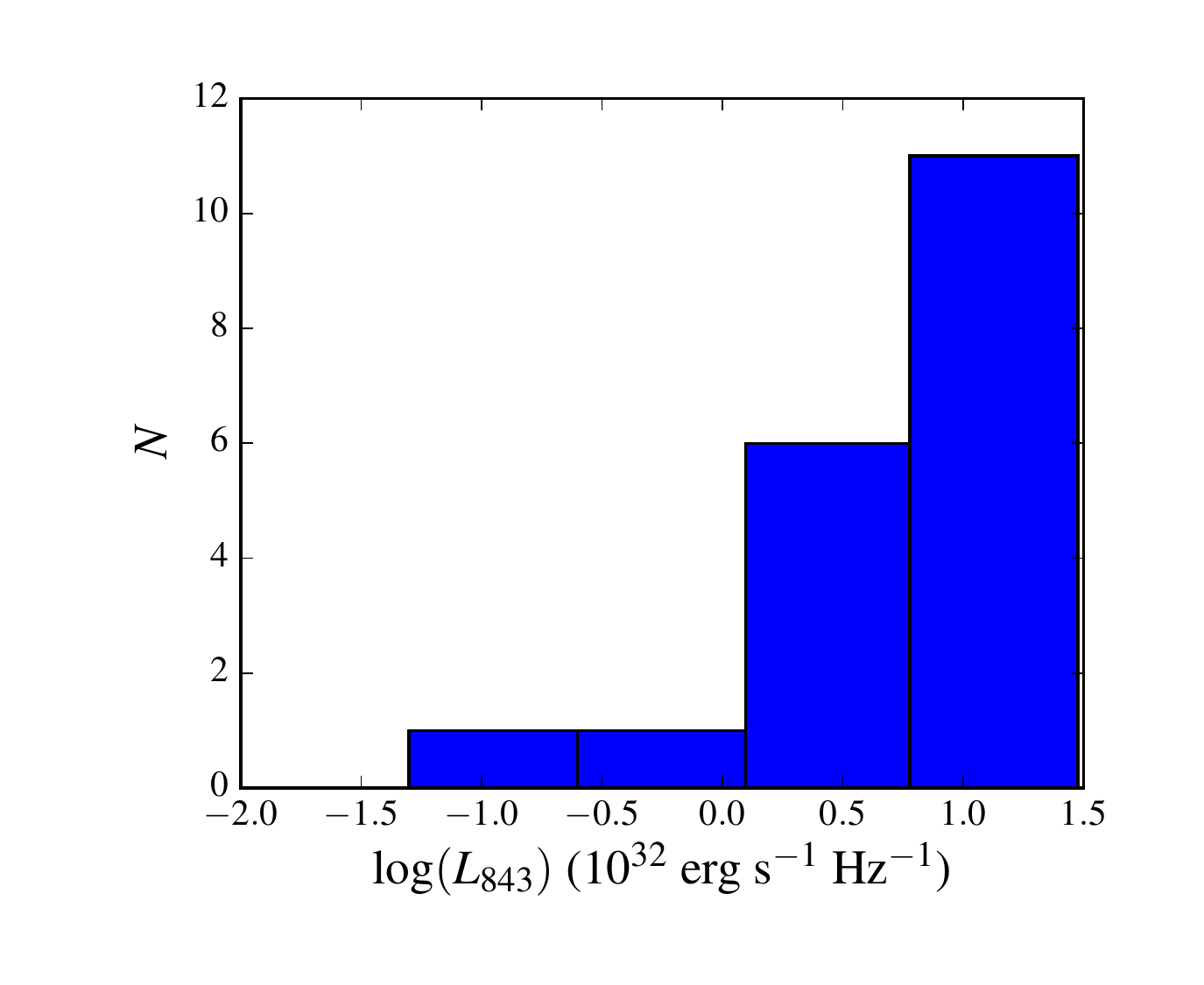} \\ \end{tabular}
\caption{The number of sources with a cRS of a given radio luminosity at 843 MHz
for $z<0.6$ (left panel) and $z>0.6$ (right panel).}
\label{F:crs_hist}
\end{figure*}

In Figure \ref{F:crs_hist} we show the number of detected cRSs as a function of the
radio luminosity at 843~MHz. Generally, systems at $z<0.6$ have lower luminosities than those
at $z>0.6$. This difference is partly due to the flux-limited nature of the radio surveys that we have used
(which means that lower-luminosity sources cannot be detected at high redshifts),
but it is also due to an increased incidence of powerful sources at higher redshifts (see Section~\ref{disc:rlf}).

For the 20 CF systems with  $t_{\rm{cool}}$(10kpc)~$\lesssim~2 \times 10^{9}$~yr
and $t_{\rm cool}/t_{\rm ff} \lesssim~10$, 13 of which have a central radio
source, the X-ray morphology can be described as small, round and compact, with
a peaked core (with a few exceptions: e.g., the Phoenix cluster, SPT-CL
J2106-5845, El Gordo, ACT-CL J0304-4921, SPT-CL J0232-4420*, ACT-CL J2129-0005*,
SPT-CL J2011-5725*\footnote{The systems with an asterisk were detected in
one or more X-ray surveys and are at $z \approx 0.3$ (see Table
\ref{lum_table}).}). Only a couple of these systems show evidence for cavities,
e.g., the Phoenix cluster \citep[see][]{mcdo15}. The two systems with
$t_{\rm cool}/t_{\rm ff} \lesssim~10$, but
$t_{\rm{cool}}$(10kpc)~$>2 \times 10^{9}$~yr are large, bright and elongated,
e.g., SPT-CL J2248-4431 (see Figure  \ref{images_no_cav}).

The four systems with
$t_{\rm{cool}}$(10 kpc) $> 2 \times 10^{9}$~yr, but $\eta_{\rm min} \lesssim 10$
(see Section \ref{Sec:cf}) do not have a cRS above the detection limit (see Table \ref{lum_table}). For the eight
intermediate systems, with $t_{\rm{cool}}$(10kpc)~$\lesssim~2 \times 10^{9}$~yr
and $t_{\rm cool}/t_{\rm ff} > 10$, only two have a cRS, and both have a
significant separation between the X-ray peak and BCG location (see Section
\ref{merging}). However, some of these have low
central temperatures and peaked SB profiles and hence might be classified as
cooling flow systems in deeper \emph{Chandra} observations (e.g., SPT-CL
J2352-4657).

Of the remaining 65 NCF systems, 32 have a cRS. The NCF systems have a
variety of X-ray morphologies, but mostly they can be described as disturbed and often
show evidence of merging activity, such as two BCGs (e.g., SPT-CL J2035-5251,
SPT-CL J2337-5942, see Figure \ref{images_no_cav}), sharp edges (e.g., SPT-CL
J2258-4044, SPT-CL J2233-5339), or an elongated tail-like appearance
(e.g., ACT-CL J0235-5121, see Figure \ref{images_no_cav}).

Notably, in contrast to local samples (e.g., the B55 sample), some of the NCFs
in our sample host powerful cRSs (up to $L_{843} \sim 1.9 \times 10^{33}$ erg
s$^{-1}$ Hz$^{-1}$). Presumably, this radio activity is
unrelated to feedback but is instead triggered by other factors. Additionally,
this difference points to evolution in the cRSs in NCFs, and we discuss
this possibility further in Section \ref{disc:rlf}.

\subsection{Star Formation Rates}\label{SF}

We use the SFRs of \citet{mcdo16}, who
computed SFRs from UV, OII, and infrared data. The
data used in \citet{mcdo16} to calculate the SFRs comes from
photometric-redshift follow-up campaigns \citep{song12,blee15} plus
\emph{U}-band imaging from \citet{mcle15}. Of the three SFRs listed in
\citet{mcdo16}, we use the infrared-derived SFR, unless one of the other values was a
detection and the infrared-derived SFR was only an upper limit or when there was no
IR-derived SFR listed for that system. If a system had two or more detections, we
use the average of the detected rates as the SFR for that system. The SFRs are
given in Table \ref{lum_table}.

\section{Results and Discussions}\label{S:results}

\subsection{X-ray vs.\ Radio Luminosity\label{disc:lx_lrad}}

\begin{figure*} \begin{tabular}{@{}cc}
\includegraphics[width=84mm]{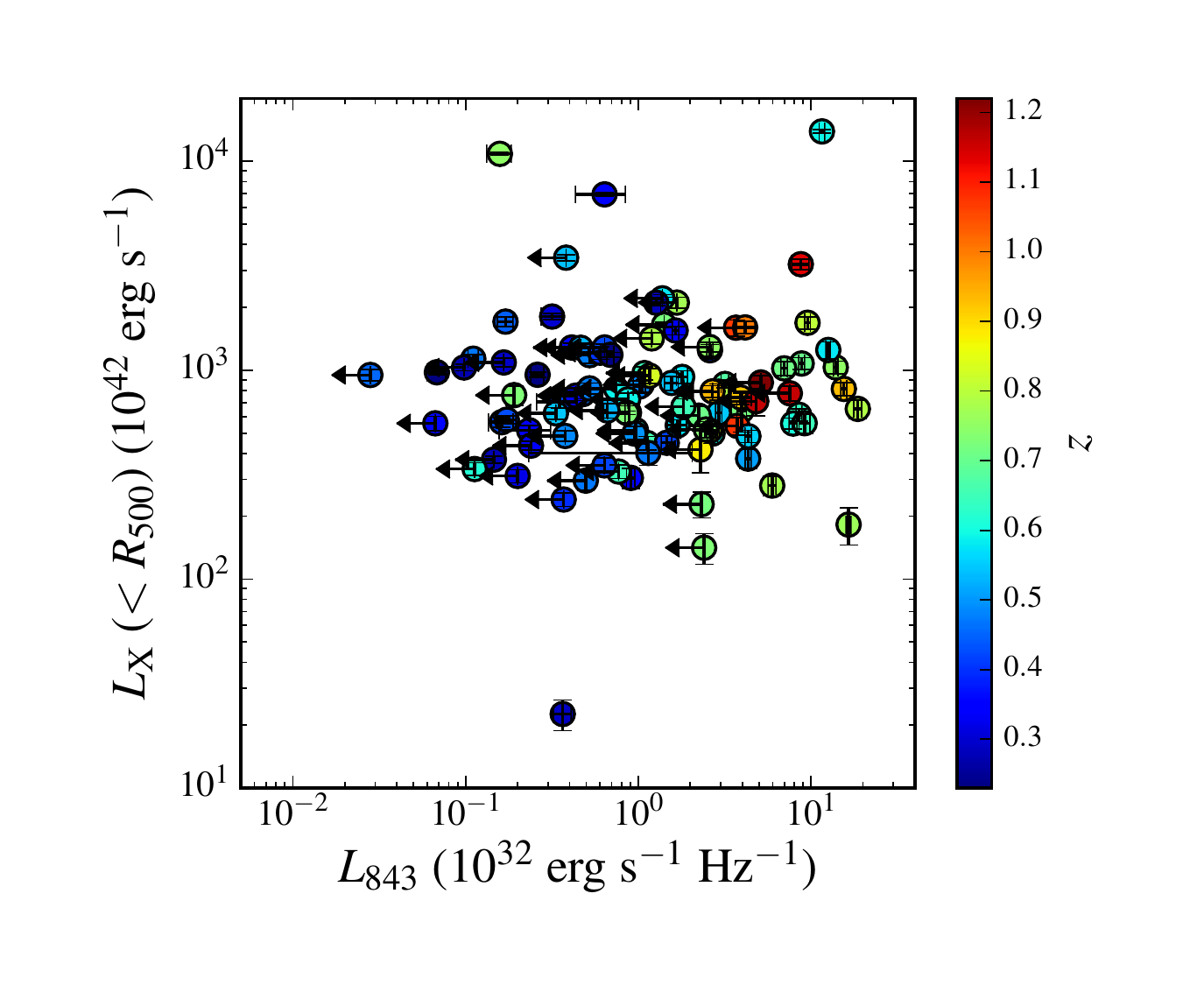} &
\includegraphics[width=84mm]{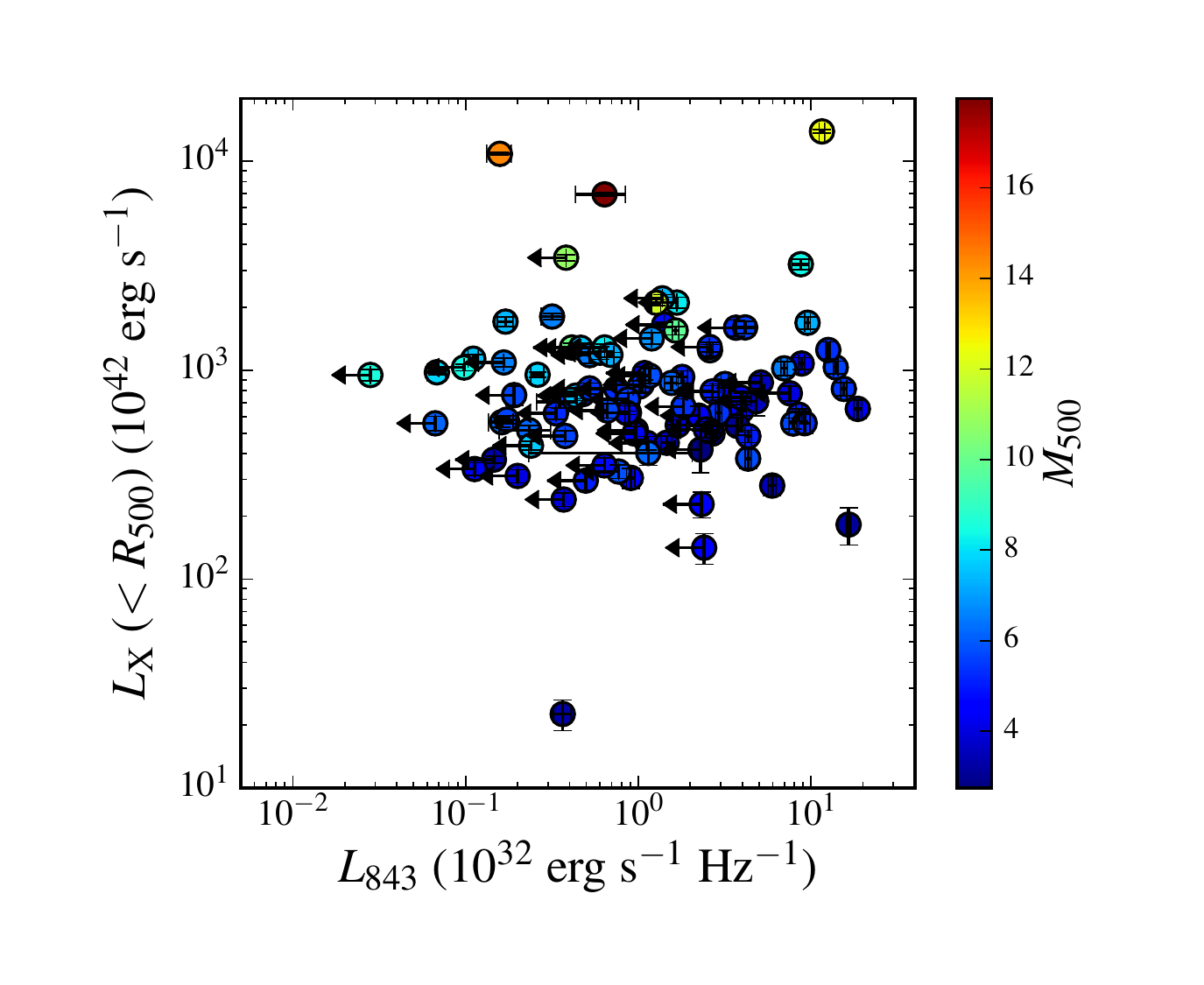} \\ \end{tabular}
\caption{Total bolometric X-ray luminosity inside the $R_{500}$
region, $L_{\rm X}(<R_{500})$, versus the rest-frame 843 MHz monochromatic radio
luminosity for the central radio source, $L_{843}$. The colour denotes the
redshift in the left panel and $M_{500}$ in the right panel.}
\label{F:LX500__vs_lrad}
\end{figure*}

In Figure \ref{F:LX500__vs_lrad}, we plot the bolometric X-ray luminosity inside the
$R_{500}$ region, $L_{\rm X}(<R_{500})$, versus the rest-frame 843 MHz monochromatic radio
luminosity for the central radio source, $L_{843}$. The highest monochromatic
radio luminosity in our sample is $L_{843}$ $\sim$ 1.9 $\times$ 10$^{33}$ erg
s$^{-1}$ Hz$^{-1}$ for SPT-CL J0449-4901 at a redshift of $z=0.79$, one of the
systems with two BCGs (see Table \ref{lum_table} and Figure
\ref{images_no_cav}). Based on the X-ray data this system was classified as a
NCF (see Table \ref{Xray_table}). Overall, the range in radio luminosity at 1400 MHz
(extrapolated from 843 MHz assuming $\alpha = 1.0$) is $L_{1400} \sim 0.04$--10 $\times$
$10^{32}$ erg s$^{-1}$ Hz$^{-1}$. This range is considerably smaller than that
of the B55 and HIFLUGCS samples, where the 1400 MHz luminosity ranges from
$10^{28}$ to $10^{35}$ erg s$^{-1}$ Hz$^{-1}$ \citep{birz12}. Additionally, the
lower limit of this range is above the threshold that separates CFs and NCFs in
local samples of $L_{1400} \sim 2.5 \times$ 10$^{30}$ erg s$^{-1}$ Hz$^{-1}$
\citep[see][]{birz12}. However, since we mainly used shallow survey data
\citep[e.g., SUMSS,][]{mauc03} the sensitivity of our radio data is generally
insufficient to detect the low-luminosity sources.

Furthermore, in our sample there are no central radio sources with $L_{843} >
10^{34}$ erg s$^{-1}$ Hz$^{-1}$, such as those that appear in local X-ray
flux-limited cluster samples (e.g., Cygnus A). This lack can be partly explained
by the radio-luminosity bias of SZ surveys \citep{lin15}, which tends to select
against such sources, and by the fact that such luminous sources are rarely
found in massive clusters (of which the SZ samples are predominantly comprised).

However, our sample does have a number of powerful radio sources, with
luminosities above the canonical FRI-FRII dividing line. The commonly used
luminosity separation between FRI and FRII sources \citep{fana74} corresponds to
$L_{178} \sim$ 2 $\times$ $10^{32}$ erg s$^{-1}$ Hz$^{-1}$ Sr$^{-1}$ at the rest frame
frequency of 178 MHz, or $L_{1400} \sim$ $10^{31.5}$ erg s$^{-1}$ Hz$^{-1}$ Sr$^{-1}$ at
1.4 GHz  (which corresponds to a monochromatic luminosity at 843 MHz of
$L_{843}$ $\sim$ 5 $\times$ 10$^{32}$ erg s$^{-1}$ Hz$^{-1}$, assuming a
spectral index of $\alpha=1.0$). The more powerful FRII sources are found to
preferentially avoid clusters at lower redshifts \citep{owen91} and be present
in rich clusters around $z=0.5$ or higher \citep{yee87,hill91,bels07}.
Morphologically, FRI sources show two-sided jet-dominated emission that smoothly
extends into the ICM and at kpc scales forms large-scale lobes of diffuse radio
emission, whereas FRII sources have lobe-dominated emission with collimated jets
on kpc scale that terminate in hot spots.\footnote{Recently, a new class was
introduced to describe the radio sources which lack extended emission
\citep[FR0,][]{sadl14,bald15}.}

There is a large overlap between sources classified as
FRI/FRII and those classified as LERG/HERG
\citep{evan06,hard07,best12,heck14,ming14}, although some FRIs are HERGs (e.g.,
Perseus, M87) and some FRIIs are LERGs \citep{gend13}. Recently, \citet{turn15}
developed a model which incorporated both FRII \citep{kais97} and FRI models
\citep{luo10} and supports the conclusion that differences between LERGs and
HERGs is due to differences in the accretion mechanism.

Using a monochromatic luminosity of $L_{843} \sim 5 \times 10^{32}$ erg s$^{-1}$
Hz$^{-1}$ to classify our sources as FRI or FRII, we find that in our
sample there are 15 sources with FRII-like radio power (see Table
\ref{lum_table} and Figure \ref{F:LX500__vs_lrad}). Because of the large beam
size of SUMSS images of $\approx 40$ arcsec \citep{mauc03}, the cRSs in our
sample are mostly unresolved and we cannot therefore distinguish between FRI or
FRII sources based on their radio morphology. The only clear case when the cRS
is resolved is in SPT-CL J0542-4100 (or RDCS J0542-4100 at $z=0.64$), where the
central source is $\sim$ 600 kpc across, but the image is inconclusive.

All of the high-power sources are
also higher-redshift sources ($z \gtrsim 0.6$). Four of them are hosted by
clusters classified as CFs (the Phoenix cluster, SPT-CL J2106-5845, SPT-CL
J0000-5748, and ACT-CL J0616-5227), all of which have possible cavities
\citep[this paper and][]{hlav15,mcdo15}.
The remaining 11
high-power sources are in clusters classified as NCFs, some of which show signs
of merging activity (e.g.,  SPT-CL J2245-6206, SPT-CL J0449-4901, SPT-CL
J2218-4519, see Figure \ref{images_no_cav}). We will discuss the relation between
the cluster state and the radio properties in more detail in Section~\ref{disc:tcool_lrad}.

Figure \ref{F:LX500__vs_lrad} \emph{left} shows that the higher redshift sources
($z>0.6$) have on average higher radio luminosity than lower redshift sources.
As we noted above, all the FRII-like cRSs are at $z\gtrsim 0.6$, and there
are only few sources with high radio luminosity, $L_{843}$ $>$ 2.7 $\times$
10$^{32}$ erg s$^{-1}$ Hz$^{-1}$, at lower redshift, $z<0.6$.
This is the case for SPT-CL J2344-4242 (the Phoenix cluster) at $z=0.595$, SPT-CL J2245-6207
at $z=0.58$ (with a radio luminosity above 10$^{33}$ erg s$^{-1}$ Hz$^{-1}$,
they  are part of the 15 most powerful radio sources discussed above), ACT-CL
J0215-5212 at $z=0.48$, SPT-CL J0456-5116 at $z=0.562$ with a radio luminosity
above $4 \times$ 10$^{32}$ erg s$^{-1}$ Hz$^{-1}$, and ACT-CL J0346-5438 at
$z=0.53$, SPT-CL J0307-6225 at $z=0.59$ and SPT-CL J0234-5831 at $z=0.415$ with
a radio luminosity above $2.7 \times$ 10$^{33}$ erg s$^{-1}$ Hz$^{-1}$.

An explanation for the higher radio luminosity in the higher-redshift systems might be
that, due to the fact that these systems tend to be younger and less relaxed than
similar systems at lower redshift, there is more merging activity that
contributes to the triggering of the radio activity \citep[see Section
\ref{disc:rlf} and][]{bran06}. This increase of the merging activity with redshift is
also supported by the commensurate increase in the SFRs at higher redshift \citep[see Section
\ref{disc:SFR} and][]{mcdo16}.

\subsection{Radio luminosity functions\label{disc:rlf}}

The radio luminosity function (RLF) encapsulates the fraction of sources in a
sample that possess a radio source of a given luminosity, and as such is a
sensitive tool for detecting evolutionary effects in the radio properties
\citep[e.g.,][]{bran06,gral11,somm11,prac16}. In this work, as we are interested only in the
central source (that responsible for feedback), we calculate the differential
RLF for only the radio source associated with the BCG that lies at the X-ray
core, in contrast to, e.g., \citet{ledl96} and \citet{bran06} who calculate it for all
radio sources within a distance $R < 0.2 R_{\rm A}$, where $R_{\rm A}$ is the
Abell radius. Therefore, our RLF may be interpreted as being the number of
central radio sources per cluster per luminosity bin.

We calculate the RLF following the approach of \citet{bran06}.
For each of the 44 clusters with a detected
central radio source, we calculate its contribution to the RLF as:
\begin{equation} W_{\textrm{RLF,} i} = 1 / N_{\textrm{cl,} i}, \end{equation} where
$W_{\textrm{RLF,} i}$ is the contribution of source $i$ and $N_{\textrm{cl,} i}$ is the
number of clusters in which source $i$ could have been detected, given its peak
flux density and the sensitivity limit of the radio observations (see Section
\ref{radio_props} for details of the radio data). As in \citet{bran06}, we
adjust the bin size of the lowest-luminosity bin so that all bins contain at
least 2 sources, and we scale the normalisation of this bin so that it matches
the other bins (which have a size of 0.5 dex). Furthermore, to allow a direct
comparison between the results of \citet{bran06} and our results, we
recalculated the RLF of \citet{bran06} using our method of considering only the
central radio source, rather than all sources within $R/R_{\rm A} < 0.2$.

\begin{figure} \includegraphics[width=84mm]{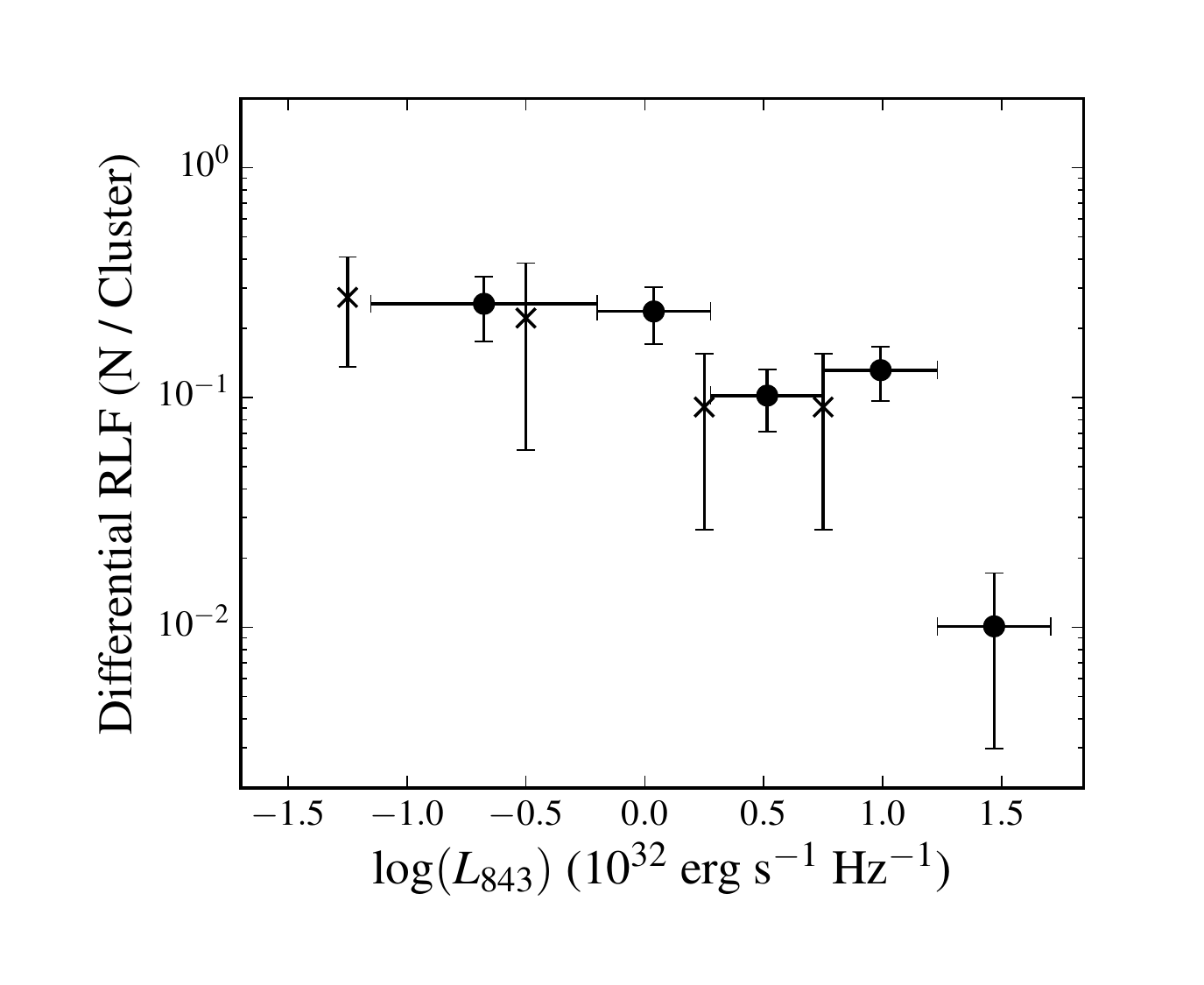} \caption{The
differential RLF at 843 MHz for our SPT sample (filled circles) and the sample
of \citet{bran06} of $0.3 < z < 0.8$ X-ray selected NEP clusters (crosses),
adjusted for our method of calculating the RLF.} \label{F:rlf_all} \end{figure}

In Figure \ref{F:rlf_all}, we plot the RLF at 843 MHz for our sample and that of
\citet{bran06}. The \citet{bran06} sample of 18 clusters was constructed by
selecting all clusters with $z > 0.3$ from the ROSAT North Ecliptic Pole (NEP) catalog of
\citet{gioi03}. Our results agree fairly well with those of \citet{bran06} over
the range of luminosities for which there is overlap. \citet{bran06} found that
their RLF did not possess the high-luminosity break seen in local RLFs
\citep[e.g.,][]{ledl96,sadl02,best05,mauc07}. In contrast, we do see evidence
for a break around a luminosity of $L_{843} \approx 10^{33}$ erg s$^{-1}$
Hz$^{-1}$, considerably above the highest luminosity probed by \citet{bran06}
and higher than the break luminosity seen in local RLFs of $L_{843} \approx
10^{32}$ erg s$^{-1}$ Hz$^{-1}$ \citep[after converting from 1400 MHz to 843
MHz;][]{ledl96}.

\begin{figure*} \begin{tabular}{@{}cc}
\includegraphics[width=84mm]{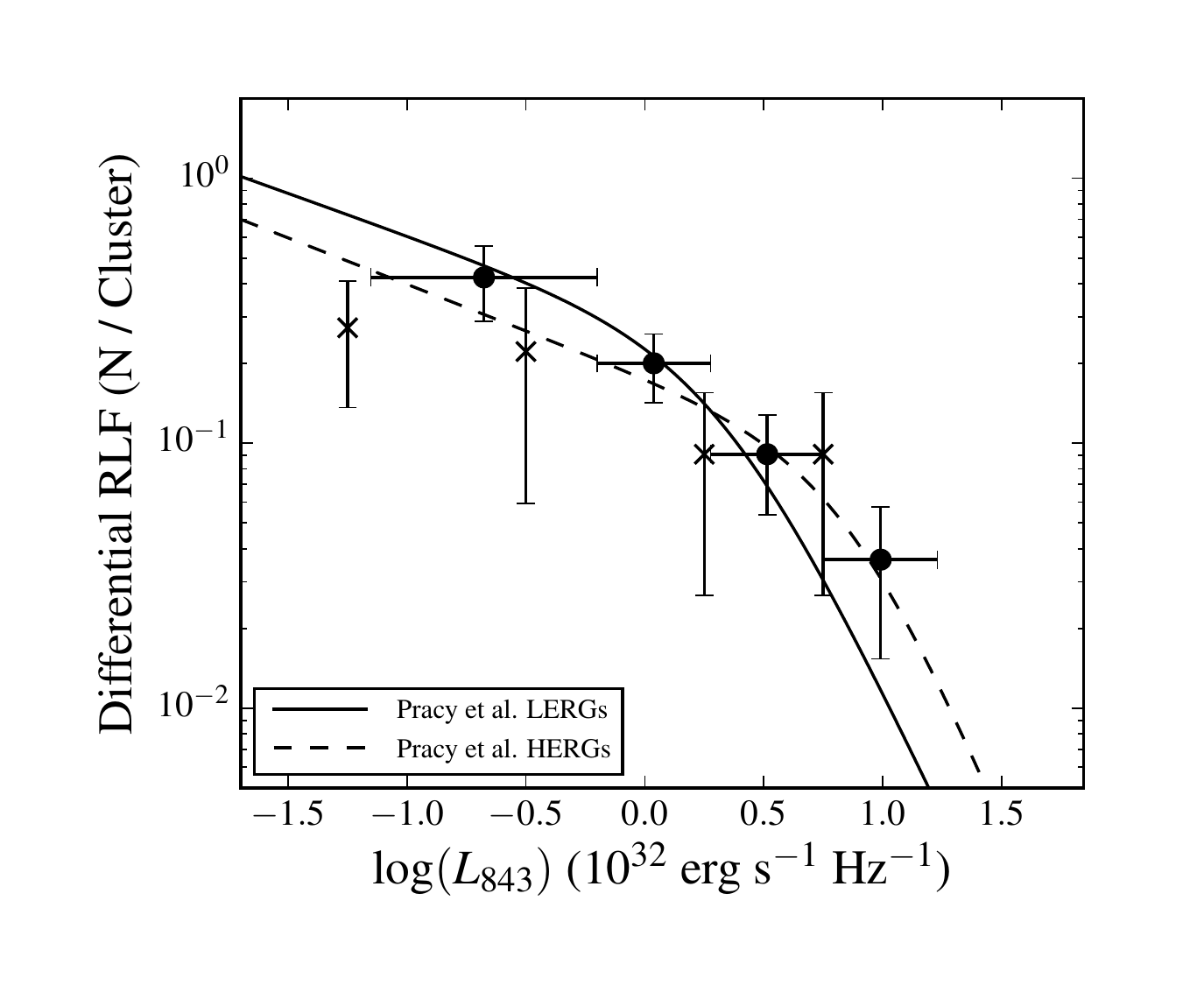} &
\includegraphics[width=84mm]{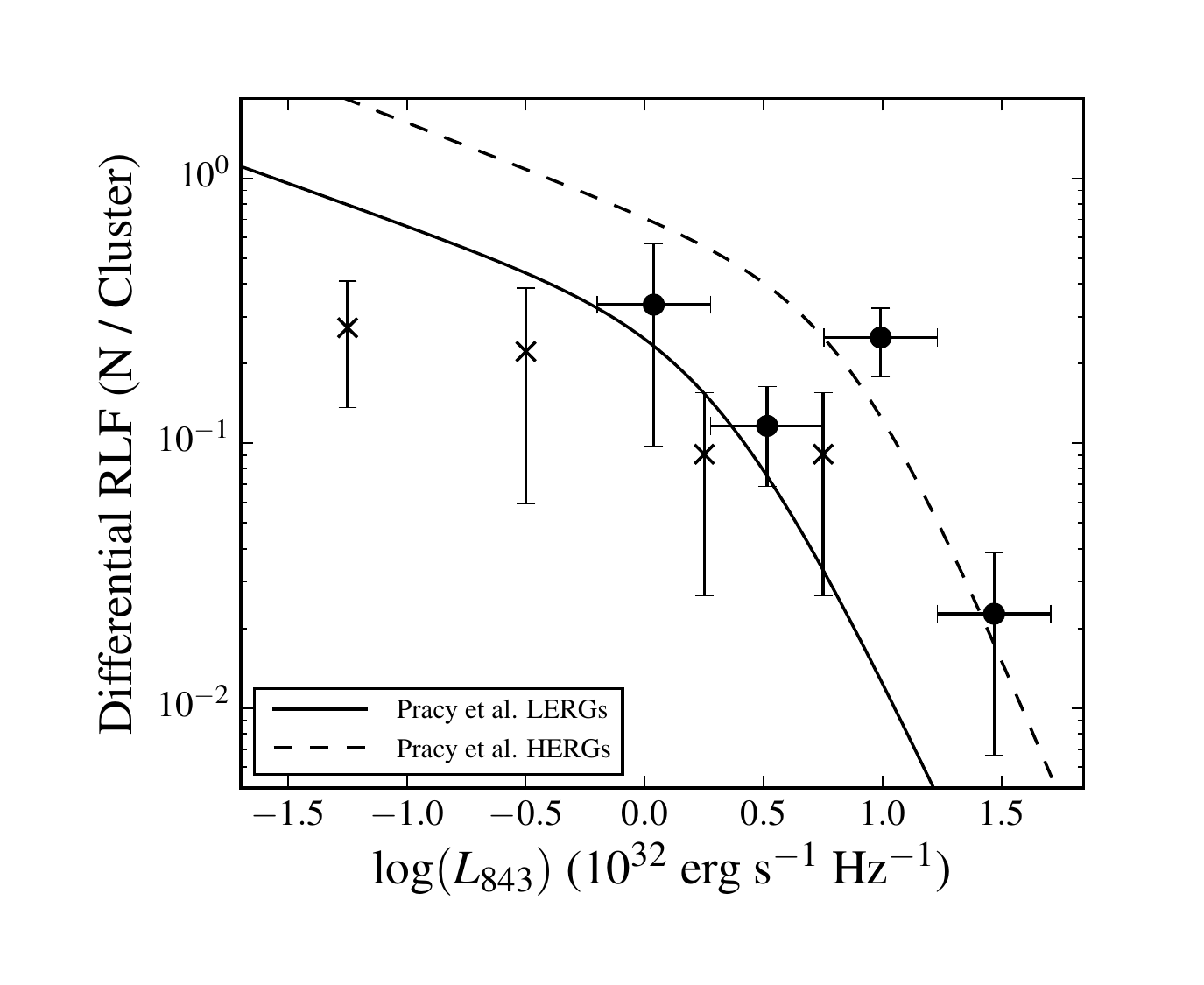} \\ \end{tabular} \caption{The
differential RLF for $z < 0.6$ SPT clusters (\emph{left}) and $z > 0.6$ SPT
clusters (\emph{right}). Symbols are the same as those in Figure
\ref{F:rlf_all}. The lines show the RLFs of \citet{prac16} for LERGS (\emph{solid
lines}) and HERGs (\emph{dashed lines}), calculated with the median redshift of
the sample and normalised so that the low-$z$ LERG relation matches the value of
the SPT-ACT RLF in the lowest-luminosity bin.} \label{F:rlf_zlt_zgt}
\end{figure*}

To better understand the origin of this break, we plot in Figure
\ref{F:rlf_zlt_zgt} the RLF in two redshift ranges ($z < 0.6$ and $z > 0.6$).
For comparison to the overall population of radio sources, we plot the
parameterisations of the RLFs determined by \citet{prac16} for LERGs and HERGs
for pure luminosity evolution, calculated for the median redshifts of the two
samples. To account for the different normalisations between our RLF and that of
\citet{prac16}, we normalise the relation of \citet{prac16} so that their LERG
relation matches the value in the lowest-luminosity bin of our $z < 0.6$
sample.\footnote{A single normalisation is appropriate, since the per-source
volume and magnitude weights used in \citet{prac16} are constant across our
(volume-limited) sample \citep[see, e.g.,][]{yuan16b}.} We choose this point for
the normalisation because the systems in this luminosity bin should be comprised
almost entirely of LERGs (alternatively, we could use the highest-luminosity bin
for the normalisation and normalise the HERG relation to fit it instead of the
LERG one, but this would give a similar value). We use a single normalisation
factor across all four relations (low- and high-redshift LERGs and HERGs) and
calculate the relations using the median redshift of each sample. One can see
from the lower-redshift plot of Figure \ref{F:rlf_zlt_zgt} that the lower
luminosity bins of our RLF follow the LERG relation well, while the
higher-luminosity ones match the HERG relation well, suggesting that there is a
transition between the two source types that blurs the break seen in local
samples \citep[e.g.,][]{ledl96}.

At higher redshifts ($z > 0.6$), shown in Figure \ref{F:rlf_zlt_zgt} right, we
see further support for this LERG-HERG dichotomy and evidence for strong
evolution in the HERG population. At these redshifts, the fraction of clusters
with a central radio source in the $L_{843} \approx 10^{33}$ erg s$^{-1}$
Hz$^{-1}$ bin is a factor of $\approx 7$ larger than the fraction in the same
bin at lower redshifts, a $\approx 3$-$\sigma$ difference. Furthermore, the RLF
in the highest luminosity bins match well the expected values from the HERG
relation of \citet{prac16}, calculated for the median redshift of the sources in
our sample at $z > 0.6$. Therefore, it appears that the RLF of the central BCG
agrees with that of the overall radio source population in the same redshift
range \citep[see][]{dunl90,will01,grim04,sadl07,best12,best14,prac16}, perhaps because
powerful, radio-loud sources in general tend to be located in cluster cores
\citep[see][]{hill91,mand09,wyle13}. Our results are also consistent with those of
Sommer at al. (2011), who found evidence for strong luminosity evolution in cluster radio galaxies

Therefore, the more powerful sources in our sample, which appear preferentially at higher
redshifts, are consistent with being HERGs. Merging may play an important role
in triggering these HERGs, as many of the high-redshift systems show signs of
recent minor mergers in both the SPT sample \citep{mcdo16}, and the NEP sample
\citep{bran06}. Additionally, it might be that minor mergers are more effective
in coupling the AGN to the cold gas than in the local universe
\citep{kavi15,shab17}.

\subsection{Cooling Time/Thermal Stability Parameters and Radio
Luminosity\label{disc:tcool_lrad}}

\begin{figure*}
\begin{tabular}{@{}cc}
\includegraphics[width=84mm]{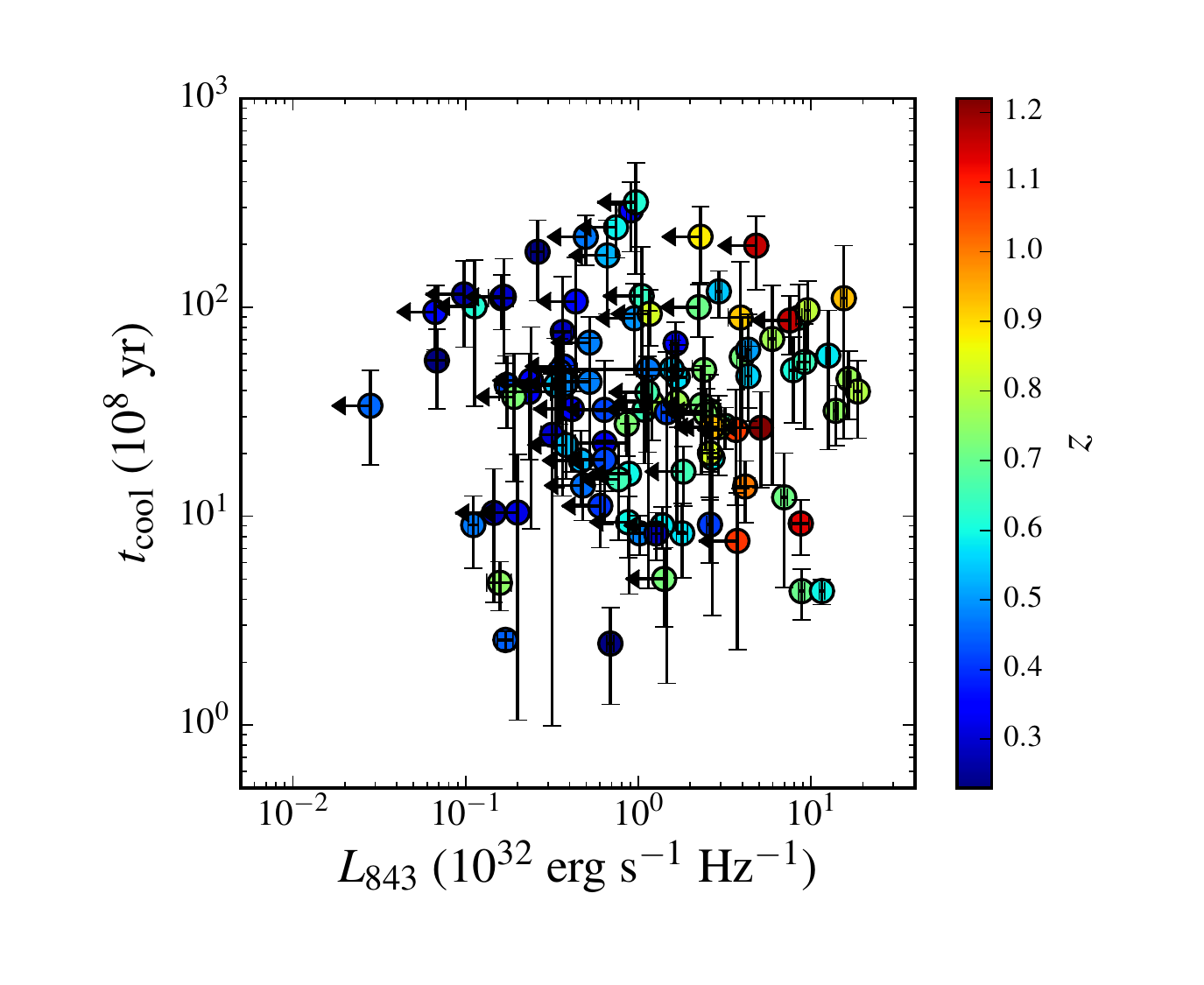} &
\includegraphics[width=84mm]{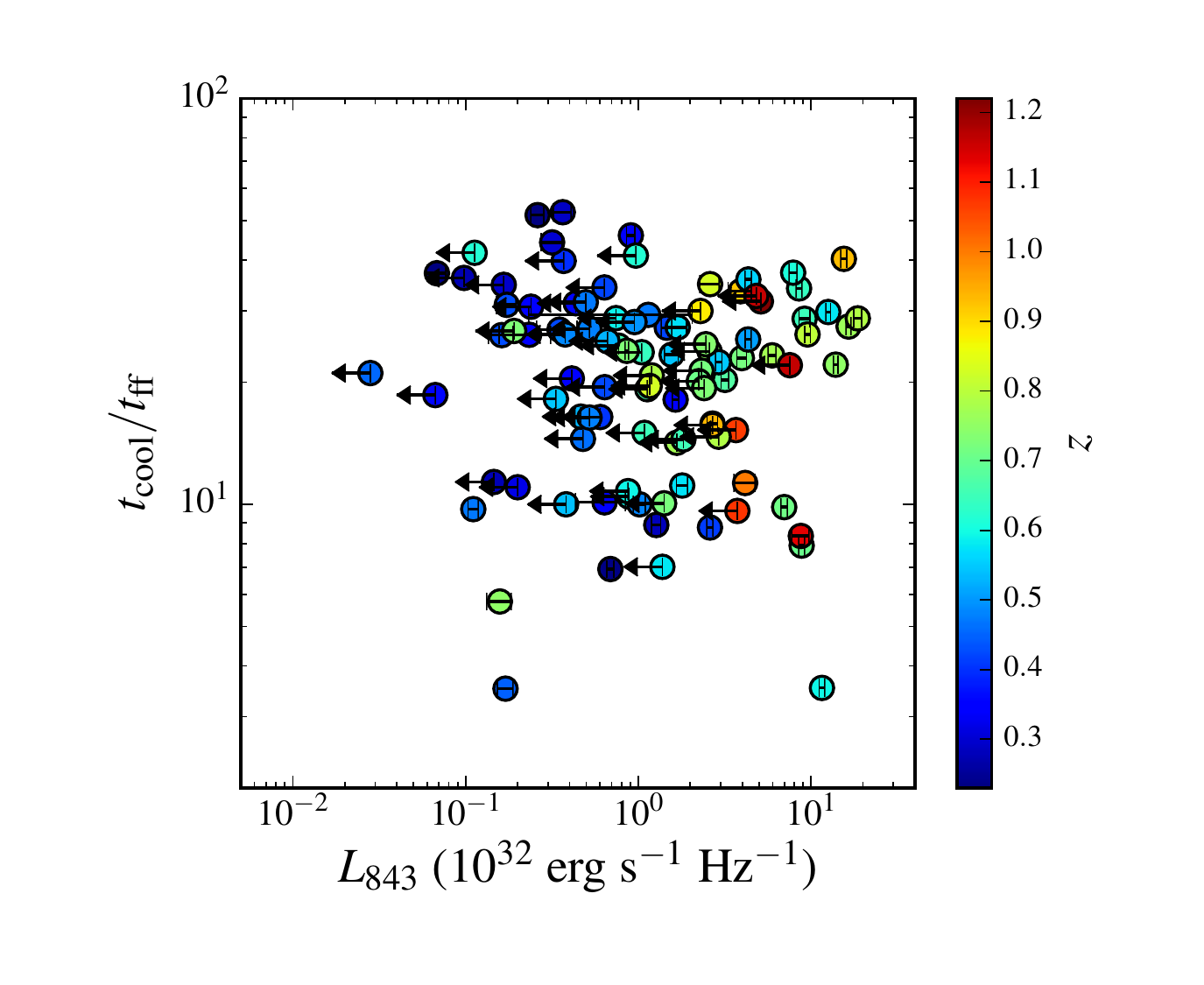} \\
\end{tabular}
\caption{\emph{Left}: Central cooling time (at 10 kpc) versus the rest-frame monochromatic
843 MHz radio luminosity. \emph{Right}: The ratio of central
cooling time and free fall time versus the radio
luminosity. The colour indicates the redshift in both panels.}
\label{F:tcool_tff}
\end{figure*}

\begin{figure}
\includegraphics[width=84mm]{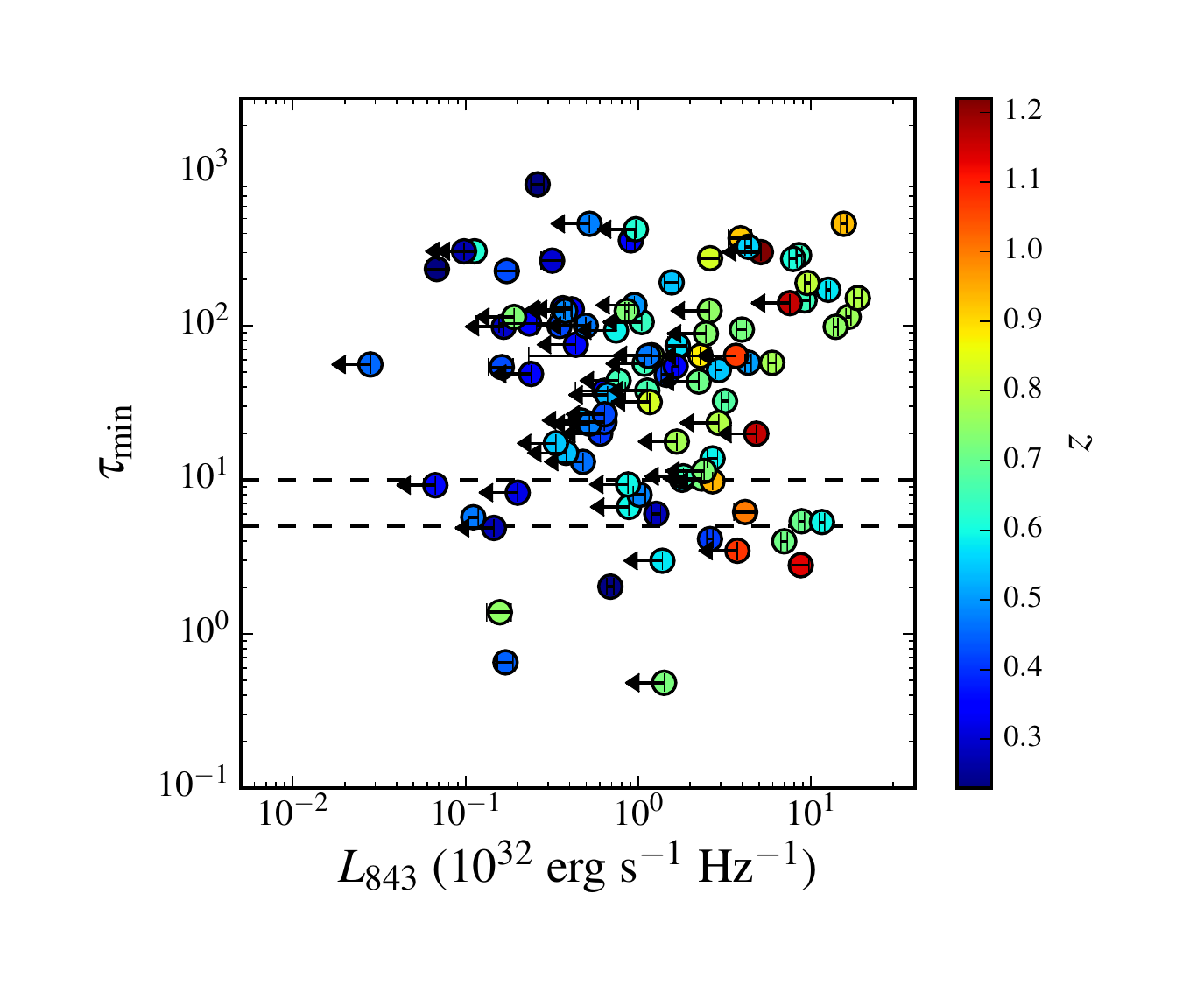}
\caption{Thermal stability parameter versus the rest-frame
monochromatic 843 MHz radio luminosity.  The colour indicates the redshift. The lines denote values of five and ten.}
\label{F:tcool_instab_vs_lrad}
\label{F:instab}
\end{figure}

In  Figure \ref{F:tcool_tff} and Figure \ref{F:instab} we plot the central cooling time
at 10 kpc  (Figure \ref{F:tcool_tff} \emph{left} panel), the ratio of central cooling time versus the
free fall time
(Figure \ref{F:tcool_tff} \emph{right} panel) and the minimum instability criterium  (Figure \ref{F:instab})
versus the monochromatic 843 MHz radio luminosity (see Table
\ref{Xray_table}). In the cases for which the central
cooling time extrapolation did not work (e.g., the surface brightness profile is
too noisy or there is significant substructure that is inconsistent with the
deprojection method, see Section \ref{Sec:cf}), we plot the cooling time of the
inner bin derived from the deprojection.  As expected from the mostly
flux-limited nature of the radio data, the lower-luminosity half of the plots is
dominated by the lower-redshift systems ($z<0.6$) and the higher-luminosity half
by the higher-redshift systems ($z>0.6$).

The sampled radio luminosities cover only part of the range sampled by the B55
and HIFLUGS samples \citep{birz12}, since, due to the generally higher redshifts
and lower sensitivity of the radio data (e.g., SUMSS), we cannot probe the behavior of the
lower-luminosity sources in our sample.\footnote{In the B55/HIFLUGCS samples we
found that the lower-luminosity systems tend to be found in NCFs and lower-mass
systems (ellipticals and groups).} Therefore, our sample only probes radio
luminosities above the threshold seen by \citet{birz12} for NCFs of $L_{843} \sim 4
\times 10^{30}$ erg s$^{-1}$ Hz$^{-1}$ (see Section \ref{disc:lx_lrad}). Strikingly,
in contrast to nearby samples \citep{birz12}, there is
no apparent difference in the distribution of radio luminosities between CF and
NCF systems above this threshold. The reason for this is the presence of powerful radio sources in
the NCFs in our sample, which are generally lacking in local samples.

\subsection{Thermal Stability and Star Formation\label{disc:SFR}}

\begin{figure*} \begin{tabular}{@{}cc}
 \includegraphics[width=84mm]{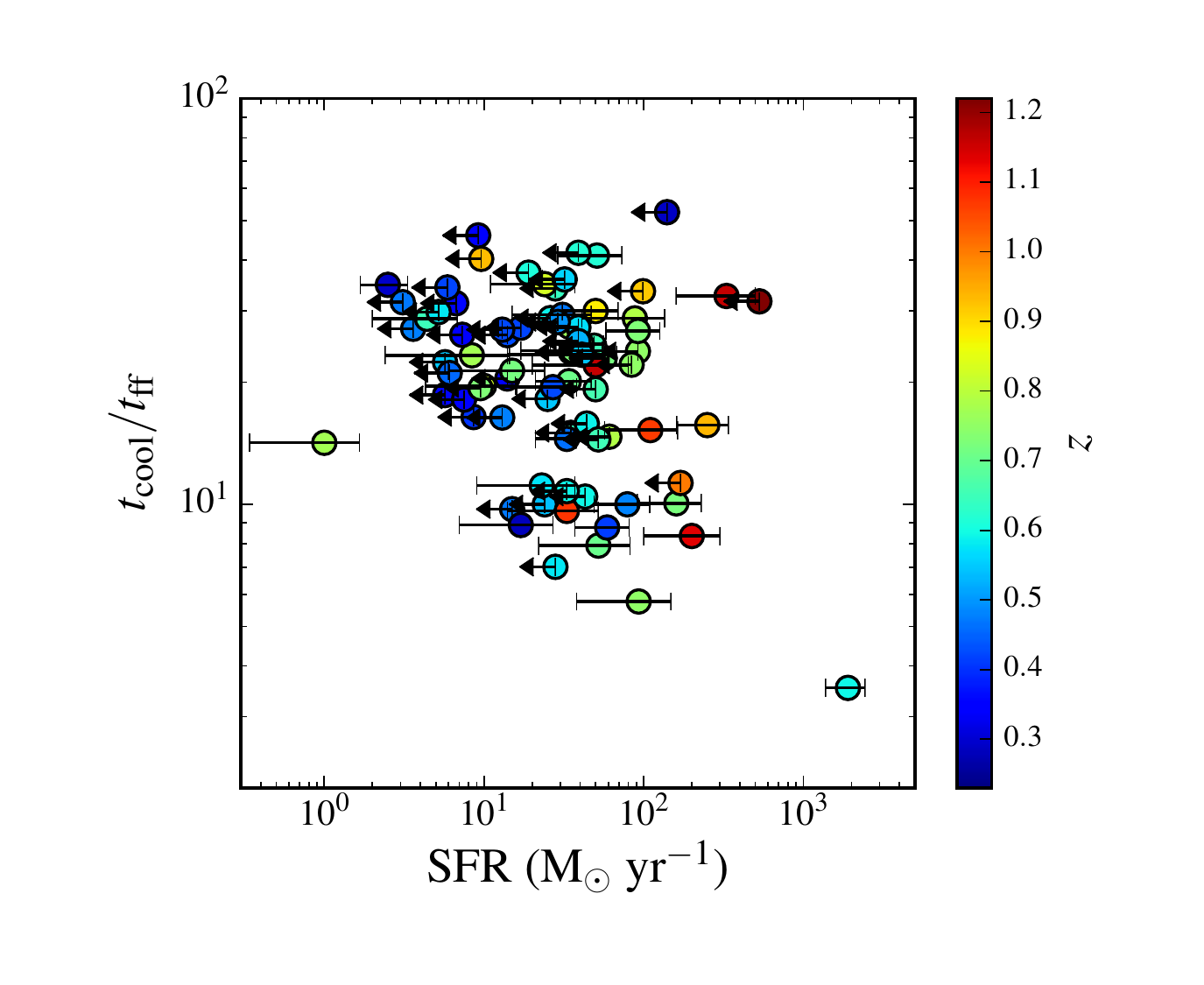} &
 \includegraphics[width=84mm]{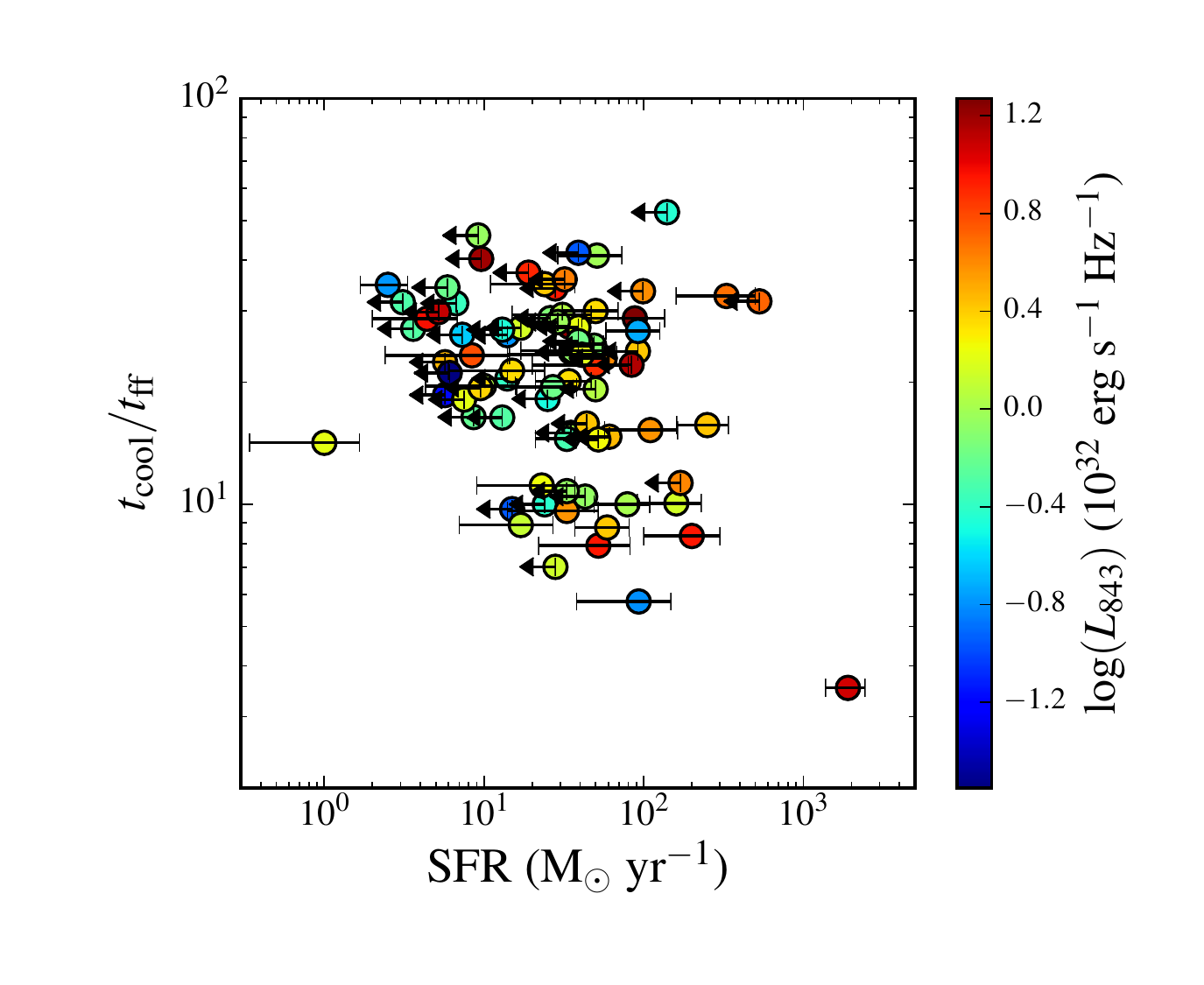} \\ \end{tabular}
 \caption{The ratio of cooling time and free fall time versus the SFR.
 The colour represents the redshift (\emph{left}) and
 radio luminosity (\emph{right}).} \label{F:tff_SF} \end{figure*}

Figure \ref{F:tff_SF} shows the thermal stability parameter (the cooling time
versus the free-fall time) versus the SFR. There is no clear thermal-stability
threshold below which star formation occurs, as is seen in nearby systems
\citep{raff08,voit08}. However, the cooling flow systems tend to have higher SFR
rates in general ($SFR>20$ M$_{\odot}$ yr$^{-1}$). Additionally, systems with
higher SFRs tend to lie at higher redshift.

The systems with the highest SFRs are found both in CF systems (e.g., the
Phoenix cluster, SPT-CL J2106-5845, and SPT-CL J2043-5035) and NCF systems
(e.g., SPT-CL J0446-5849, SPT-CL J2345-6406, and SPT-CL J0547-5345), and as the
left panel of Figure \ref{F:tff_SF} shows, they are some of the highest-redshift
systems. For example, the highest SFRs in NCF systems are in SPT-CL J0446-5849
at $z=1.16$ and  SPT-CL J2345-6406 at $z=0.94$.  SPT-CL
J2345-6406 shows some struczure in the X-ray image (see Figure \ref{images_no_cav}),
which could be due either
to cavities or merging activity; and the lower central temperature (kT $\sim$ 4)
and peaked SB are indications that some of these high-redshift systems with high
SFRs might be in a cooling stage but lack sufficient counts to be securely
detected as a cool core (see also Section \ref{disc:tcool_lrad}). In a similar
situation are SPT-CL J0534-5005 at $z=0.881$ and SPT-CL J2236-4555 at $z=1.16$.

The main question that arises from Figure \ref{F:tff_SF} is why we do not see
the SFR threshold as in nearby samples. A possible explanation is that the
systems with high SFRs do have shorter cooling times but the X-ray data were
insufficient to detect them. However, among the most likely such systems (see
Section \ref{morph} and \ref{disc:tcool_lrad}), only a few have detections of
SFR $>20$ M$_{\odot}$ yr$^{-1}$ (e.g., SPT-CL J0509-5342 and SPT-CL J0058-6145).
Additionally, it is possible that minor mergers are more common at higher
redshifts and these trigger bursts of star formation \citep{mcdo16}. Between the
systems with detected high SFRs and possible evidence of interactions are SPT-CL
J0547-5345 at $z=1.067$, SPT-CL J0406-4804 at $z=0.737$ and SPT-CL J2035-5251 at
$z=0.424$ (see Figure \ref{images_no_cav}).

Additionally, in right panel of Figure \ref{F:tff_SF}, there is no clear dependence
of the SFR or cooling state on the radio luminosity. For example, the
highest-luminosity systems are distributed fairly randomly and appear in both
CFs (e.g., SPT-CL J2106-5845, SPT-CL J0000-5748, and SPT-CL J2344-4242) and NCFs
(e.g., SPT-CL J0449-4901-with 2 BCGs, see Figure \ref{images_no_cav}, and SPT-CL
J0058-6145, with possible cavities, see Figure \ref{images_cav}). Interestingly,
a few of the strongest CF clusters have the highest SFRs in our sample
and harbour some of the most powerful cRSs (e.g; the Phoenix cluster and SPT-CL J2106-5845).

\subsection{Quenching cooling\label{agn_feedback}}

For systems with visible cavities in \emph{Chandra} images,  it was found that
AGN heating is sufficient to balance cooling losses from the X-ray emitting gas
within the cooling radius in at least 50 per cent of the systems \citep{raff06}. This
sample was mostly composed of nearby systems, but 10 out of a total of 33
systems were at $z>0.1$ (the highest at $z=0.545$). Furthermore, \citet{hlav12}
increased the number of higher-redshift ($0.3<z<0.5$) sources using the MACS
luminous cluster sample and found that, as in nearby cluster samples, AGN
feedback supplies enough energy to balance the cooling in the inner regions of
the cluster. To investigate whether AGN feedback is enough to balance cooling
out to a redshift $z=1.2$, we use the monochromatic radio luminosity to estimate
a cavity power using the radio-to-jet power scaling relations of \citet{cava10}.
We adopt a spectral index of $\alpha = 1.0$ to transform our 843
MHz radio powers to 1400 MHz powers.

\begin{figure}
\includegraphics[width=84mm]{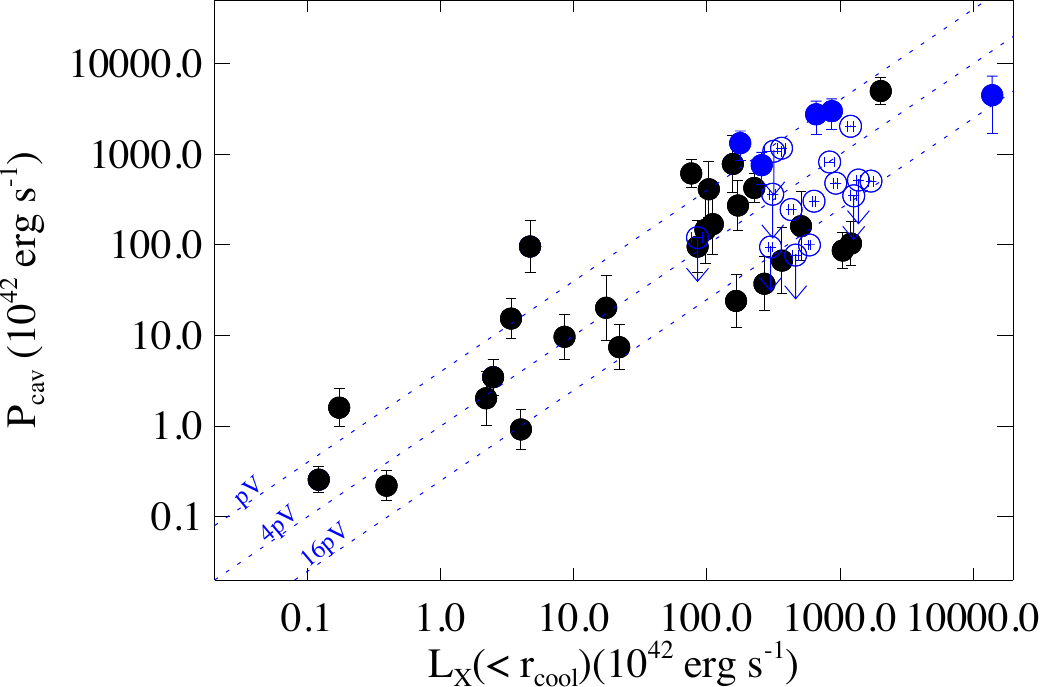} \\
\caption{Cavity power of the central AGN versus the X-ray luminosity of the
ICM inside the cooling region. The filled symbols are the
measured cavity power from the HIFLUGCS sample \citep[\emph{black} symbols;][]{birz12} and
SPT sample \citep[\emph{blue} symbols;][]{hlav15,mcdo15}, and the open
symbols are the estimated cavity power using the scaling relations between
monochromatic 1.4 GHz radio luminosity and the cavity power of
 \citet{cava10}. The diagonal lines indicate $P_{\rm{cav}}=L_{\rm{X}}$ assuming
 $pV$, $4pV$ or $16pV$ as the energy deposited.}
\label{F:feedback} \end{figure}

The \citet{cava10} relation is an extension of the \citet{birz08} relation, with
the addition of 21 nearby elliptical systems from \citet{nuls09} at the
lower-luminosity end, and is given by:
\begin{equation}
\log(P_{\rm{cav}})=1.91+0.75 \times \log(L_{1400}),
\end{equation}
where $P_{\rm{cav}}$ is in units of 10$^{42}$ erg s$^{-1}$, and $L_{1400}$ in
units of 10$^{40}$ erg s$^{-1}$ ($\approx$ 10$^{24}$ W Hz$^{-1}$). We plot in
Figure \ref{F:feedback} the measured or predicted cavity power versus the total
X-ray luminosity of the intracluster gas within the cooling radius (see Section
\ref{analysis}) for the 20 CF systems in our sample. For comparison with previous
results from the the nearby universe, these predicted values are overplotted
with the measured values from the HIFLUGCS sample \citep[see Figure
\ref{F:feedback} and][]{birz12}. In general, the measured and predicted jet
powers are sufficient to balance cooling in $\approx 50$ per cent of the high-redshift CF systems in our
sample, similar to local samples \citep{birz12}.

We also investigated the radio-to-jet power scaling relation of \citet{godf16}.
This relation was computed using a multivariant regression between cavity power
($P_{\rm{cav}}$), radio luminosity at 1.4 GHz ($L_{1400}$) and distance
($D_{_{L}}$) using combined cavity samples of \citet{birz08}, \citet{cava10} and
\citet{osul11}. Generally, it predicts more jet power at low radio luminosities
and less jet power at higher luminosities. For our sample, the predicted jet
powers from this relation are a factor of $\sim 6$--8 times less than those
predicted by the relation of \citet{cava10} or measured directly \citep[][and
this paper]{hlav15,mcdo15}.

As pointed out by \citet{godf16}, both the scaling relations of \citet{cava10}
and \citet{godf16} are inconsistent with models of radio-source evolution: one
expects a slope of 0.5 from buoyancy arguments and a slope of 0.8 from FRII
expansion models \citep{will99,ines17}.  However, \citet{birz08} found that accounting
for the effects of spectral aging on the observed radio luminosity gives a slope
of $\sim 0.5$, bringing the observed scaling between jet power and radio
luminosity into agreement with the buoyancy models. Basically, these aging
effects result in a steepening of the relation, as lower-luminosity sources tend
to have more spectral aging than the higher-luminosity sources. \footnote{The
monochromatic scaling relations at lower frequencies, such as at 327 MHz
\citep{birz08} or 140 MHz \citep{koko17} also have a slope of
$\sim 0.5$. Although low-frequency observations are generally less sensitive to
spectral aging effects than higher-frequency observations, they still suffer
from the overall dimming of the radio emission. In order to account for the
effects of this dimming and the spectral shape on the scaling relations,
information on the spectral age (e.g., the break frequency) is required.}

\section{Summary and Discussion \label{summ}}

In this paper we use a large sample of  SZ-selected
clusters from the southern hemisphere and galactic plane (the SPT and ACT
samples) to investigate AGN feedback to high redshift ($0.3<z<1.2$).
To identify the CF systems in our sample, we use the instability criterion whereby
systems with $t_{\rm cool}/t_{\rm ff} \lesssim~10$ are considered unstable to
cooling (resulting in 22 CF systems; see Section \ref{morph} for a description of
these systems). In this case the instability criterion of 10 corresponds roughly
to a inner cooling time of 2 $\times$ 10$^{9}$ yr (with 2 exceptions, see
\ref{Sec:cf}), but not all the systems with $t_{\rm{cool}}$(inner)~$\lesssim~2
\times 10^{9}$~yr have $t_{\rm cool}/t_{\rm ff} \lesssim~10$ (these are the
intermediate systems described in Section \ref{morph}). The reason for such a
high cooling time value for the separation between CF systems and NCF systems is
the fact that for such a high redshift sample we are not able to probe the
densest gas close to the core. As a result, the separation based on cooling time
does not work as well as for the nearby systems (see the left panel of Figure
\ref{F:tcool_tff}).

We investigate whether the locally-derived scaling relation between cavity power
and radio luminosity \citep[e.g.,][]{birz04} holds at higher redshifts. Direct
measurement of AGN feedback power through cavities is possible for only few
systems in our sample, e.g., the Phoenix cluster \citep{mcdo13b}
and SPT-CL J2031-4037 (this paper, see Section \ref{cav}). There is evidence of
X-ray structure in some other systems, which might be due to cavities (e.g.,
SPT-CL J2135-5726, SPT-CL J2222-4834, among many others, see Figure
\ref{images_cav} and Section \ref{cav}), but because of the shallow nature of
the X-ray and radio data, many of these potential X-ray cavities are uncertain.
We therefore use radio data from the SUMSS survey
\citep{bock99} and SFRs from \citet{mcdo16}, in addition to the \emph{Chandra}
X-ray data, to investigate AGN feedback. As discussed in the introduction, the SFR
is an important ingredient in the feedback process and traces the imperfect
balance between heating and cooling \citep[see the discussion in][]{voit16}.

Below we summarise our main results:

\begin{enumerate}
\item We find evidence in the RLF of the cRS,
calculated at 843 MHz, for strong evolution in the higher-luminosity sources,
such that the fraction of sources hosting a cRS of $L_{843}$ $\sim$ 10$^{33}$
erg s$^{-1}$ Hz$^{-1}$ is $\approx 7$ times higher in our sample of $z>0.6$
sources than in the $z<0.6$ sources. This evolution is consistent with other
studies \citep[e.g.,][]{prac16} of the general galaxy population that find that
high-power HERGs are much more common at high redshifts than at lower redshifts.
We argue that the underlying cause of the break is therefore the same as that in
the general population, namely it is due to differences of the accretion
mechanism onto the SMBH in the low- and high-luminosity sources
\citep{best14,ming14,fern15}. We postulate that mergers likely have an important
influence on this accretion mechanism \citep[see also][]{hard07,ramo12}, since many of the high-redshift systems
appear to be going through minor mergers \citep{mcdo16}, and minor mergers are
thought to be more effective in coupling the AGN to the cold gas than in the
local universe \citep{kavi15,shab17}.

In support of this scenario, we found that there are 15 sources with FRII-like
radio power in our sample (see Figure \ref{F:LX500__vs_lrad}), all of which lie
at $z>0.6$ and many of which are located in likely NCF clusters, which generally
show increased merging activity compared to CF clusters. Furthermore, the
increased SFRs at higher redshifts seen in our sample  \citep[see Section
\ref{disc:SFR} and][]{mcdo16} are consistent with a commensurate increase in the
merging activity.

\item We do not find a clear separation between CF and NCF systems based on the
radio luminosity or SFR as is observed in the nearby universe, where NCF systems
tend to have low  radio luminosities \citep{sun09,birz12} and little to no star
formation \citep{raff08,cava08,voit08}. However, the CF systems in our sample do
tend to have higher SFRs in general than the NCF systems (SFR$>20$
M$_{\odot}$ yr$^{-1}$).

\item We find that the predicted degree to which AGN feedback can balance the
cooling losses in our sample depends on the relation used to transform from
radio luminosity to jet power, with the relation of \citet{cava10} implying that
the high-redshift CF systems from SPT/ACT sample have enough energy power to
balance cooling, and the relation of \citet{godf16} implying that these
high-redshift systems  have at least four times too little energy to balance
cooling. However, the few direct measurements of cavities that exist for our
sample \citep[e.g., the Phoenix cluster,][]{mcdo15} agree with the values
predicted by the \citet{cava10} relation. Further cavity measurements in these
systems are required to determine which relation is the more correct one.

Additionally, many of these high redshift systems are luminous radio sources,
lying at the transition between FRI and FRII sources. Furthermore,
\citet{mcdo15} shows that the Phoenix cluster lies at the transition between
radio-mode and quasar mode AGN activity \citep{chur05,russ13,hlav13}. Therefore,
we cannot exclude the possibility that, for these powerful radio sources, one
needs to use a different radio-to-jet power relation \citep[e.g., the FRII
relation with a slope of 0.8;][]{will99}. In particular, a relation that
includes spectral information, such as that derived by \citet{birz08}, would
help to mitigate the effects of inhomogeneous radio source populations and
spectral aging on the predicted jet power.

\end{enumerate}

\begin{landscape}
\begin{table*}
\caption{X-ray Properties}
\label{Xray_table}
\scalebox{0.7}{
\begin{tabular}{@{}lccccccccccccc}
 & Obs. ID$^b$ & $t^c$ &  \multicolumn{2}{c}{X-Ray Core (J2000)}$^d$ & $kT$$^e$ & $n_{\rm e}$$^f$  & $r^g$ & $\Delta (kT)$$^h$  &
$t_{\rm cool}$$^i$  & $t_{\rm cool}$ (10 kpc)$^j$ & $\frac{t_{\rm cool}}{t_{\rm ff}}$$^k$ & $\eta_{\rm min}$$^l$  \\
\cline{4-5}
System$^a$ & & (ks) & RA & DEC &  (keV) & (cm$^-3$ ) & (kpc) &  & (Gyr)  & (Gyr) &  &  \\
 \hline
SPT-CL J0000-5748 & 9335 & 29.6 & 00 00 59.97 & -57 48 33.66 & $5.46^{+1.46}_{-0.96}$ & $0.040^{+0.002}_{-0.002}$  & 24.6 &  $1.65^{+1.22}_{-0.74}$ & $1.08^{+0.40}_{-0.29}$  &  $0.44^{+0.60}_{-0.12}$  &  7.92 & 5.40  \\
SPT-CL J0013-4906 & 13462 & 11.5 & 00 13 19.65 & -49 06 38.11 & $5.63^{+4.12}_{-1.93}$ & $0.0056^{+0.0002}_{0.0004}$  & 72.0 & \ldots & $9.27^{+4.11}_{-3.31}$  & $1.12^{+0.51}_{-0.41}$ & 16.41  & 20.03  \\
ACT-CL J0014-0056 & 16228 & 28.3 & 00 14 54.47* & -00 57 04.9* &  $5.96^{+2.88}_{-1.61}$ & $0.013^{+0.001}_{-0.001} $ & 43.5 & $1.88^{+1.35}_{-1.05}$ & $3.96^{+1.67}_{-1.45}$  & $1.85^{+0.81}_{-0.71}$ & 16.45  &  24.36  \\
SPT-CL J0014-4952 & 13471 & 46.5 & 00 14 50.1* & -49 52 54.24* & $7.90^{+4.23}_{-2.52}$* & $0.037^{+0.001}_{-0.001}$ & 83.1 &  \ldots & $10.96^{+6.30}_{-4.47}$  & $3.19^{+2.02}_{-1.56}$ & 23.83 & 125.68      \\
ACT-CL J0022-0036 & 16226 & 53.5 & 00 22 13.35 & -00 36 35.13 & $10.55^{+10.14}_{-3.87}$ & $0.0069^{+0.0005}_{-0.0005}$ & 66.6 & \ldots & $9.69^{+6.78}_{-3.67}$  & $3.73^{+2.67}_{-1.51}$ &  26.22 &  190.34  \\
SPT-CL J0033-6326 & 13483 & 17.8 & 00 33 52.68* & -63 26 39.84* & $3.89^{+1.63}_{-0.62}$ & $0.0065^{+0.0011}_{-0.0017}$ & 67.3 & $2.67^{+2.57}_{-1.49}$ & $3.90^{+5.27}_{-2.60}$  & $1.60^{+2.22}_{-1.17}$ & 10.45 & 6.68 \\
SPT-CL J0040-4407 &  13395 & 6.4 & 00 40 50.59* & -44 07 52.44* & $10.8^{+13.3}_{-3.4}$* & $0.0092^{+0.0001}_{-0.0001}$ & 60.7 & \ldots & $6.87^{+5.27}_{-2.75}$  & $3.26^{+2.29}_{-1.63}$ & 20.42 & 128.20    \\
SPT-CL J0058-6145 & 13479 & 44.6 & 00.58 20.9 & -61 46 01.8 & $9.96^{+14.02}_{-4.52}$ & $0.0057^{+0.0005}_{-0.0004}$ &63.6 & \ldots & $12.27^{+9.03}_{-4.83}$ & $2.00^{+1.48}_{-0.81}$ & 34.88 & 275.44   \\
ACT-CL J0059-0049 & 16227 & 36.6 & 00 59 08.86* & -00 50 06.18 &  $7.90^{+3.25}_{-2.08}$ & $0.0063^{+0.0001}_{-0.0001}$ & 82.6 & $2.94^{+2.42}_{-1.62}$ & $9.50^{+2.07}_{-2.34}$ & $3.50^{+1.13}_{-1.20}$ & 20.78  & 64.51  \\
SPT-CL J0102-4603 & 13485 & 54.1 & 01:02:42.5* & -46:04:19.6* & $3.044^{+1.18}_{-0.89}$ & $0.0046^{+0.0006}_{-0.0005}$ & 69.3 & $1.50^{+0.95}_{-0.84}$ & $8.19^{+2.76}_{-3.62}$ & $3.40^{+1.54}_{-1.82}$ & 21.34  & 10.23  \\
ACT-CL J0102-4915* & 14022 & 145.9 & 01 02 58.078* & -49 16 29.23* & $3.55^{+0.60}_{-0.40}$ & $0.067^{+0.004}_{-0.004}$ & 15.2 &  $4.08^{+1.09}_{-1.13}$ & $0.461^{+0.150}_{-0.120}$ & $0.480^{+0.158}_{-0.127}$ & 5.76 & 1.39   \\
SPT-CL J0106-5943 & 13468 & 16.1 & 01:06:28.19 & -59:43:12.5 &  $7.60^{+5.56}_{-1.94}$ & $0.0079^{+0.0006}_{-0.0006}$ & 47.2 & \ldots & $6.834^{+4.50}_{-2.24}$ & $3.94^{+3.06}_{-2.08}$ & 26.17  & 103.80 \\
SPT-CL J0123-4821 & 13491 & 59.9 & 01:23:12.054 & -48:21:25.25 & $8.79^{+6.49}_{-2.98}$ & $0.0037^{+0.0002}_{-0.0002}$ & 76.8 & \ldots & $17.79^{+10.61}_{-4.30}$ & $10.11^{+8.73}_{-6.75}$ & 41.68  & 305.70  \\
SPT-CL J0142-5032 & 13467 & 23.3 & 01:42:10.63* & -50:32:23.28* &  $8.20^{+8.18}_{-3.63}$ & $0.0050^{+0.0002}_{-0.0005}$ & 82.2 & \ldots & $12.16^{+8.24}_{-5.55}$ & $3.73^{+2.93}_{-2.26}$ & 26.74 & 114.28  \\
SPT-CL J0151-5954 & 13480,14380 & 46.1,29.4 & 01:51:26.9* & -59:54:29.4* & $3.79^{+1.38}_{-0.77}$ & $0.0019^{+0.0004}_{-0.0003}$ & 54.6 & \ldots & $15.84^{+14.79}_{-8.36}$ & $7.62^{+8.66}_{-6.37}$ & 52.46  & 130.48    \\
ACT-CL J0152-0100* & 3580 & 17.7 & 01 52 42.5* & 01 00 42.8* & $7.96^{+8.84}_{-3.23}$ & $0.010^{+0.001}_{-0.001}$ & 27 & \ldots & $5.57^{+4.37}_{-2.29}$ & $4.42^{+4.31}_{-3.14}$ & 37.1 &  234.2  \\
SPT-CL J0156-5541 & 13489 & 68.1 & 01:56:10.55* & -55:41:54* & $11.56^{+18.82}_{-5.44}$ & $0.0079^{+0.0009}_{-0.0007}$ & 55.2 & \ldots & $9.66^{+8.26}_{-4.58}$ & $2.65^{+2.28}_{-1.29}$ & 31.63  & 301.84  \\
SPT-CL J0200-4852 & 13487 & 20.7 & 02:00:34.9* & -48:52:10.8* & $7.84^{+13.29}_{-3.14}$ & $0.0057^{+0.0006}_{-0.0005}$ & 56.9 & $1.73^{+1.80}_{-3.02}$ & $8.86^{+11.22}_{-4.09}$ & \ldots  & 28.13  & 136.68  \\
ACT-CL J0206-0114 & 16229 & 26.9 & 02 06 13.64* & -01 15 13.37* & $5.83^{+1.98}_{-1.32}$  & $0.0043^{+0.0003}_{-0.0002}$ & 110.9 & $2.85^{+2.53}_{-1.60}$ & $12.44^{+5.11}_{-2.19}$ & $2.71^{+1.18}_{-0.61} $& 20.26  & 32.53 \\
SPT-CL J0212-4656 & 13464 & 25.2 & 02:12:23.6* & -46:57:15.1*  &  $4.86^{+3.14}_{-1.33}$ & $0.0032^{+0.0002}_{-0.0005}$ & 82.1 & \ldots & $11.18^{+10.75}_{-5.30}$ & $1.50^{+1.45}_{-0.74}$ & 24.60  & 44.18 \\
ACT-CL J0215-5212 & 12268 & 16.9 & 02 15 11.86* & -52 12 21.51*  &  $5.85^{+2.23}_{-1.20}$* & $0.0020^{+0.0001}_{-0.0001}$ & 160.1 & \ldots &  $23.33^{+8.62}_{-5.73}$ & $6.28^{+3.45}_{-2.98}$ & 25.49  & 56.76  \\
ACT-CL J0217-5245 & 12269 & 18.0 &  02 17 15.05* & -52 45 43.82*  & $7.56^{+6.29}_{-2.16}$ & $0.0016^{+0.0000}_{-0.0000}$ & 114.4 & \ldots &  $29.13^{+23.03}_{-10.68}$ & \ldots  & 46.0  & 359.73    \\
SPT-CL J0232-4420* & 4993 & 11.1 & 02 32 18.65 & -44 20 47.37 & $4.97^{+0.87}_{-0.64}$ & $0.039^{+0.002}_{-0.002}$ & 29.5 & $1.86^{+0.41}_{-0.40}$  & $0.99^{+0.29}_{-0.23}$ &  $0.83^{+0.26}_{-0.21}$ & 8.89 & 6.02 \\
ACT-CL J0232-5257 & 12263 & 18.0 & 02 32 48.81* & -52 57 12.84* & $7.03^{+2.96}_{-1.93}$ & $0.0024^{+0.0002}_{-0.0001}$ & 147.4 &  \ldots & $24.15^{+6.65}_{-6.89}$ & \ldots  & 28.76  & 93.52   \\
SPT-CL J0234-5831 &  13403 &  9.2 & 02 34 41.93 & -58 31 24.11  & $ 4.60^{+1.71}_{-1.03}$ & $0.012^{+0.001}_{-0.001}$ & 28.4 & $4.15^{+3.26}_{-3.86}$ & $1.38^{+0.47}_{-0.46}$  & $0.912^{+0.317}_{-0.318}$ & 8.76 & 4.13  \\
ACT-CL J0235-5121 & 12262 & 17.5 & 02 35 44.61* & -51 20 59.18*  &   $5.11^{+1.97}_{-1.18}$ & $0.0035^{+0.0002}_{-0.0002}$ & 57.1 & $1.36^{+0.44}_{-0.57}$ & $11.06^{+4.95}_{-3.23}$ & \ldots  & 26.32 & 54.09  \\
ACT-CL J0237-4939 & 12266 & 36.5 &  02 37 01.46 & -49 38 09.37  & $5.90^{+1.49}_{-1.27}$ & $0.0037^{+0.0002}_{-0.0001}$ & 57.7 & \ldots & $14.26^{+4.78}_{-2.42}$ & $5.22^{+2.09}_{-1.44}$ & 39.80  & 128.30   \\
SPT-CL J0243-5930 &  13484,15573 & 23.5,16.5 & 02 43 27.109* & -59 31 03.02* & $7.24^{+5.35}_{-1.03}$ & $0.012^{+0.0005}_{-0.0008}$ & 42.6 & $1.43^{+0.86}_{-0.73}$ & $5.00^{+2.56}_{-1.65}$ & $3.23^{+1.92}_{-1.45}$ & 15 & 56.83  \\
ACT-CL J0245-5302* & 12260 & 18.4 & 02 45 24.734 & -53 01 45.9  & $7.20^{+49.46}_{-4.23}$ & $0.0057^{+0.0000}_{-0.0000}$ & 37.2 & $2.13^{+1.06}_{-0.65}$ & $9.11^{+23.70}_{-8.64}$ & $2.45^{+6.38}_{-2.35}$ &  44.20  & 265.66  \\
SPT-CL J0252-4824 & 13494 & 25.8 & 02:52:48.38 & -48:24:44.53 & $2.90^{+3.95}_{-1.32}$ & $0.0032^{+0.0010}_{-0.0006}$ & 60.0 & $2.45^{+1.55}_{-3.40}$ & $11.36^{+10.25}_{-7.83}$ & $3.23^{+2.98}_{-2.32}$ & 34.23  & 23.79   \\
SPT-CL J0256-5617 &  13481,14448 & 19.0,22.5 &  02 56 26.23* &  -56 17 49.11*  & $ 6.92^{+38.48}_{-3.05}$ & $0.0030^{+0.0000}_{-0.0000}$ & 86.3 & \ldots & $11.33^{+36.41}_{-8.18}$ & \ldots  & 23.71  & 106.05   \\
SPT-CL J0304-4401 & 13402 & 12.8 &  03 04 16.608 & -44 01 31.94 & $6.23^{+4.01}_{-2.07}$* &  $0.0034^{+0.0000}_{-0.0000}$ & 98.7 & $1.53^{+0.90}_{-1.06}$ & $11.51^{+7.31}_{-4.89}$ & $3.38^{+2.29}_{-1.62}$ &  21.06  & 56.19  \\
ACT-CL J0304-4921 & 12265 & 18.2  & 03 04 16.24 & -49 21 25.2 &  $4.15^{+1.16}_{-0.73}$ & $0.019^{+0.002}_{-0.002}$ & 27.4 & $1.86^{+0.89}_{-0.68}$ & $1.64^{+0.87}_{-0.61}$ & $0.91^{+0.48}_{-0.34}$ & 9.73 & 5.73  \\
SPT-CL J0307-5042 & 13476 & 33.3 & 03 07 50.87 & -50 42 07.68 &  $4.36^{+2.51}_{-1.20}$ & $0.0060^{+0.0006}_{-0.0007}$ & 64.7 & $2.94^{+2.16}_{-1.97}$ & $6.51^{+4.28}_{-2.99}$ & $4.24^{+3.49}_{-2.87}$ & 18.19  &  17.32  \\
SPT-CL J0307-6225 &  12191 & 21.1 & 03 07 15.6 & -62 26 50.7 &  $ 3.34^{+1.26}_{-0.69}$ & $0.0012^{+0.0004}_{-0.0004}$ & 176.2 & $2.26^{+1.60}_{-1.23}$ & $154.19^{+287.58}_{-122.33}$ & $1.90^{+3.57}_{-1.57}$ & 15.81  & 13.81  \\
SPT-CL J0310-4646 & 13492 & 32.3 & 03 10 32.3* & -46 47 08.7* &  $6.56^{+2.36}_{-1.65}$ & $0.0052^{+0.0004}_{-0.0003}$ & 90.1 & $1.14^{+3.42}_{-1.17}$ & $10.026^{+3.35}_{-2.83}$ & \ldots  & 20.11  & 43.55   \\
SPT-CL J0324-6236 &  12181,13137 &  18.5,21.6 & 03 24 12.48 & -62 35 57.18 & $8.85^{+2.71}_{-2.04}$ & $0.0075^{+0.0003}_{-0.0003}$ & 69.3 &  \ldots & $87.95^{+18.37}_{-19.61}$ & $5.78^{+3.11}_{-3.15}$  & 22.93  & 94.74     \\
  .. & 13213 & 10.9 & & & & & & & \\
ACT-CL J0326-0043* & 5810 & 9.4 & 03 26 49.9 & -00 43 51.9 & $3.91^{+1.19}_{-0.85}$ & $0.138^{+0.10}_{-0.008}$ & 11.3 & $1.79^{+0.53}_{-0.61}$ & $0.28^{+0.10}_{-0.03}$ & $0.256^{+0.090}_{-0.025}$ & 3.51 & 0.66 \\
SPT-CL J0334-4659 & 13470 & 22.7 & 03 34 11.03 & -46 59 45.6 &  $4.61^{+0.95}_{-0.48}$ &  $0.020^{+0.001}_{-0.001}$ & 34.0 & $1.25^{+0.36}_{-0.34}$ & $2.31^{+0.46}_{-0.48}$ & $0.83^{+0.17}_{-0.18}$ & 10  & 8.01  \\
ACT-CL J0346-5438 & 12270,13155 & 17.3,15.2 & 03 46 56 & -54 38 54.09 &  $6.35^{+2.43}_{-1.42}$ &  $0.0042^{+0.0002}_{-0.0002}$ & 94.4 & \ldots & $11.92^{+4.34}_{-3.01}$ & \ldots  & 22.40  & 51.70   \\
SPT-CL J0348-4514 & 13465 & 12.3 & 03 48 16.9* & -45 15 05.13* & $3.33^{+1.09}_{-0.70}$ & $0.0041^{+0.0003}_{-0.0005}$ & 92.5 & $3.66^{+3.74}_{-1.75}$ & $9.51^{+3.04}_{-3.24}$ & \ldots  & 18.51 & 9.25    \\
SPT-CL J0352-5647 &  13490 & 30.4 & 03 52 57.8 & -56 47 50.3 &  $5.70^{+2.51}_{-1.22}$ & $0.0033^{+0.0000}_{-0.0000}$ & 109.8 & \ldots & $11.68^{+5.14}_{-3.16}$ & $3.92^{+2.33}_{-1.90}$ & 19.22  & 38.03  \\
SPT-CL J0406-4804 & 13477 & 22.7 & 04 06 55.7* & -48 04 47.75* & $7.60^{+3.72}_{-1.96}$ & $0.0025^{+0.0001}_{-0.0001}$ &  168.5 & \ldots & $23.14^{+8.93}_{-6.39}$ & $3.08^{+1.30}_{-1.00}$ & 24.80  & 89.13  \\
SPT-CL J0411-4819 & 16355,17536 & 24.4,31.0 & 04 11 16.181 & -48 18 55.47 & $5.44^{+1.21}_{-1.15}$  &  $0.0150^{+0.0008}_{-0.0007}$ & 31.5 & $1.85^{+1.06}_{-1.04}$ & $3.39^{+0.57}_{-0.75}$ & $1.86^{+0.35}_{-0.44}$ &  19.44 & 26.70  \\
SPT-CL J0417-4748 & 13397 & 20.2 & 04 17 23.49 & -47 48 49.939 & $4.53^{+0.75}_{-0.56}$ & $0.026^{+0.001}_{-0.001}$ & 37.3 & $2.70^{+0.83}_{-0.97}$ & $1.45^{+0.35}_{-0.30}$ & $0.90^{+0.24}_{-0.21}$ & 7.02 &  2.99  \\
SPT-CL J0426-5455 & 13472 & 28.3 & 04 26 05.32 & -54 54 57.16  &   $10.37^{+6.88}_{-5.47}$ & $0.0022^{+0.0000}_{-0.0000}$ & 140.2 & \ldots & $31.83^{+10.77}_{-17.42}$ & \ldots  & 41.02  & 424.77  \\
ACT-CL J0438-5419 & 12259 & 18.1 & 04 38 17.32 & -54 19 23.06 &  $9.22^{+3.28}_{-2.11}$ & $0.0150^{+0.0008}_{-0.0008}$ & 42.2  & $1.33^{+0.74}_{-0.42}$ & $3.86^{+1.54}_{-1.05}$ & $2.19^{+0.94}_{-0.69}$ & 10 & 15  \\
SPT-CL J0441-4854 & 13475 & 22.2 & 04 41 48.38* & -48 55 26.8* &  $6.60^{+3.75}_{-1.83}$* & $0.0132^{+0.0000}_{-0.0000}$ & 49.7 & $1.09^{+0.58}_{-0.67}$ & $4.03^{+1.77}_{-1.26}$ & $2.67^{+1.37}_{-1.10}$ & 14.68  & 23.50   \\
SPT-CL J0446-5849 &  13482,15560 & 28.3,18.6  & 04 46 04.7* & -58 49 55.5* & $5.77^{+2.88}_{-1.93}$* & $0.0025^{+0.001}_{-0.001}$ & 179.3 &  \ldots & $19.74^{+5.54}_{-7.56}$ & \ldots &  19.97 & 32.62  \\
SPT-CL J0449-4901 & 13473 & 42.8 & 04 49 06.26* & -49 01 36.42* &  $8.57^{+6.29}_{-2.90}$ & $0.0032^{+0.0002}_{-0.0002}$ & 115.8 & \ldots & $18.41^{+11.37}_{-6.86}$ & $3.95^{+2.52}_{1.59}$ & 28.72  & 151.80  \\
SPT-CL J0456-5116 & 13474 & 44.6 & 04 56 27.64* & -51 16 43.7* &  $10.24^{+6.76}_{-3.06}$ & $0.0054^{+0.0004}_{-0.0004}$ & 62.2 & \ldots & $12.32^{+6.53}_{-4.26}$ & $4.69^{+2.63}_{-1.83}$ & 35.82  & 326.68  \\
SPT-CL J0509-5342 &  9432 & 26.0 & 05 09 21.35 & -53 42 12.74 & $4.65^{+1.48}_{-0.83}$ & $0.012^{+0.001}_{-0.001}$ & 40.3 & $12.16^{+7.11}_{-14.24}$ & $3.23^{+1.44}_{-1.00}$ & $1.40^{+0.64}_{-0.45}$ & 14.50 & 13.13  \\
SPT-CL J0517-5430 & 15099 & 17.4 & 05 16 34.23* & -54 31 37.76* &  $5.44^{+5.63}_{-1.83}$ & $0.0038^{+0.0006}_{-0.0005}$ & 58.5 & $2.26^{+1.43}_{-2.41}$ & $11.24^{+11.20}_{-5.86}$ & \ldots  & 34.74  & 98.83    \\
SPT-CL J0528-5300 &  9341,10862 & 14.1,11.2 &  05 28 04.98* & -52 59 45.86* &  $8.60^{+5.50}_{-3.60}$* & $0.0030^{+0.0002}_{-0.0002}$ & 200.0 & \ldots & $16.56^{+6.10}_{-4.78}$ & $4.54^{+1.93}_{-1.62}$ & 27.43  & 114.54  \\
& 11747,1187 & 16.6,28.1 & & & & & & &  \\
& 11996,12092  & 9.7,19.3 & & & & & & &  \\
& 13126 & 17.3 & & & & & & &  \\
SPT-CL J0534-5005 & 11748,12001 & 24.7,18.0 & 05 33 37.97*  & -50 05 46.31*  & $5.47^{+3.00}_{-1.54}$* & $0.0023^{+0.0002}_{-0.0002}$ & 131.4 & \ldots &  $21.78^{+9.03}_{-8.67}$ & \ldots  & 29.95  & 63.94  \\
& 12002 & 25.5  & & & & & & &  \\
SPT-CL J0542-4100* & 914 & 45.0 & 05 42 49.5* & -41 00 01* &  $10.24^{+10.12}_{-4.36}$ & $0.0077^{+0.0004}_{-0.0007}$ & 47.4 & \ldots & $ 8.93^{+5.87}_{-3.90}$ & \ldots   & 34.05  & 288.60 \\
\hline
\end{tabular}
}
\end{table*}
\end{landscape}

\begin{landscape}
\begin{table*}
\begin{minipage}{200mm}
\contcaption{}
\label{tcool_table}
\scalebox{0.7}{
\begin{tabular}{@{}lcccccccccccc}
 & Obs. ID$^b$ & $t^c$ & \multicolumn{2}{c}{X-Ray Core (J2000)}$^d$ & $kT$$^e$ & $n_{\rm e}$$^f$ & $r^g$ & $\Delta (kT)^h$ &
$t_{\rm cool}$$^i$ & $t_{\rm cool}$ (10 kpc)$^j$ & $\frac{t_{\rm cool}}{t_{\rm ff}}$$^k$ & $\eta_{\rm min}$$^l$  \\
\cline{4-5}
System$^a$& & (ks) & RA & DEC &  (keV) & (cm$^-3$) & (kpc) &  & (Gyr) & (Gyr) &  &  \\
 \hline
SPT-CL J0547-5345 & 9332,9336 & 13.2,26.1 & 05 46 36.97 & -53 45 37.26 &  $9.68^{+7.18}_{-3.28}$ & $0.013^{+0.001}_{-0.001}$ & 48.0 & \ldots & $4.05^{+3.36}_{-1.71}$ & $2.59^{+2.36}_{-1.46}$ & 15.24  & 63.97   \\
& 10851,1173 & 7.8,12.0 & & & & & & &  \\
SPT-CL J0552-5709 & 11743,11871 & 16.1,18.6 &  05 51 34.61* & -57 08 44.9* & $4.33^{+5.93}_{-1.35}$ & $0.0038^{+0.0009}_{-0.0008}$ & 54.7 & $3.20^{+1.48}_{-1.18}$ & $8.24^{+16.78}_{-5.79}$ & $3.15^{+6.71}_{-3.00}$ & 27.24  & 48.14  \\
SPT-CL J0555-6405 & 13404 & 9.6 & 05 55 29.57* & -64 06 03.94* &  $4.69^{+5.74}_{-2.02}$ & $0.0045^{+0.0007}_{-0.0006}$ & 61.4 & $2.48^{+2.02}_{-3.12}$ & $10.42^{+7.72}_{-5.98}$ & $4.47^{+4.16}_{-3.60}$ & 30.70  & 48.81   \\
ACT-CL J0559-5249 & 12264,13116 & 37.7,21.3 & 05 59 42.95 & -52 49 47.91 & $8.96^{+5.66}_{-2.78}$ & $0.0057^{+0.0004}_{-0.0004}$ & 53.0 & \ldots & $10.92^{+5.58}_{-3.52}$ & $5.00^{+2.95}_{-2.20}$ &  37.23  & 272.52   \\
& 13117 & 39.6 & & & & & & & \\
ACT-CL J0616-5227 &  12261,13127 & 25.1, 8.641 & 06 16 34.37* & -52 27 09.61* &  $3.63^{+2.63}_{-1.29}$ & $0.016^{+0.008}_{-0.008}$ & 36.6 & \ldots & $2.02^{+4.86}_{-1.21}$ & $1.23^{+2.97}_{-7.76}$ & 9.84 & 3.99   \\
SPT-CL J0655-5234 & 13486 & 18.7 & 06 55 53.9* & -52 34 18.5* & $6.61^{+2.47}_{-1.47}$ & $0.0026^{+0.0001}_{-0.0002}$ & 124.8 & \ldots & $21.75^{+5.92}_{-5.81}$ & \ldots  & 31.48  & 100.20    \\
ACT-CL J0707-5522 & 12271 & 18.3 & 07 07 04.73* & -55 23 21.14* &  $10.04^{+6.26}_{-3.31}$ &  $0.004^{+0.000}_{-0.000}$ & 77.1 & \ldots & $16.82^{+5.67}_{-5.54}$ & $4.25^{+1.63}_{-1.61}$ & 31.02  & 227.45  \\
SPT-CL J2011-5725* & 4995 & 20.7 & 20 11 26.8 & -57 25 12.68 &  $3.14^{+1.19}_{-0.82}$ & $0.027^{+0.005}_{-0.005}$ &  16.6 & $1.63^{+0.48}_{-0.64}$ & $1.20^{+0.85}_{-0.74}$ & $1.04^{+0.74}_{-0.65}$ & 11.35 & 4.88 \\
SPT-CL J2023-5535 & 15108 & 16.2 & 20 23 21.84* & -55 35 47.78* & $11.73^{+12.24}_{-4.70}$ & $0.0041^{+0.0002}_{-0.0003}$ & 64.6 & \ldots & $18.44^{+11.37}_{-7.69}$ & \ldots  & 51.60  & 830.07  \\
SPT-CL J2031-4037 & 13517 & 8.7 & 20 31 52.6* & -40 37 27.45* & $8.00^{+3.45}_{-1.99}$ & $0.0082^{+0.0004}_{-0.0004}$ & 67.0  & $1.86^{+0.81}_{-0.70}$ & $6.72^{+2.68}_{-1.80}$ & \ldots  & 18.12  & 54.80   \\
SPT-CL J2034-5936 &  12182 & 54.6 & 20 34 09.11*  & -59 36 16.58* & $10.54^{+33.67}_{-5.31}$ & $0.006^{+0.001}_{-0.002}$ & 48.2 & \ldots & $8.93^{+27.72}_{-7.58}$ & \ldots   & 33.50  & 371.32  \\
SPT-CL J2035-5251 & 13466 & 16.3 & 20 35 11.19* & -52 51 22.23* & $4.28^{+5.53}_{-0.88}$ & $0.0018^{+0.0009}_{-0.0003}$ & 87.1 & $2.52^{+1.88}_{-1.34}$ & $14.12^{+30.70}_{-10.86}$ & $5.05^{+11.26}_{-4.60}$ & 29.30 & 63.92   \\
SPT-CL J2043-5035 &  13478 & 71.6 & 20 43 17.74 & -50 35 32.32 &  $4.13^{+2.05}_{-1.20}$ & $0.09^{+0.01}_{-0.01}$ & 8.9 & $2.09^{+0.50}_{-0.43}$ & $0.496^{+0.191}_{-0.205}$ & $0.501^{+0.192}_{-0.206}$  & 10.06 & 0.482  \\
SPT-CL J2106-5845 & 12189 & 43.4 & 21 06 05.3 & -58 44 31.2 &   $3.95^{+1.37}_{-0.93}$ & $0.015^{+0.000}_{-0.000}$ & 60.6 & $3.31^{+2.56}_{-1.54}$ & $2.804^{+0.574}_{-0.789}$ & $0.923^{+0.204}_{-0.271}$ & 8.36 & 2.80 \\
ACT-CL J2129-0005* & 9370 & 28.8 & 21 29 39.9 & 00 05 21.6 & $2.91^{+0.42}_{-0.30}$ & $0.076^{+0.013}_{-0.012}$ & 6.4 & $4.82^{+2.58}_{-1.50}$ & $0.25^{+0.18}_{-0.12}$ & \ldots & 6.93 & 2.03  \\
SPT-CL J2135-5726 &  13463 &  14.9 & 21 35 38.373 & -57 26 27.55 & $7.73^{+3.43}_{-2.12}$ & $0.0055^{+0.0003}_{-0.0003}$  & 74.2 & \ldots & $11.06^{+3.03}_{-2.82}$ & $4.92^{+2.22}_{-2.16}$ & 26.94  & 99.93  \\
SPT-CL J2145-5644 & 13398 &  12.0 & 21 45 52.418 & -56 44 48.95 & $17.93^{+17.13}_{-6.59}$ & $0.0055^{+0.0005}_{-0.0006}$ & 82.2 & \ldots & $12.33^{+11.75}_{-6.62}$ & $5.28^{+5.28}_{-3.26}$ & 27.08  & 460.98  \\
SPT-CL J2146-4632 & 13469 &  71.7 & 21 46 34.67* & -46 32 58.5* & $10.55^{+10.20}_{-3.75}$ & $0.0027^{+0.0003}_{-0.0003}$ & 104.5 & \ldots & $23.30^{+20.63}_{-11.55}$ & $11.06^{+11.91}_{-8.72}$ & 40.27  & 460.98   \\
SPT-CL J2148-6116 & 13488 & 29.1 & 21 48 44.43* & -61 16 41.63* & $6.40^{+2.43}_{-1.72}$ & $0.0035^{+0.0000}_{-0.0000}$ & 102.8 & $3.12^{+2.94}_{-2.11}$ &  $15.53^{+3.00}_{-3.22}$ & $4.63^{+1.92}_{-1.97}$ & 27.30  & 74.50 \\
ACT-CL J2154-0049 & 16230 & 55.7 & 21 54 32.2 & -00 48 59.6 & $8.24^{+5.59}_{-2.56}$ &  $0.0137^{+0.0008}_{-0.0007}$ & 29.6 & \ldots & $4.30^{+2.22}_{-0.96}$ & $4.22^{+2.47}_{-1.50}$ & 26.22 & 125.24  \\
SPT-CL J2218-4519 & 13501 & 31.2 & 22 18 59.06* & -45 18 55.4* & $8.12^{+12.08}_{-2.41}$ & $0.0030^{+0.0002}_{-0.0002}$ & 112.5 & \ldots & $17.85^{+19.85}_{-6.48}$ & $5.48^{+6.44}_{-2.88}$ & 28.66  & 146.61   \\
SPT-CL J2222-4834 & 13497 & 27.9 & 22 22 50.9 & -48 34 36.14 & $4.59^{+1.47}_{-1.12}$ & $0.015^{+0.001}_{-0.001}$ & 40.1 & $1.90^{+0.96}_{-0.86}$ & $3.14^{+0.78}_{-0.94}$ & $1.64^{+0.44}_{-0.52}$ & 14.46 & 10.57 \\
SPT-CL J2232-6000 & 13502 & 29.8 & 22 32 33.7 & -59 59 53.89 & $4.87^{+0.99}_{-0.84}$ & $0.0129^{+0.0009}_{-0.0005}$ & 44.2 & $1.51^{+0.45}_{-0.49}$ & $2.64^{+0.85}_{-0.83}$ & $0.94^{+0.31}_{-0.31}$ & 10.78 & 9.35 \\
SPT-CL J2233-5339 & 13504 & 15.8 & 22 33 16.8* & -53 39 07.4* &  $5.52^{+2.35}_{-1.30}$ & $0.0063^{+0.0004}_{-0.0004}$ & 74.9 & $1.57^{+1.13}_{-0.87}$ & $6.78^{+3.28}_{-2.22}$ & \ldots  & 16.36  & 23.22  \\
SPT-CL J2236-4555 & 13507,15266 & 41.3,32.4 &  22 36 52.18* & -45 55 51.7* & $11.15^{+5.36}_{-3.46}$* & $0.0084^{+0.0004}_{-0.0005}$ & 71.0 & \ldots & $8.65^{+3.14}_{-2.70}$ & \ldots  & 22.02  & 140.72   \\
SPT-CL J2245-6207 & 13499 & 26.2 & 22 45 01.8* & -62 07 44.85* & $8.26^{+5.51}_{-2.67}$ & $0.0026^{+0.0002}_{-0.0003}$ & 119.7 & $3.82^{+6.02}_{-3.18}$ &  $19.73^{+14.93}_{-8.77}$ & $4.40^{+4.40}_{-2.72}$ & 29.78  & 171.87  \\
SPT-CL J2248-4431* & 4966 &  21.2 & 22 48 44.5* & -44 31 48.5* &   $15.39^{+11.20}_{-4.36}$ & $0.027^{+0.002}_{-0.002}$ & 25.6 & \ldots & $2.65^{+1.70}_{-0.96}$ & $2.24^{+1.49}_{-0.93}$ & 10.11  & 37.93   \\
SPT-CL J2258-4044 & 13495 & 48.2 & 22 58 49.89* & -40 44 20.07* &  $5.81^{+2.31}_{-1.45}$ & $0.0054^{+0.0004}_{-0.0003}$ & 86.0 & $4.23^{+2.25}_{-1.73}$ & $9.32^{+3.21}_{-2.80}$ & \ldots  & 19.57  & 32.09   \\
SPT-CL J2259-6057 & 13498 & 57.3 & 22 59 01.04 & -60 57 38.44 & $9.34^{+3.87}_{-2.57}$ & $0.0139^{+0.0005}_{-0.0006}$ & 39.7 & \ldots & $4.85^{+135}_{-1.22}$ & $3.20^{+1.10}_{-1.02}$ & 22.09 & 98.70   \\
SPT-CL J2301-4023 & 13505 & 53.4 & 23 01 53.13 & -40 23 06.51 &  $9.68^{+3.91}_{-2.78}$ & $0.0093^{+0.0004}_{-0.0004}$ & 55.4 & \ldots & $7.32^{+2.22}_{-2.02}$ & $2.77^{+0.90}_{-0.83}$ & 23.89  & 125.02  \\
SPT-CL J2306-6505 & 13503 & 21.5 &  23 06 55.9* & -65 05 15.8* & $4.33^{+2.19}_{-1.03}$ & $0.0020^{+0.0003}_{-0.0003}$ & 126.9 & \ldots & $17.73^{+13.83}_{-8.42}$ & \ldots  & 25.24  & 35.71  \\
SPT-CL J2325-4111* &  13405 & 7.4  & 23 25 11.47* & -41 12 13.3* &  $5.57^{+10.40}_{-2.40}$* & $0.0045^{+0.0006}_{-0.0009}$ & 61.6 & $2.73^{+2.20}_{-5.16}$ & $10.68^{+14.21}_{-6.73}$ & \ldots  & 31.30  & 75.55    \\
SPT-CL J2331-5051 & 9333 & 26.9 &  23 31 51.307 & -50 51 53.77 &  $5.57^{+2.79}_{-1.59}$ & $0.032^{+0.003}_{-0.002}$ & 24.2 & $2.59^{+2.44}_{-1.24}$ & $1.49^{+0.67}_{-0.57}$ & $0.828^{+0.372}_{-0.319}$ & 11.13 & 10.04   \\
SPT-CL J2335-4544 & 13496 & 18.0 & 23 35 08.42 & -45 44 23.67 & $10.63^{+4.12}_{-3.13}$ & $0.0048^{+0.0002}_{-0.0002}$ & 99.1 & \ldots & $15.01^{+3.93}_{-3.71}$ & $5.07^{+1.96}_{-1.91}$ & 27.37  & 191.87   \\
ACT-CL J2337-0016* & 11728 & 15.7 & 23 37 58.0 & 00 16 11.17 & $9.4^{+14.6}_{-2.9}$ & $0.0051^{+0.0005}_{-0.0006}$ & 58 & \ldots & $11.54^{+13.37}_{-5.10}$ & \ldots & 36.13 & 305.57  \\
SPT-CL J2337-5942 & 11859 & 17.5 & 23 37 25.09* & -59 42 20.59* & $5.64^{+5.27}_{-1.82}$ & $0.0064^{+0.0008}_{-0.0006}$ & 87.7 & $3.09^{+3.45}_{-3.18}$ & $6.88^{+5.76}_{-3.23}$ & $3.51^{+3.61}_{-2.66}$ & 14.18  & 17.76   \\
SPT-CL J2341-5119 & 9345,11799 & 28.2,47.0 & 23 41 12.49 & -51 19 44.17 & $4.47^{+0.91}_{-1.19}$ & $0.029^{+0.003}_{-0.002}$ & 25.6 & $3.41^{+2.16}_{-1.39}$ & $1.60^{+0.30}_{-0.45}$ & $1.39^{+0.35}_{-0.45}$ & 11.29 & 6.19   \\
SPT-CL J2343-5411 & 11741,11870 & 58.1,16.7 & 23 42 46.27 & -54 11 05.83 & $3.36^{+1.64}_{-0.81}$ & $0.020^{+0.003}_{-0.005}$ & 28.0 & $2.05^{+4.24}_{-1.12}$ & $1.49^{+1.82}_{-1.04} $ & $0.762^{+0.930}_{-0.532}$ & 9.63 & 3.479 \\
& 12014,12091 & 53.0,35.4 & & & & & & &  \\
SPT-CL J2344-4242* & 13401 & 11.1 & 23 44 43.96 & -42 43 12.7 &  $12.65^{+1.48}_{-1.70}$ & $0.087^{+0.001}_{-0.001}$ & 41.0 & $1.03^{+0.22}_{-0.19}$ & $0.7991^{+0.0983}_{-0.1080}$ & $0.438^{+0.0550}_{-0.0603}$ & 3.53  & 5.33    \\
SPT-CL J2345-6406 & 13500 & 55.1 & 23 45 00.16* & -64 05 49.0* & $4.09^{+1.33}_{-0.98}$ & $0.005^{+0.0005}_{-0.0004}$ & 95.0 & $5.58^{+2.50}_{-4.49}$ & $8.25^{+3.88}_{-1.75}$ & $2.67^{+1.45}_{-0.92}$ & 15.68  & 9.81  \\
SPT-CL J2352-4657 & 13506 & 68.7 & 23 52 16.25* & -46 57 35.97* &  $3.57^{+0.77}_{-0.68}$ & $0.0047^{+0.0004}_{-0.0003}$ &  81.0 & $2.55^{+3.70}_{-1.30}$ & $8.61^{+3.53}_{-1.45}$ & $5.02^{+2.88}_{-2.19}$ & 19.34  & 11.45  \\
SPT-CL J2355-5056 &  11746,11998 & 11.2,9.6 & 23 55 47.39 &  -50 55 40.58 &  $3.26^{+0.58}_{-0.41}$ & $0.007^{+0.002}_{-0.003}$ & 34.3 & $1.53^{+0.42}_{-0.34}$ & $2.09^{+4.25}_{-1.75}$ & $1.04^{+2.14}_{-0.94}$ & 11.02 & 8.27  \\
SPT-CL J2359-5009 & 9334,11742 & 24.3,21.0 & 23 59 43.56* & -50 10 15.87* & $5.13^{+5.88}_{-1.64}$ & $0.004^{+0.001}_{-0.001}$ & 54.8 & \ldots & $7.06^{+16.95}_{-5.66}$ & \ldots  & 23.29  & 57.64   \\
&  1864,11997 &  18.5,58.6  & & & & & & &  \\
\hline
\end{tabular}
}
\tiny{
\\$^a$Alternative names for ACT-CL J0102-4915 (El Gordo);  ACT-CL J0152-0100
(A267); ACT-CL J0245-5302 (AS0295); ACT-CL J0326-0043 \\ﬂ(MACS J0326-0043); SPT-CL
J0542-4100 (RDCS J0542-4100); SPT-CL J2011-5725 (RXCJ2011.3-5725);  ACT-CL
J2129-0005 (RXJ2129.6+0005); \\SPT-CL J2248-4431 (AS1063); SPT-CL J2325-4111
(ACOS1121); ACT-CL J2337-0016 (A2631); SPT-CL J2344-4242 (Phoenix cluster).\\
$^b$Some
systems have multiple \emph{Chandra} observations:e.g.; SPT-CL J0151-5954 (Obs.
ID 13480, 14380).\\
$^c$The time on source after reprocessing the data.\\
$^d$The X-ray core position from this work; the asterisks mark the systems with uncertain
core positions.\\
$^e$Central-bin temperature (innermost region used for
deprojection). The systems marked with asterisk are the ones with only 2 bins
for deprojection (see text).\\
$^f$Central-bin electron density (innermost
region used for deprojection).\\
$^g$The radius for the innermost
region used for spectral deprojection.\\
$^h$The temperature drop is calculated
as the ratio between the highest and the lowest temperatures of the profile when
the profile is rising upward smoothly. \\For the systems which do not have a
statistically significant temperature drop or their temperature profile is rising
inwardly there is no entry for the temperature drop.\\
$^i$The cooling time
of the innermost region (in Gyr).\\
$^j$The cooling time at 10 kpc derived using
the deprojected surface brightness profiles (in Gyr). In some cases the surface
brightness deprojection did not work well \\(due, e.g., to the surface brightness
profile dropping or flattening towards the centre). In these cases \\there is no entry for the
cooling time at 10 kpc.\\
$^k$The ratio of the cooling time to free fall time.\\
$^l$Thermal stability parameter from \citet{voit08}; $\rm Inst.=\rm min(kT/
\Lambda n_{e} n_{H} r^{2}$). For some of the systems the instability profile is
still rising,  therefore there is no \\minimum (e.g., A399, A3158, A754, A2163).
See text for details.}
\end{minipage}
\end{table*}
\end{landscape}

\begin{table*}
\begin{minipage}{145mm}
\caption{Cluster and cRS Properties}
\label{lum_table}
\scalebox{0.8}{
\begin{tabular}{@{}lccccccccc}
\hline
&  & $L_{\rm X}(<r_{\rm cool})$  & $r_{\rm cool}$ & $M^{SZ}_{500}$ & $L_{\rm X}(<R_{500})$ & $R_{500}$ &
$L_{\rm 843MHz}$$^b$ & SFR$^c$   \\ System$^a$ & $z$  & (10$^{42}$ erg s$^{-1}$)  &
(kpc) & (10$^{14}$ M$_{\odot}$) & (10$^{42}$ erg s$^{-1}$) & (kpc) & (10$^{32}$ erg
s$^{-1}$Hz$^{-1}$) &  (M$_{\odot}$ yr$^{_1}$)  \\
\hline
 \multicolumn{9}{c}{CF Sample} \\
 \hline
SPT-CL J0000-5748 & 0.7019(1) & $868^{+67}_{-48}$   & 130 & 4.29 $\pm$ 0.71 (5) &
1078$^{+61}_{-71}$ & 679 & 8.83 $\pm$ 0.33 &  $52^{+59}_{-30}$ &  \\
SPT-CL J0033-6326 & 0.597 (5) &  $178^{+17}_{-18}$  & 101 & 4.72 $\pm$ 0.88 (1) &
$627^{+59}_{-65}$ & 766 & $<$ 0.88  & $<43$  \\
ACT-CL J0102-4915*  & 0.87 (5) & $589^{+10}_{-9}$  & 152 & 14.43 $\pm$ 2.1 (1) &
$10867^{+188}_{-156}$ & 894  & 0.16 $\pm$ 0.03  & $93^{+120}_{-55}$   \\
SPT-CL J0232-4420* & 0.284 (5) & $930^{+33}_{-29}$ & 125 & 12.01 $\pm$ 1.80 (1) &
$2105^{+76}_{-69}$ & 1333 & 1.27 $\pm$ 0.08 & $17^{+19}_{-10}$  \\
SPT-CL J0234-5831  & 0.415(1) & $836^{+86}_{-67}$ & 130 & 7.64 $\pm$ 1.5 (5) &
$1246^{+86}_{-81}$ & 1040 & 2.60 $\pm$ 0.10 & $59^{+35}_{-22}$  \\
ACT-CL J0304-4921 & 0.392 (5) &  $464^{+29}_{-24}$ & 100 & 7.57 $\pm$ 1.2 (1) &
$1136^{+57}_{-55}$ & 1053 & 0.11 $\pm$ 0.01*  & $<15$  \\
ACT-CL J0326-0043* & 0.448 (4) & $1709^{+90}_{-75}$ & 660 & 7.4 $\pm$ 1.4 (2) &
$1709^{+77}_{-95}$ & 1000 & 0.17 $\pm$ 0.02  &  \ldots  \\
SPT-CL J0334-4659 & 0.485 (5) &  $429^{+16}_{-16}$ & 125 & 5.52 $\pm$ 0.95 (1) &
$836^{+43}_{-54}$ & 881 & 1.10 $\pm$ 0.08 &  $79^{+45}_{-39}$  \\
SPT-CL J0417-4748 & 0.581 (5) &  $1268^{+50}_{-62}$ & 142 & 7.41 $\pm$ 1.15 (1) &
$2209^{+137}_{-92}$ & 901 & $<$ 1.38 & $<28$  \\
ACT-CL J0616-5227 & 0.684 (6) &   $258^{+9}_{-22}$ & 107 & 6.8 $\pm$ 2.9 (3) &
$1021^{+76}_{-77}$ & 807 & 6.99 $\pm$ 0.31 &  \ldots    \\
SPT-CL J2011-5725* & 0.2786 (1) &  $301^{+12}_{-10}$ & 131 & 3.18 $\pm$ 0.89 (5) &
$373^{+15}_{-14}$ & 869 & $<$ 0.15 &  \ldots   \\
SPT-CL J2043-5035 & 0.7234 (1) &   $1375^{+35}_{-38}$ & 287 & 4.71 $\pm$ 1.0 (5) &
$1653^{+59}_{-50}$ & 696 & $<$ 1.41 & $160^{+123}_{-69}$   \\
SPT-CL J2106-5845 & 1.132 (1) &  $1201^{+55}_{-60}$ & 146 & 8.36 $\pm$ 1.71 (5)  &
$3218^{+219}_{-193}$ & 610 & 8.73  $\pm$ 0.99 & $200^{+240}_{-100}$  \\
ACT-CL J2129-0005* & 0.234 (4) & $638^{+15}_{-15}$ & 101 & 7.3 $\pm$ 1.6 (2) &
$1186^{+33}_{-32}$ & 1171 & 0.690 $\pm$ 0.03  &  \ldots  \\
SPT-CL J2232-6000 & 0.594 (5) &  $312^{+19}_{-15}$ & 110 & 5.55 $\pm$ 0.97 (1) &
$726^{+48}_{-49}$ & 810  & $<$ 0.87 & $<33$  \\
SPT-CL J2331-5051 & 0.576 (1) & $662^{+42}_{-36}$ & 116 & 5.14 $\pm$ 0.71 (5) &
$929^{+75}_{-66}$ & 801 & 1.80 $\pm$ 0.134 & $23^{+36}_{-14}$    \\
SPT-CL J2341-5119 &  1.003 (1) & $364^{+30}_{-22}$ & 77 & 5.61 $\pm$ 0.82 (5) &
$1604^{+88}_{-107}$ & 588 & 4.15 $\pm$ 0.58 & $<170$     \\
SPT-CL J2343-5411 & 1.075 (1) & $318^{+44}_{-48}$ & 116 & 3.0 $\pm$ 0.5 (5) &
$542^{+53}_{-54}$ & 452 & $<$ 3.75 & $ 33^{+42}_{-19}$  \\
SPT-CL J2344-4242*  & 0.595 (2) & $13913^{+273}_{-318}$ & 173 & 12.5 $\pm$ 1.57 (6)
& $13913^{+335}_{-270}$ & 1061 & 11.57 $\pm$ 0.41 & $1900^{+926}_{-525}$ \\
SPT-CL J2355-5056 & 0.3196 (1) &   $85^{+6}_{-8}$ & 93 & 4.07 $\pm$ 0.57 (5) &
$312^{+21}_{-23}?$ & 908 & $<$ 0.20  &  \ldots   \\
\hline
 \multicolumn{9}{c}{NCF Sample} \\
 \hline

SPT-CL J0013-4906 & 0.406 (5) &  $205^{+19}_{-15}$  & 67 & 7.08 $\pm$ 1.15(1) &
$1214^{+81}_{-86}$ &  1019 & $<$ 0.60 & $<8.6$ \\
ACT-CL J0014-0056  & 0.533 (4) & $270^{+16}_{-24}$  & 80 & 7.6 $\pm$ 1.4 (2) &
$1281^{+58}_{-62}$ &  944 & $<$ 0.47 &  \ldots   \\
SPT-CL J0014-4952 & 0.752 (2) & $123^{+95}_{-48}$  & 78 & 5.31 $\pm$ 0.92 (1) &
$1292^{+70}_{-72}$ & 704 & $<$ 2.59   &  $<92$ \\
ACT-CL J0022-0036 & 0.805 (4) & $76^{+9}_{-10}$  & 47 & $7.3 \pm 1.2$ (2) &
$1688^{+134}_{-112}$ & 750 & $9.55 \pm 0.36$ & \ldots   \\
SPT-CL J0040-4407 & 0.35 (2) &  $349^{+26}_{-26}$  & 61 & 10.18 $\pm$ 1.32 (6) &
$1285^{+68}_{-62}$ & 1200 & $<$ 0.41 &  $<14$  \\
SPT-CL J0058-6145 &  0.83 (5) & $84^{+8}_{-15}$ & 57 & 4.36 $\pm$ 0.81 (1) &
$514^{+48}_{-39}$ & 619 & 2.60 $\pm$ 0.33 & $24^{+34}_{-13}$   \\
ACT-CL J0059-0049 & 0.786 (4) &  $142^{+13}_{-17}$ & 69   & 6.9 $\pm$ 1.2 (2) &
$1421^{+90}_{-90}$ & 747 & $<$ 1.20 & \ldots   \\
SPT-CL J0102-4603 & 0.72 (5) &  $21^{+5}_{-3}$ & 66 & 4.49 $\pm$ 0.85 (1) &
$229^{+36}_{-32}$ &  683 & $<$ 2.33 &  $15^{+20}_{-9}$  \\
SPT-CL J0106-5943 & 0.348 (5) &  $86^{+8}_{-8}$ & 58 & 6.23 $\pm$ 1.05 (1) &
$516^{+32}_{-29}$ & 1620 & 0.23 $\pm$ 0.08 &  $< 7.3$  \\
SPT-CL J0123-4821 & 0.62 (5) & $12^{+4}_{-2}$ & 32 & 4.46 $\pm$ 0.87 (1) &
$337^{+27}_{-24}$ & 738 & $<$ 0.113* &  $< 39$   \\
SPT-CL J0142-5032 & 0.73 (5) &  $49^{+39}_{-19}$ & 40 & 5.75 $\pm$ 0.95 (1) &
$760^{+84}_{-80}$ &  735 & $<$ 0.19* &  $92^{+54}_{-34}$  \\
SPT-CL J0151-5954 & 0.29 (5) & $0.80^{+0.89}_{-0.52}$ & 17 & 3.24 $\pm$ 0.90 (1) &
$22^{+5}_{-4}$ & 857 & 0.37 $\pm$ 0.04 & $<140$    \\
ACT-CL J0152-0100* & 0.23 (4) & $133^{+5}_{-4}$ & 67 & 7.9 $\pm$ 1.6 (2) &
$978^{+35}_{-29}$ & 1206 & 0.068 $\pm$ 0.009  & \ldots  \\
SPT-CL J0156-5541 &  1.22 (5) & $242^{+27}_{-18}$  & 58 & 3.63 $\pm$ 0.70 (1) &
$874^{+72}_{-72}$ & 432 & $<$ 5.12 & $<530$  \\
SPT-CL J0200-4852 & 0.498 (5) &  $106^{+29}_{-42}$  & 49 & 4.76 $\pm$ 0.90 (1) &
$498^{+32}_{-39}$ & 830 & $<$ 0.95 &  $<29$  \\
ACT-CL J0206-0114 & 0.676 (4) &  $340^{+18}_{-26}$  & 76 & 5.7 $\pm$ 1.1 (2) &
$860^{+47}_{-39}$ & 766 & 3.18 $\pm$ 0.17  & \ldots  \\
SPT-CL J0212-4657 & 0.655 (5) &  $50^{+6}_{-6}$ & 66 & 5.88 $\pm$ 0.98 (1) &
$327^{+20}_{-24}$ & 787 & $<$ 0.77* & $<49$  \\
ACT-CL J0215-5212 & 0.48 (6) & $96^{+17}_{-27}$ & 69 & 5.8 $\pm$ 1.7 (3) &
$378^{+49}_{-39}$ & 900 & 4.33 $\pm$ 0.17 &  \ldots  \\
ACT-CL J0217-5245 & 0.34 (3) & \ldots & \ldots & 4.42 $\pm$ 0.89 (1) &
$305^{+47}_{-34}$ & 916 & 0.91 $\pm$ 0.05 & $<9.2$     \\
ACT-CL J0232-5257 & 0.556 (5) & \ldots & \ldots & 5.36 $\pm$ 0.94 (1) &
$815^{+74}_{-88}$ & 825 & $<$ 0.74 & $<26 $  \\
ACT-CL J0235-5121 & 0.278 (5) &  \ldots & \ldots & 6.41 $\pm$ 1.08 (1) &
$564^{+30}_{-30}$ & 1086 & 0.16 $\pm$ 0.03  & $<14$  \\
ACT-CL J0237-4939 & 0.334 (5) &  \ldots & \ldots & 3.99 $\pm$ 0.86 (1) &
$241^{+21}_{-18}$ & 889 & $<$ 0.370 & \ldots   \\
SPT-CL J0243-5930  & 0.65(1) & $232^{+17}_{-17}$ & 90 & 4.18 $\pm$ 0.89 (5) &
$969^{+77}_{-70}$ & 709 & $<$ 1.09 & $<35$   \\
ACT-CL J0245-5302* & 0.3 (3) & $69^{+6}_{-7}$ & 46 & 6.6 $\pm$ 1.0 (4) &
$1807^{+57}_{-63}$ & 1079 & 0.317 $\pm$ 0.043 & \ldots     \\
SPT-CL J0252-4824 & 0.421 (5) &  $8^{+2}_{-2}$ & 34 & 4.79 $\pm$ 0.93 (1) &
$350^{+24}_{-29}$ & 884 & $<$ 0.64 & $<5.9$  \\
SPT-CL J0256-5617 & 0.64(1) &   \ldots  & \ldots & 4.25 $\pm$ 0.89 (5) &
$852^{+80}_{-77}$ & 718 & $<$ 1.05 &  $<35$   \\
SPT-CL J0304-4401 &  0.458 (5) & $64^{+10}_{-7}$  & 53 & 8.55 $\pm$ 1.32 (1) &
$949^{+56}_{-61}$ & 1042 & $<$ 0.028* &  $<6.1$  \\
SPT-CL J0307-5042 & 0.55 (5) &  $77^{+8}_{-4}$ & 74 & 5.26 $\pm$ 0.93 (1) &
$622^{+41}_{-40}$ & 824 & $<$ 0.334* & $<25$  \\
SPT-CL J0307-6225 & 0.59(1) &  $73^{+16}_{-17}$ & 97  & 4.68 $\pm$ 0.96 (5) &
$497^{+49}_{-51}$ & 772 & 2.69 $\pm$ 0.19 & $<44$   \\
SPT-CL J0310-4646 &  0.709 (5) & $117^{+11}_{-18}$ & 69 & 4.31 $\pm$ 0.83 (1) &
$606^{+86}_{-67}$ &  679 & $<$ 2.24 & $34^{+21}_{-13}$  \\
SPT-CL J0324-6236 &  0.72(1) & \ldots & \ldots & 4.68 $\pm $ 0.86 (5) &
$642^{+320}_{-136}$ & 696 & 3.98 $\pm$ 0.25   & $<57$     \\
ACT-CL J0346-5438 & 0.53 (5) &  $58^{+11}_{-11}$ & 49 & 5.47 $\pm$ 0.94 (1) &
$625^{+56}_{-69}$ & 848 & 2.93 $\pm$ 0.12 & $<5.7$  \\
SPT-CL J0348-4514 & 0.358 (5) & $64^{+8}_{-6}$ & 80 & 6.17 $\pm$ 1.03 (1) &
$557^{+49}_{-45}$ & 1010 & $<$ 0.0668* & $<5.7$  \\
SPT-CL J0352-5647 &  0.66(1) &  $80^{+11}_{-9}$ & 74 & 4.00 $\pm$ 0.86 (5) &
$448^{+31}_{-34}$  & 693 & $<$ 1.13 & $<50$  \\
SPT-CL J0406-4804 & 0.737 (5) &  $82^{+25}_{-25}$ & 61 & 4.61 $\pm$ 0.83 (1) &
$520^{+61}_{-48}$ &  679 & $<$ 2.46 & $41^{+26}_{-15}$  \\
SPT-CL J0411-4819 & 0.424 (5) &  $202^{+8}_{-9}$ & 74 & 8.18 $\pm$ 1.27 (1) &
$1286^{+62}_{-52}$ & 1054 & $<$ 0.639* & $27^{+5}_{-11}$ \\
SPT-CL J0426-5455 & 0.62(1) &  \ldots & \ldots & 4.93 $\pm$ 1.00 (5) & $517^{+48}_{-51}$ &
766 & $<$ 0.97 & $51^{+41}_{-22}$  \\
ACT-CL J0438-5419 & 0.421 (5) &  $909^{+33}_{-33}$ & 119 & 10.8 $\pm$ 1.62 (1) &
$3460^{+132}_{-116}$ & 1159 & $<$ 0.38 & $<24$  \\
SPT-CL J0441-4854 & 0.79 (5) &  $265^{+21}_{-24}$ & 89 & 4.74 $\pm$ 0.83 (1) &
$789^{+83}_{-71}$ & 657 & $<$ 2.92 & $<61$  \\
SPT-CL J0446-5849 &  1.16 (5) & \ldots & \ldots & $3.68 \pm 0.82$ (5) &
$711^{+178}_{-107}$ & 440 & $<$ 4.82 & $330^{+270}_{-170}$  \\
SPT-CL J0449-4901 & 0.79 (2) &  \ldots & \ldots & 4.57 $\pm$ 0.86 (6) &
$653^{+60}_{-68}$ & 649 & 18.66 $\pm$ 0.96 & $88^{+110}_{-47}$  \\
SPT-CL J0456-5116 & 0.562 (5) &  $38^{+6}_{-7}$ & 47 & 5.09 $\pm$ 0.89 (1) &
$483^{+36}_{-33}$ & 807 & 4.34 $\pm$ 0.19 &  $<32$  \\
SPT-CL J0509-5342 & 0.4626(1) & $207^{+13}_{-11}$ & 90 & 5.36 $\pm$ 0.71 (5) &
$769^{+50}_{-38}$ & 309 & $<$ 0.48 & $33^{+30}_{-12}$  \\
SPT-CL J0517-5430 & 0.295(1) &  $13^{+12}_{-5}$ & 36  & 6.46 $\pm$ 1.32 (5) &
$1090^{+47}_{-52}$ & 1081 & $<$ 0.17 & $2.5^{+1.3}_{-0.82}$  \\
SPT-CL J0528-5300 & 0.7648(1) &  \ldots & \ldots & 3.18 $\pm$ 0.61 (5) &
$182^{+36}_{-37}$ & 589 & 16.50 $\pm$ 0.54  & $<34$  \\
SPT-CL J0534-5005 & 0.881(1) &  \ldots & \ldots & 2.68 $\pm$ 0.61 (5) &
$417^{+127}_{-95}$ & 510 & $<$ 2.29  & $50^{+29}_{-19}$  \\
SPT-CL J0542-4100* & 0.642 (5) & $78^{+24}_{-25}$ & 41 & 5.16 $\pm$ 0.94 (1) &
$612^{+41}_{-43}$ & 761 & 8.53 $\pm$ 0.33  &  $<28$   \\
\hline
\end{tabular}
}
\end{minipage} \end{table*}

\begin{table*}
\begin{minipage}{145mm}
\contcaption{}
%\caption{Cluster and cRS Properties -- cont.}
%\label{lum_table}
\scalebox{0.8}{
\begin{tabular}{@{}lcccccccc} \hline & & $L_{\rm X}(<r_{\rm
cool})$ & $r_{\rm cool}$ & $M^{SZ}_{500}$ & $L_{\rm X}(<R_{500})$ & $R_{500}$ &
$L_{\rm 843MHz}$$^b$ & SFR$^c$  \\ System$^a$ & $z$ & (10$^{42}$ erg s$^{-1}$) & (kpc)
& (10$^{14}$ M$_{\odot}$) & (10$^{42}$ erg s$^{-1}$)  & (kpc) & (10$^{32}$ erg s$^{-1}$Hz$^{-1}$) &
(M$_{\odot}$ yr$^{_1}$)   \\
\hline
SPT-CL J0547-5345 & 1.067 (1) &  $453^{+52}_{-58}$ & 90 & 5.25 $\pm$ 0.75 (5) &
$1599^{+106}_{-141}$ & 549 & $<$ 3.68 & $110^{+89}_{-53}$ \\
SPT-CL J0552-5709 & 0.423 (1) & $212^{+11}_{-13}$ & 51 & 3.75 $\pm$ 0.54 (5) &
$450^{+34}_{-35}$ & 818.329 & 1.46 $\pm$ 0.10 & $<17$   \\
SPT-CL J0555-6405 & 0.345 (5) &  $21^{+8}_{-5}$ & 39 & 7.69 $\pm$ 1.22 (1) &
$436^{+20}_{-22}$ & 1097 & $<$ 0.24 & \ldots    \\
ACT-CL J0559-5249 & 0.609 (5) &  $32^{+15}_{-14}$ & 31 & 5.78 $\pm$ 0.95 (1) &
$556^{+34}_{-35}$ & 811 & 7.88 $\pm$ 0.42 & $<19$   \\
SPT-CL J0655-5234 & 0.47 (5) &  \ldots & \ldots & 5.1 $\pm$ 0.93 (1) &
$296^{+29}_{-31}$ & 869 & $<$ 0.50 & $<3.1$   \\
ACT-CL J0707-5522 & 0.296 (7) &  \ldots & \ldots & 5.7 $\pm$ 1.7 (7) &
$580^{+30}_{-26}$ & 1030 & 0.17 $\pm$ 0.02 &  \ldots   \\
SPT-CL J2023-5535 & 0.232 (1) &  $39^{+20}_{-17}$ & 29 & 7.86 $\pm$ 1.24 (1) &
$952^{+31}_{-27}$ & 1202 & 0.26 $\pm$ 0.02 &  \ldots  \\
SPT-CL J2031-4037 &  0.342 (5) & $289^{+25}_{-24}$ & 78 & 9.83 $\pm$ 1.5 (1) &
$1550^{+62}_{-73}$ & 1193 & 1.645 $\pm$ 0.070 & $<7.5$   \\
SPT-CL J2034-5936 & 0.92 (1) &    $132^{+16}_{-17}$ & 71 & 4.32 $\pm$ 0.89 (5) &
$741^{+36}_{-54}$ &  577 & 3.91 $\pm$ 0.60 & $<99$   \\
SPT-CL J2035-5251 & 0.47 (1) &  $271^{+23}_{-22}$ & 47 & 6.18 $\pm$ 1.25 (5) &
$402^{+77}_{-51}$ & 930 & 1.14 $\pm$ 0.91 & $31^{+29}_{-16}$ \\
SPT-CL J2135-5726 & 0.427 (1) & \ldots   & \ldots & 5.68 $\pm$ 1.11 (5) &
$705^{+50}_{-46}$ & 935 & 0.35 $\pm$ 0.09   & $<13$   \\
SPT-CL J2145-5644 & 0.48 (1) &  $112^{+13}_{-17}$ & 48  & 6.39 $\pm$ 1.25 (5) &
$1181^{+106}_{-105}$ & 933 & $<$ 0.52 & $<3.6$  \\
SPT-CL J2146-4632 & 0.933 (1) &  \ldots & \ldots & 5.36 $\pm$ 1.07 (5) &
$814^{+69}_{-59}$ & 614 & 15.49 $\pm$ 0.62  & $<9.6$ \\
SPT-CL J2148-6116 & 0.571 (1) &  \ldots & \ldots & 4.04 $\pm$ 0.89 (5) &
$546^{+35}_{-34}$ & 746 & 1.69 $\pm$ 0.21 & $<39$  \\
ACT-CL J2154-0049 & 0.488 (4) & $100^{+6}_{-9}$ & 61 & 5.7 $\pm$ 1.3 (2) &
$485^{+25}_{-28}$ & 889 & $<$ 0.38  &  \ldots   \\
SPT-CL J2218-4519 & 0.65 (5) & \ldots & \ldots & 5.31 $\pm$ 0.92 (1) &
$558^{+55}_{-59}$ & 763 & 9.18 $\pm$ 0.67 & $4.4^{+5}_{-2.4}$   \\
SPT-CL J2222-4834 & 0.652 (5) &  $175^{+10}_{-15}$ & 84 & 5.42 $\pm$ 0.93 (1) &
$670^{+70}_{-59}$ & 767 & $<$ 1.82 & $<52$   \\
SPT-CL J2233-5339 & 0.48 (5) &  $312^{+20}_{-17}$ & 85 & 5.48 $\pm$ 0.98 (1) &
$819^{+55}_{-49}$ & 883 & $<$ 0.52 & $<13$  \\
SPT-CL J2236-4555 & 1.16 (5) &  \ldots & \ldots & 4.02 $\pm$ 0.74 (1) &
$776^{+70}_{-64}$ & 467 & $<$ 7.54  &  $50^{+60}_{-30}$ \\
SPT-CL J2245-6206 & 0.58 (5) & $46^{+21}_{-14}$ &  55 & 5.4  $\pm$ 0.94 (1) &
$1250^{+86}_{-118}$ & 812 & 12.59 $\pm$ 0.40 & $<5.2$   \\
SPT-CL J2248-4431* &  0.351 (2) & $2370^{+43}_{-52}$ & 142 & 17.97 $\pm$ 2.18 (6) &
$6951^{+128}_{-114}$ & 1449 & 0.64 $\pm$ 0.21  &  \ldots    \\
SPT-CL J2258-4044 & 0.83 (5) & $ 85^{+15}_{-12}$ & 73 & 5.88 $\pm$ 0.95 (1) &
$941^{+83}_{-78}$ & 684 & $<$ 5.82* & $10^{+12}_{-5.7}$  \\
SPT-CL J2259-6057 & 0.75 (5) &  $235^{+18}_{-17}$ & 74 & 5.61 $\pm$ 0.94 (1) &
$1032^{+80}_{-74}$ & 718 & 13.90 $\pm$ 0.44  & $<84$ \\
SPT-CL J2301-4023 & 0.73 (5) &  $114^{+8}_{-7}$ & 59 & 4.81 $\pm$ 0.86 (1) &
$625^{+3}_{-12}$ &  693 & 0.86 $\pm$ 0.02* & $36^{+53}_{-19}$  \\
SPT-CL J2306-6505 & 0.53 (5) & $20^{+16}_{-8}$ & 36 & 5.73 $\pm$ 0.98 (1) &
$642^{+64}_{-52}$ & 861 & $<$ 0.66 & $<39$  \\
SPT-CL J2325-4111* & 0.358 (5) &  $25^{+7}_{-5}$ & 40 & 7.55 $\pm$ 1.2 (1) &
$756^{+43}_{-39}$ & 1080 & $<$ 0.43 & $<6.7$  \\
SPT-CL J2335-4544 & 0.547 (5) & \ldots & \ldots & 6.17 $\pm$ 1.02 (1) &
$870^{+65}_{-67}$ & 871 & 1.56 $\pm$ 0.14  & $41^{+61}_{-27}$  \\
ACT-CL J2337-0016* & 0.275 (4) & $38^{+6}_{-4}$ & 43 & 8.4 $\pm$ 1.7 (2) &
$1031^{+48}_{-38}$ & 1191 & $<$ 0.098  &  \ldots  \\
SPT-CL J2337-5942 & 0.775 (1) &  $387^{+35}_{-34}$ & 98  & 8.14 $\pm$ 1.14 (5) &
$2108^{+128}_{_130}$ & 799 & $<$ 1.67 & $1.0^{+1.5}_{-0.7}$  \\
SPT-CL J2345-6406 & 0.94 (5) &  $106^{+8}_{-15}$ & 89 & 5.1 $\pm$ 0.86 (1) &
$792^{+72}_{-51}$ & 599 & $<$ 2.59 & $250^{+130}_{-89}$   \\
SPT-CL J2352-4657 & 0.73 (5) &  \ldots & \ldots & 4.42 $\pm$ 0.83 (1) &
$141^{+40}_{-24}$ & 674  & $<$ 2.40 & $<9.5$  \\
SPT-CL J2359-5009 & 0.775 (1) &  $35^{+14}_{-8}$ & 62 & 3.54 $\pm$ 0.54 (5) &
$281^{+41}_{-30}$ & 605 & 5.96 $\pm$ 0.31 & $8.4^{+13}_{-6.0}$    \\
\hline
\end{tabular}}
\\ References: (1) \citet{blee15}; (2) \citet{hass13}; (3) \citet{hilt13}; (4)
\citet{marr11}; (5) \citet{reic13}; (6) \citet{ruel14};
(7) \citet{sifo13}.\\
$^a$Alternative names for ACT-CL J0102-4915 (El Gordo); ACT-CL J0152-0100
(A267); SPT-CL J0232-4420 (RXCJ0232.2-4420); ACT-CL J0245-5302 (AS0295); ACT-CL
J0326-0043 (MACS J0326-0043); SPT-CL J0542-4100 (RDCS J0542-4100); SPT-CL
J2011-5725 (RXCJ2011.3-5725); ACT-CL J2129-0005 (RXJ2129.6+0005); SPT-CL
J2248-4431 (AS1063); SPT-CL J2325-4111 (ACOS1121); ACT-CL J2337-0016 (A2631);
SPT-CL J2344-4242 (phoenix). The asterisk marks systems with uncertain
core positions.\\
$^b$Rest-frame monochromatic radio luminosity at 843 MHz using the flux densities from
SUMSS \citep{bock99}, except  ACT-CL J0014-0056, ACT-CL J0022-0036, ACT-CL
J0059-0049, ACT-CL J0152-0100, ACT-CL J0206-0114, ACT-CL J2129-0005, ACT-CL
J2154-0049, and ACT-CL J2337-0016 where NVSS flux densities were used \citep{cond98};
ACT-CL J0326-0043 where the FIRST flux density was used \citep{helf15}; and for
ACT-CL J0102-4915 and ACT-CL J0152-0100 where the GMRT flux density at
610 MHz from \citet{lind14} and \citet{kale13}, respectively, was used. The systems marked with asterisk
are the ones for which we have GMRT data at 325 MHz (Intema et al. in preparation). For SPT-CL
J0106-5943, SPT-CL J2135-5726 and SPT-CL J2248-4431, we measured the flux
densities from SUMSS images ($5.7 \pm 1.9$ mJy, $5.3 \pm 2.5$ mJy and $15.4 \pm
4.9$ mJy, respectively). The numbers without errors are the upper limit using the
noise in the SUMSS or NVSS image: 6--10 mJy beam$^{-1}$ (depending on the
declination) for SUMSS \citep{mauc03} and 2.5 mJy beam$^{-1}$ for NVSS
\citep{cond98}.\\
$^c$Star formation rates from \citet{mcdo16}.
\end{minipage} \end{table*}

\clearpage

\section*{Acknowledgements}

The scientific results reported in this article are based on data obtained from
the Chandra Data Archive.
This research has made use of software provided by the
Chandra X-ray Center (CXC) in the application packages, such as CIAO.
The authors thank the referee for constructive comments which
improved the clarity and the flow of the paper.

\bibliographystyle{mnras}
\bibliography{/Users/Laura/Documents/Bibliography/master_references}

\begin{thebibliography}{}
\makeatletter
\relax
\def\mn@urlcharsother{\let\do\@makeother \do\$\do\&\do\#\do\^\do\_\do\%\do\~}
\def\mn@doi{\begingroup\mn@urlcharsother \@ifnextchar [ {\mn@doi@}
  {\mn@doi@[]}}
\def\mn@doi@[#1]#2{\def\@tempa{#1}\ifx\@tempa\@empty \href
  {http://dx.doi.org/#2} {doi:#2}\else \href {http://dx.doi.org/#2} {#1}\fi
  \endgroup}
\def\mn@eprint#1#2{\mn@eprint@#1:#2::\@nil}
\def\mn@eprint@arXiv#1{\href {http://arxiv.org/abs/#1} {{\tt arXiv:#1}}}
\def\mn@eprint@dblp#1{\href {http://dblp.uni-trier.de/rec/bibtex/#1.xml}
  {dblp:#1}}
\def\mn@eprint@#1:#2:#3:#4\@nil{\def\@tempa {#1}\def\@tempb {#2}\def\@tempc
  {#3}\ifx \@tempc \@empty \let \@tempc \@tempb \let \@tempb \@tempa \fi \ifx
  \@tempb \@empty \def\@tempb {arXiv}\fi \@ifundefined
  {mn@eprint@\@tempb}{\@tempb:\@tempc}{\expandafter \expandafter \csname
  mn@eprint@\@tempb\endcsname \expandafter{\@tempc}}}

\bibitem[\protect\citeauthoryear{{Ahoranta}, {Finoguenov}, {Pinto}, {Sanders},
  {Kaastra}, {de Plaa}  \& {Fabian}}{{Ahoranta} et~al.}{2016}]{ahor16}
{Ahoranta} J.,  {Finoguenov} A.,  {Pinto} C.,  {Sanders} J.,  {Kaastra} J.,
  {de Plaa} J.,   {Fabian} A.,  2016, \mn@doi [\aap]
  {10.1051/0004-6361/201527523}, \href
  {http://adsabs.harvard.edu/abs/2016A%26A...592A.145A} {592, A145}

\bibitem[\protect\citeauthoryear{{Antognini}, {Bird}  \& {Martini}}{{Antognini}
  et~al.}{2012}]{anto12}
{Antognini} J.,  {Bird} J.,   {Martini} P.,  2012, \mn@doi [\apj]
  {10.1088/0004-637X/756/2/116}, \href
  {http://adsabs.harvard.edu/abs/2012ApJ...756..116A} {756, 116}

\bibitem[\protect\citeauthoryear{{Baldi}, {Forman}, {Jones}, {Kraft}, {Nulsen},
  {Churazov}, {David}  \& {Giacintucci}}{{Baldi} et~al.}{2009}]{bald09}
{Baldi} A.,  {Forman} W.,  {Jones} C.,  {Kraft} R.,  {Nulsen} P.,  {Churazov}
  E.,  {David} L.,   {Giacintucci} S.,  2009, \mn@doi [\apj]
  {10.1088/0004-637X/707/2/1034}, \href
  {http://adsabs.harvard.edu/abs/2009ApJ...707.1034B} {707, 1034}

\bibitem[\protect\citeauthoryear{{Baldi}, {Capetti}  \& {Giovannini}}{{Baldi}
  et~al.}{2015}]{bald15}
{Baldi} R.~D.,  {Capetti} A.,   {Giovannini} G.,  2015, \mn@doi [\aap]
  {10.1051/0004-6361/201425426}, \href
  {http://adsabs.harvard.edu/abs/2015A%26A...576A..38B} {576, A38}

\bibitem[\protect\citeauthoryear{{Banerjee} \& {Sharma}}{{Banerjee} \&
  {Sharma}}{2014}]{bane14}
{Banerjee} N.,  {Sharma} P.,  2014, \mn@doi [\mnras] {10.1093/mnras/stu1179},
  \href {http://adsabs.harvard.edu/abs/2014MNRAS.443..687B} {443, 687}

\bibitem[\protect\citeauthoryear{{Belsole}, {Worrall}, {Hardcastle}  \&
  {Croston}}{{Belsole} et~al.}{2007}]{bels07}
{Belsole} E.,  {Worrall} D.~M.,  {Hardcastle} M.~J.,   {Croston} J.~H.,  2007,
  \mn@doi [\mnras] {10.1111/j.1365-2966.2007.12298.x}, \href
  {http://adsabs.harvard.edu/abs/2007MNRAS.381.1109B} {381, 1109}

\bibitem[\protect\citeauthoryear{{Best} \& {Heckman}}{{Best} \&
  {Heckman}}{2012}]{best12}
{Best} P.~N.,  {Heckman} T.~M.,  2012, \mn@doi [\mnras]
  {10.1111/j.1365-2966.2012.20414.x}, \href
  {http://adsabs.harvard.edu/abs/2012MNRAS.421.1569B} {421, 1569}

\bibitem[\protect\citeauthoryear{{Best}, {Kauffmann}, {Heckman}, {Brinchmann},
  {Charlot}, {Ivezi{\'c}}  \& {White}}{{Best} et~al.}{2005}]{best05}
{Best} P.~N.,  {Kauffmann} G.,  {Heckman} T.~M.,  {Brinchmann} J.,  {Charlot}
  S.,  {Ivezi{\'c}} {\v Z}.,   {White} S.~D.~M.,  2005, \mn@doi [\mnras]
  {10.1111/j.1365-2966.2005.09192.x}, \href
  {http://adsabs.harvard.edu/abs/2005MNRAS.362...25B} {362, 25}

\bibitem[\protect\citeauthoryear{{Best}, {Kaiser}, {Heckman}  \&
  {Kauffmann}}{{Best} et~al.}{2006}]{best06}
{Best} P.~N.,  {Kaiser} C.~R.,  {Heckman} T.~M.,   {Kauffmann} G.,  2006,
  \mn@doi [\mnras] {10.1111/j.1745-3933.2006.00159.x}, \href
  {http://adsabs.harvard.edu/abs/2006MNRAS.368L..67B} {368, L67}

\bibitem[\protect\citeauthoryear{{Best}, {Ker}, {Simpson}, {Rigby}  \&
  {Sabater}}{{Best} et~al.}{2014}]{best14}
{Best} P.~N.,  {Ker} L.~M.,  {Simpson} C.,  {Rigby} E.~E.,   {Sabater} J.,
  2014, \mn@doi [\mnras] {10.1093/mnras/stu1776}, \href
  {http://adsabs.harvard.edu/abs/2014MNRAS.445..955B} {445, 955}

\bibitem[\protect\citeauthoryear{{B{\^\i}rzan}, {Rafferty}, {McNamara}, {Wise}
  \& {Nulsen}}{{B{\^\i}rzan} et~al.}{2004}]{birz04}
{B{\^\i}rzan} L.,  {Rafferty} D.~A.,  {McNamara} B.~R.,  {Wise} M.~W.,
  {Nulsen} P.~E.~J.,  2004, \mn@doi [\apj] {10.1086/383519}, \href
  {http://adsabs.harvard.edu/abs/2004ApJ...607..800B} {607, 800}

\bibitem[\protect\citeauthoryear{{B{\^\i}rzan}, {McNamara}, {Nulsen}, {Carilli}
   \& {Wise}}{{B{\^\i}rzan} et~al.}{2008}]{birz08}
{B{\^\i}rzan} L.,  {McNamara} B.~R.,  {Nulsen} P.~E.~J.,  {Carilli} C.~L.,
  {Wise} M.~W.,  2008, \mn@doi [\apj] {10.1086/591416}, \href
  {http://adsabs.harvard.edu/abs/2008ApJ...686..859B} {686, 859}

\bibitem[\protect\citeauthoryear{{B{\^\i}rzan}, {Rafferty}, {Nulsen},
  {McNamara}, {R{\"o}ttgering}, {Wise}  \& {Mittal}}{{B{\^\i}rzan}
  et~al.}{2012}]{birz12}
{B{\^\i}rzan} L.,  {Rafferty} D.~A.,  {Nulsen} P.~E.~J.,  {McNamara} B.~R.,
  {R{\"o}ttgering} H.~J.~A.,  {Wise} M.~W.,   {Mittal} R.,  2012, \mn@doi
  [\mnras] {10.1111/j.1365-2966.2012.22083.x}, \href
  {http://adsabs.harvard.edu/abs/2012MNRAS.427.3468B} {427, 3468}

\bibitem[\protect\citeauthoryear{{Blanton}, {Sarazin}, {McNamara}  \&
  {Wise}}{{Blanton} et~al.}{2001}]{blan01}
{Blanton} E.~L.,  {Sarazin} C.~L.,  {McNamara} B.~R.,   {Wise} M.~W.,  2001,
  \mn@doi [\apjl] {10.1086/323269}, \href
  {http://adsabs.harvard.edu/abs/2001ApJ...558L..15B} {558, L15}

\bibitem[\protect\citeauthoryear{{Bleem} et~al.,}{{Bleem}
  et~al.}{2015}]{blee15}
{Bleem} L.~E.,  et~al., 2015, \mn@doi [\apjs] {10.1088/0067-0049/216/2/27},
  \href {http://adsabs.harvard.edu/abs/2015ApJS..216...27B} {216, 27}

\bibitem[\protect\citeauthoryear{{Bock}, {Large}  \& {Sadler}}{{Bock}
  et~al.}{1999}]{bock99}
{Bock} D.,  {Large} M.~I.,   {Sadler} E.~M.,  1999, \mn@doi [\aj]
  {10.1086/300786}, \href {http://adsabs.harvard.edu/abs/1999AJ....117.1578B}
  {117, 1578}

\bibitem[\protect\citeauthoryear{{B{\"o}hringer} et~al.,}{{B{\"o}hringer}
  et~al.}{2004}]{bohr04}
{B{\"o}hringer} H.,  et~al., 2004, \mn@doi [\aap] {10.1051/0004-6361:20034484},
  \href {http://adsabs.harvard.edu/abs/2004A%26A...425..367B} {425, 367}

\bibitem[\protect\citeauthoryear{{Botteon}, {Gastaldello}, {Brunetti}  \&
  {Kale}}{{Botteon} et~al.}{2016}]{bott16}
{Botteon} A.,  {Gastaldello} F.,  {Brunetti} G.,   {Kale} R.,  2016, \mn@doi
  [\mnras] {10.1093/mnras/stw2089}, \href
  {http://adsabs.harvard.edu/abs/2016MNRAS.463.1534B} {463, 1534}

\bibitem[\protect\citeauthoryear{{Branchesi}, {Gioia}, {Fanti}, {Fanti}  \&
  {Perley}}{{Branchesi} et~al.}{2006}]{bran06}
{Branchesi} M.,  {Gioia} I.~M.,  {Fanti} C.,  {Fanti} R.,   {Perley} R.,  2006,
  \mn@doi [\aap] {10.1051/0004-6361:20053767}, \href
  {http://adsabs.harvard.edu/abs/2006A%26A...446...97B} {446, 97}

\bibitem[\protect\citeauthoryear{{Brighenti}, {Mathews}  \& {Temi}}{{Brighenti}
  et~al.}{2015}]{brig15}
{Brighenti} F.,  {Mathews} W.~G.,   {Temi} P.,  2015, \mn@doi [\apj]
  {10.1088/0004-637X/802/2/118}, \href
  {http://adsabs.harvard.edu/abs/2015ApJ...802..118B} {802, 118}

\bibitem[\protect\citeauthoryear{{Brunetti} \& {Jones}}{{Brunetti} \&
  {Jones}}{2014}]{brun14}
{Brunetti} G.,  {Jones} T.~W.,  2014, \mn@doi [International Journal of Modern
  Physics D] {10.1142/S0218271814300079}, \href
  {http://adsabs.harvard.edu/abs/2014IJMPD..2330007B} {23, 1430007}

\bibitem[\protect\citeauthoryear{{Carlstrom} et~al.,}{{Carlstrom}
  et~al.}{2011}]{carl11}
{Carlstrom} J.~E.,  et~al., 2011, \mn@doi [\pasp] {10.1086/659879}, \href
  {http://adsabs.harvard.edu/abs/2011PASP..123..568C} {123, 568}

\bibitem[\protect\citeauthoryear{{Cavagnolo}, {Donahue}, {Voit}  \&
  {Sun}}{{Cavagnolo} et~al.}{2008}]{cava08}
{Cavagnolo} K.~W.,  {Donahue} M.,  {Voit} G.~M.,   {Sun} M.,  2008, \mn@doi
  [\apjl] {10.1086/591665}, \href
  {http://adsabs.harvard.edu/abs/2008ApJ...683L.107C} {683, L107}

\bibitem[\protect\citeauthoryear{{Cavagnolo}, {McNamara}, {Nulsen}, {Carilli},
  {Jones}  \& {B{\^\i}rzan}}{{Cavagnolo} et~al.}{2010}]{cava10}
{Cavagnolo} K.~W.,  {McNamara} B.~R.,  {Nulsen} P.~E.~J.,  {Carilli} C.~L.,
  {Jones} C.,   {B{\^\i}rzan} L.,  2010, \mn@doi [\apj]
  {10.1088/0004-637X/720/2/1066}, \href
  {http://adsabs.harvard.edu/abs/2010ApJ...720.1066C} {720, 1066}

\bibitem[\protect\citeauthoryear{{Cavagnolo}, {McNamara}, {Wise}, {Nulsen},
  {Br{\"u}ggen}, {Gitti}  \& {Rafferty}}{{Cavagnolo} et~al.}{2011}]{cava11}
{Cavagnolo} K.~W.,  {McNamara} B.~R.,  {Wise} M.~W.,  {Nulsen} P.~E.~J.,
  {Br{\"u}ggen} M.,  {Gitti} M.,   {Rafferty} D.~A.,  2011, \mn@doi [\apj]
  {10.1088/0004-637X/732/2/71}, \href
  {http://adsabs.harvard.edu/abs/2011ApJ...732...71C} {732, 71}

\bibitem[\protect\citeauthoryear{{Churazov}, {Sazonov}, {Sunyaev}, {Forman},
  {Jones}  \& {B{\"o}hringer}}{{Churazov} et~al.}{2005}]{chur05}
{Churazov} E.,  {Sazonov} S.,  {Sunyaev} R.,  {Forman} W.,  {Jones} C.,
  {B{\"o}hringer} H.,  2005, \mn@doi [\mnras]
  {10.1111/j.1745-3933.2005.00093.x}, \href
  {http://adsabs.harvard.edu/abs/2005MNRAS.363L..91C} {363, L91}

\bibitem[\protect\citeauthoryear{{Condon}, {Cotton}, {Greisen}, {Yin},
  {Perley}, {Taylor}  \& {Broderick}}{{Condon} et~al.}{1998}]{cond98}
{Condon} J.~J.,  {Cotton} W.~D.,  {Greisen} E.~W.,  {Yin} Q.~F.,  {Perley}
  R.~A.,  {Taylor} G.~B.,   {Broderick} J.~J.,  1998, \mn@doi [\aj]
  {10.1086/300337}, \href {http://adsabs.harvard.edu/abs/1998AJ....115.1693C}
  {115, 1693}

\bibitem[\protect\citeauthoryear{{Daly}, {Sprinkle}, {O'Dea}, {Kharb}  \&
  {Baum}}{{Daly} et~al.}{2012}]{daly12}
{Daly} R.~A.,  {Sprinkle} T.~B.,  {O'Dea} C.~P.,  {Kharb} P.,   {Baum} S.~A.,
  2012, \mn@doi [\mnras] {10.1111/j.1365-2966.2012.21060.x}, \href
  {http://adsabs.harvard.edu/abs/2012MNRAS.423.2498D} {423, 2498}

\bibitem[\protect\citeauthoryear{{Danielson}, {Lehmer}, {Alexander}, {Brandt},
  {Luo}, {Miller}, {Xue}  \& {Stott}}{{Danielson} et~al.}{2012}]{dani12}
{Danielson} A.~L.~R.,  {Lehmer} B.~D.,  {Alexander} D.~M.,  {Brandt} W.~N.,
  {Luo} B.,  {Miller} N.,  {Xue} Y.~Q.,   {Stott} J.~P.,  2012, \mn@doi
  [\mnras] {10.1111/j.1365-2966.2012.20626.x}, \href
  {http://adsabs.harvard.edu/abs/2012MNRAS.422..494D} {422, 494}

\bibitem[\protect\citeauthoryear{{Dickey} \& {Lockman}}{{Dickey} \&
  {Lockman}}{1990}]{dick90}
{Dickey} J.~M.,  {Lockman} F.~J.,  1990, \mn@doi [\araa]
  {10.1146/annurev.aa.28.090190.001243}, \href
  {http://adsabs.harvard.edu/abs/1990ARA%26A..28..215D} {28, 215}

\bibitem[\protect\citeauthoryear{{Donahue}, {Stocke}  \& {Gioia}}{{Donahue}
  et~al.}{1992}]{dona92}
{Donahue} M.,  {Stocke} J.~T.,   {Gioia} I.~M.,  1992, \mn@doi [\apj]
  {10.1086/170914}, \href {http://adsabs.harvard.edu/abs/1992ApJ...385...49D}
  {385, 49}

\bibitem[\protect\citeauthoryear{{Donahue} et~al.,}{{Donahue}
  et~al.}{2010}]{dona10}
{Donahue} M.,  et~al., 2010, \mn@doi [\apj] {10.1088/0004-637X/715/2/881},
  \href {http://adsabs.harvard.edu/abs/2010ApJ...715..881D} {715, 881}

\bibitem[\protect\citeauthoryear{{Donahue} et~al.,}{{Donahue}
  et~al.}{2015}]{dona15}
{Donahue} M.,  et~al., 2015, \mn@doi [\apj] {10.1088/0004-637X/805/2/177},
  \href {http://adsabs.harvard.edu/abs/2015ApJ...805..177D} {805, 177}

\bibitem[\protect\citeauthoryear{{Dunlop} \& {Peacock}}{{Dunlop} \&
  {Peacock}}{1990}]{dunl90}
{Dunlop} J.~S.,  {Peacock} J.~A.,  1990, \mnras, \href
  {http://adsabs.harvard.edu/abs/1990MNRAS.247...19D} {247, 19}

\bibitem[\protect\citeauthoryear{{Dunn} \& {Fabian}}{{Dunn} \&
  {Fabian}}{2004}]{dunn04}
{Dunn} R.~J.~H.,  {Fabian} A.~C.,  2004, \mn@doi [\mnras]
  {10.1111/j.1365-2966.2004.08365.x}, \href
  {http://adsabs.harvard.edu/abs/2004MNRAS.355..862D} {355, 862}

\bibitem[\protect\citeauthoryear{{Edge}, {Stewart}, {Fabian}  \&
  {Arnaud}}{{Edge} et~al.}{1990}]{edge90}
{Edge} A.~C.,  {Stewart} G.~C.,  {Fabian} A.~C.,   {Arnaud} K.~A.,  1990,
  \mnras, \href {http://adsabs.harvard.edu/abs/1990MNRAS.245..559E} {245, 559}

\bibitem[\protect\citeauthoryear{{Evans}, {Worrall}, {Hardcastle}, {Kraft}  \&
  {Birkinshaw}}{{Evans} et~al.}{2006}]{evan06}
{Evans} D.~A.,  {Worrall} D.~M.,  {Hardcastle} M.~J.,  {Kraft} R.~P.,
  {Birkinshaw} M.,  2006, \mn@doi [\apj] {10.1086/500658}, \href
  {http://adsabs.harvard.edu/abs/2006ApJ...642...96E} {642, 96}

\bibitem[\protect\citeauthoryear{{Fabian}}{{Fabian}}{1994}]{fabi94}
{Fabian} A.~C.,  1994, \mn@doi [\araa] {10.1146/annurev.aa.32.090194.001425},
  \href {http://adsabs.harvard.edu/abs/1994ARA%26A..32..277F} {32, 277}

\bibitem[\protect\citeauthoryear{{Fabian}}{{Fabian}}{2012}]{fabi12}
{Fabian} A.~C.,  2012, \mn@doi [\araa] {10.1146/annurev-astro-081811-125521},
  \href {http://adsabs.harvard.edu/abs/2012ARA%26A..50..455F} {50, 455}

\bibitem[\protect\citeauthoryear{{Fabian} et~al.,}{{Fabian}
  et~al.}{2000}]{fabi00}
{Fabian} A.~C.,  et~al., 2000, \mnras, \href
  {http://adsabs.harvard.edu/abs/2000MNRAS.318L..65F} {318, L65}

\bibitem[\protect\citeauthoryear{{Fabian}, {Celotti}  \& {Erlund}}{{Fabian}
  et~al.}{2006}]{fabi06b}
{Fabian} A.~C.,  {Celotti} A.,   {Erlund} M.~C.,  2006, \mn@doi [\mnras]
  {10.1111/j.1745-3933.2006.00234.x}, \href
  {http://adsabs.harvard.edu/abs/2006MNRAS.373L..16F} {373, L16}

\bibitem[\protect\citeauthoryear{{Fabian} et~al.,}{{Fabian}
  et~al.}{2016}]{fabi16}
{Fabian} A.~C.,  et~al., 2016, \mn@doi [\mnras] {10.1093/mnras/stw1350}, \href
  {http://adsabs.harvard.edu/abs/2016MNRAS.461..922F} {461, 922}

\bibitem[\protect\citeauthoryear{{Fabian}, {Walker}, {Russell}, {Pinto},
  {Sanders}  \& {Reynolds}}{{Fabian} et~al.}{2017}]{fabi17}
{Fabian} A.~C.,  {Walker} S.~A.,  {Russell} H.~R.,  {Pinto} C.,  {Sanders}
  J.~S.,   {Reynolds} C.~S.,  2017, \mn@doi [\mnras] {10.1093/mnrasl/slw170},
  \href {http://adsabs.harvard.edu/abs/2017MNRAS.464L...1F} {464, L1}

\bibitem[\protect\citeauthoryear{{Fanaroff} \& {Riley}}{{Fanaroff} \&
  {Riley}}{1974}]{fana74}
{Fanaroff} B.~L.,  {Riley} J.~M.,  1974, \mnras, \href
  {http://adsabs.harvard.edu/abs/1974MNRAS.167P..31F} {167, 31P}

\bibitem[\protect\citeauthoryear{{Farage}, {McGregor}  \& {Dopita}}{{Farage}
  et~al.}{2012}]{fara12}
{Farage} C.~L.,  {McGregor} P.~J.,   {Dopita} M.~A.,  2012, \mn@doi [\apj]
  {10.1088/0004-637X/747/1/28}, \href
  {http://adsabs.harvard.edu/abs/2012ApJ...747...28F} {747, 28}

\bibitem[\protect\citeauthoryear{{Fernandes} et~al.,}{{Fernandes}
  et~al.}{2015}]{fern15}
{Fernandes} C.~A.~C.,  et~al., 2015, \mn@doi [\mnras] {10.1093/mnras/stu2517},
  \href {http://adsabs.harvard.edu/abs/2015MNRAS.447.1184F} {447, 1184}

\bibitem[\protect\citeauthoryear{{Fogarty}, {Postman}, {Connor}, {Donahue}  \&
  {Moustakas}}{{Fogarty} et~al.}{2015}]{foga15}
{Fogarty} K.,  {Postman} M.,  {Connor} T.,  {Donahue} M.,   {Moustakas} J.,
  2015, \mn@doi [\apj] {10.1088/0004-637X/813/2/117}, \href
  {http://adsabs.harvard.edu/abs/2015ApJ...813..117F} {813, 117}

\bibitem[\protect\citeauthoryear{{Forman} et~al.,}{{Forman}
  et~al.}{2005}]{form05}
{Forman} W.,  et~al., 2005, \mn@doi [\apj] {10.1086/429746}, \href
  {http://adsabs.harvard.edu/abs/2005ApJ...635..894F} {635, 894}

\bibitem[\protect\citeauthoryear{{Fowler} et~al.,}{{Fowler}
  et~al.}{2007}]{fowl07}
{Fowler} J.~W.,  et~al., 2007, \mn@doi [\ao] {10.1364/AO.46.003444}, \href
  {http://adsabs.harvard.edu/abs/2007ApOpt..46.3444F} {46, 3444}

\bibitem[\protect\citeauthoryear{{Gaspari}, {Ruszkowski}  \&
  {Sharma}}{{Gaspari} et~al.}{2012}]{gasp12}
{Gaspari} M.,  {Ruszkowski} M.,   {Sharma} P.,  2012, \mn@doi [\apj]
  {10.1088/0004-637X/746/1/94}, \href
  {http://adsabs.harvard.edu/abs/2012ApJ...746...94G} {746, 94}

\bibitem[\protect\citeauthoryear{{Gaspari}, {Ruszkowski}  \& {Oh}}{{Gaspari}
  et~al.}{2013}]{gasp13}
{Gaspari} M.,  {Ruszkowski} M.,   {Oh} S.~P.,  2013, \mn@doi [\mnras]
  {10.1093/mnras/stt692}, \href
  {http://adsabs.harvard.edu/abs/2013MNRAS.432.3401G} {432, 3401}

\bibitem[\protect\citeauthoryear{{Gaspari}, {Brighenti}  \& {Temi}}{{Gaspari}
  et~al.}{2015}]{gasp15}
{Gaspari} M.,  {Brighenti} F.,   {Temi} P.,  2015, \mn@doi [\aap]
  {10.1051/0004-6361/201526151}, \href
  {http://adsabs.harvard.edu/abs/2015A%26A...579A..62G} {579, A62}

\bibitem[\protect\citeauthoryear{{Gaspari}, {Temi}  \& {Brighenti}}{{Gaspari}
  et~al.}{2017}]{gasp17}
{Gaspari} M.,  {Temi} P.,   {Brighenti} F.,  2017, \mn@doi [\mnras]
  {10.1093/mnras/stw3108}, \href
  {http://adsabs.harvard.edu/abs/2017MNRAS.466..677G} {466, 677}

\bibitem[\protect\citeauthoryear{{Gendre}, {Best}, {Wall}  \& {Ker}}{{Gendre}
  et~al.}{2013}]{gend13}
{Gendre} M.~A.,  {Best} P.~N.,  {Wall} J.~V.,   {Ker} L.~M.,  2013, \mn@doi
  [\mnras] {10.1093/mnras/stt116}, \href
  {http://adsabs.harvard.edu/abs/2013MNRAS.430.3086G} {430, 3086}

\bibitem[\protect\citeauthoryear{{Giacintucci}, {Markevitch}, {Cassano},
  {Venturi}, {Clarke}  \& {Brunetti}}{{Giacintucci} et~al.}{2017}]{giac17}
{Giacintucci} S.,  {Markevitch} M.,  {Cassano} R.,  {Venturi} T.,  {Clarke}
  T.~E.,   {Brunetti} G.,  2017, preprint, \href
  {http://adsabs.harvard.edu/abs/2017arXiv170101364G} {} (\mn@eprint {arXiv}
  {1701.01364})

\bibitem[\protect\citeauthoryear{{Gioia}, {Henry}, {Mullis}, {B{\"o}hringer},
  {Briel}, {Voges}  \& {Huchra}}{{Gioia} et~al.}{2003}]{gioi03}
{Gioia} I.~M.,  {Henry} J.~P.,  {Mullis} C.~R.,  {B{\"o}hringer} H.,  {Briel}
  U.~G.,  {Voges} W.,   {Huchra} J.~P.,  2003, \mn@doi [\apjs]
  {10.1086/378229}, \href {http://adsabs.harvard.edu/abs/2003ApJS..149...29G}
  {149, 29}

\bibitem[\protect\citeauthoryear{{Godfrey} \& {Shabala}}{{Godfrey} \&
  {Shabala}}{2013}]{godf13}
{Godfrey} L.~E.~H.,  {Shabala} S.~S.,  2013, \mn@doi [\apj]
  {10.1088/0004-637X/767/1/12}, \href
  {http://adsabs.harvard.edu/abs/2013ApJ...767...12G} {767, 12}

\bibitem[\protect\citeauthoryear{{Godfrey} \& {Shabala}}{{Godfrey} \&
  {Shabala}}{2016}]{godf16}
{Godfrey} L.~E.~H.,  {Shabala} S.~S.,  2016, \mn@doi [\mnras]
  {10.1093/mnras/stv2712}, \href
  {http://adsabs.harvard.edu/abs/2016MNRAS.456.1172G} {456, 1172}

\bibitem[\protect\citeauthoryear{{Gralla}, {Gladders}, {Yee}  \&
  {Barrientos}}{{Gralla} et~al.}{2011}]{gral11}
{Gralla} M.~B.,  {Gladders} M.~D.,  {Yee} H.~K.~C.,   {Barrientos} L.~F.,
  2011, \mn@doi [\apj] {10.1088/0004-637X/734/2/103}, \href
  {http://adsabs.harvard.edu/abs/2011ApJ...734..103G} {734, 103}

\bibitem[\protect\citeauthoryear{{Grimes}, {Rawlings}  \& {Willott}}{{Grimes}
  et~al.}{2004}]{grim04}
{Grimes} J.~A.,  {Rawlings} S.,   {Willott} C.~J.,  2004, \mn@doi [\mnras]
  {10.1111/j.1365-2966.2004.07510.x}, \href
  {http://adsabs.harvard.edu/abs/2004MNRAS.349..503G} {349, 503}

\bibitem[\protect\citeauthoryear{{Guo} \& {Mathews}}{{Guo} \&
  {Mathews}}{2014}]{guo14}
{Guo} F.,  {Mathews} W.~G.,  2014, \mn@doi [\apj]
  {10.1088/0004-637X/780/2/126}, \href
  {http://adsabs.harvard.edu/abs/2014ApJ...780..126G} {780, 126}

\bibitem[\protect\citeauthoryear{{Gupta} et~al.,}{{Gupta}
  et~al.}{2016}]{gupt16}
{Gupta} N.,  et~al., 2016, preprint, \href
  {http://adsabs.harvard.edu/abs/2016arXiv160505329G} {} (\mn@eprint {arXiv}
  {1605.05329})

\bibitem[\protect\citeauthoryear{{Hall} \& {Green}}{{Hall} \&
  {Green}}{1998}]{hall98}
{Hall} P.~B.,  {Green} R.~F.,  1998, \mn@doi [\apj] {10.1086/306349}, \href
  {http://adsabs.harvard.edu/abs/1998ApJ...507..558H} {507, 558}

\bibitem[\protect\citeauthoryear{{Hardcastle}, {Evans}  \&
  {Croston}}{{Hardcastle} et~al.}{2006}]{hard06}
{Hardcastle} M.~J.,  {Evans} D.~A.,   {Croston} J.~H.,  2006, \mn@doi [\mnras]
  {10.1111/j.1365-2966.2006.10615.x}, \href
  {http://adsabs.harvard.edu/abs/2006MNRAS.370.1893H} {370, 1893}

\bibitem[\protect\citeauthoryear{{Hardcastle}, {Evans}  \&
  {Croston}}{{Hardcastle} et~al.}{2007}]{hard07}
{Hardcastle} M.~J.,  {Evans} D.~A.,   {Croston} J.~H.,  2007, \mn@doi [\mnras]
  {10.1111/j.1365-2966.2007.11572.x}, \href
  {http://adsabs.harvard.edu/abs/2007MNRAS.376.1849H} {376, 1849}

\bibitem[\protect\citeauthoryear{{Hasselfield} et~al.,}{{Hasselfield}
  et~al.}{2013}]{hass13}
{Hasselfield} M.,  et~al., 2013, \mn@doi [\jcap]
  {10.1088/1475-7516/2013/07/008}, \href
  {http://adsabs.harvard.edu/abs/2013JCAP...07..008H} {7, 008}

\bibitem[\protect\citeauthoryear{{Heckman} \& {Best}}{{Heckman} \&
  {Best}}{2014}]{heck14}
{Heckman} T.~M.,  {Best} P.~N.,  2014, \mn@doi [\araa]
  {10.1146/annurev-astro-081913-035722}, \href
  {http://adsabs.harvard.edu/abs/2014ARA%26A..52..589H} {52, 589}

\bibitem[\protect\citeauthoryear{{Helfand}, {White}  \& {Becker}}{{Helfand}
  et~al.}{2015}]{helf15}
{Helfand} D.~J.,  {White} R.~L.,   {Becker} R.~H.,  2015, \mn@doi [\apj]
  {10.1088/0004-637X/801/1/26}, \href
  {http://adsabs.harvard.edu/abs/2015ApJ...801...26H} {801, 26}

\bibitem[\protect\citeauthoryear{{Hicks}, {Mushotzky}  \& {Donahue}}{{Hicks}
  et~al.}{2010}]{hick10}
{Hicks} A.~K.,  {Mushotzky} R.,   {Donahue} M.,  2010, \mn@doi [\apj]
  {10.1088/0004-637X/719/2/1844}, \href
  {http://adsabs.harvard.edu/abs/2010ApJ...719.1844H} {719, 1844}

\bibitem[\protect\citeauthoryear{{Hill} \& {Lilly}}{{Hill} \&
  {Lilly}}{1991}]{hill91}
{Hill} G.~J.,  {Lilly} S.~J.,  1991, \mn@doi [\apj] {10.1086/169597}, \href
  {http://adsabs.harvard.edu/abs/1991ApJ...367....1H} {367, 1}

\bibitem[\protect\citeauthoryear{{Hillel} \& {Soker}}{{Hillel} \&
  {Soker}}{2014}]{hill14}
{Hillel} S.,  {Soker} N.,  2014, preprint, \href
  {http://adsabs.harvard.edu/abs/2014arXiv1403.5137H} {} (\mn@eprint {arXiv}
  {1403.5137})

\bibitem[\protect\citeauthoryear{{Hillel} \& {Soker}}{{Hillel} \&
  {Soker}}{2017}]{hill16}
{Hillel} S.,  {Soker} N.,  2017, \mn@doi [\mnras] {10.1093/mnrasl/slw231},
  \href {http://adsabs.harvard.edu/abs/2017MNRAS.466L..39H} {466, L39}

\bibitem[\protect\citeauthoryear{{Hilton} et~al.,}{{Hilton}
  et~al.}{2013}]{hilt13}
{Hilton} M.,  et~al., 2013, \mn@doi [\mnras] {10.1093/mnras/stt1535}, \href
  {http://adsabs.harvard.edu/abs/2013MNRAS.435.3469H} {435, 3469}

\bibitem[\protect\citeauthoryear{{Hine} \& {Longair}}{{Hine} \&
  {Longair}}{1979}]{hine79}
{Hine} R.~G.,  {Longair} M.~S.,  1979, \mnras, \href
  {http://adsabs.harvard.edu/abs/1979MNRAS.188..111H} {188, 111}

\bibitem[\protect\citeauthoryear{{Hlavacek-Larrondo}, {Fabian}, {Edge},
  {Ebeling}, {Sanders}, {Hogan}  \& {Taylor}}{{Hlavacek-Larrondo}
  et~al.}{2012}]{hlav12}
{Hlavacek-Larrondo} J.,  {Fabian} A.~C.,  {Edge} A.~C.,  {Ebeling} H.,
  {Sanders} J.~S.,  {Hogan} M.~T.,   {Taylor} G.~B.,  2012, \mn@doi [\mnras]
  {10.1111/j.1365-2966.2011.20405.x}, \href
  {http://adsabs.harvard.edu/abs/2012MNRAS.421.1360H} {421, 1360}

\bibitem[\protect\citeauthoryear{{Hlavacek-Larrondo}, {Fabian}, {Edge},
  {Ebeling}, {Allen}, {Sanders}  \& {Taylor}}{{Hlavacek-Larrondo}
  et~al.}{2013}]{hlav13}
{Hlavacek-Larrondo} J.,  {Fabian} A.~C.,  {Edge} A.~C.,  {Ebeling} H.,  {Allen}
  S.~W.,  {Sanders} J.~S.,   {Taylor} G.~B.,  2013, \mn@doi [\mnras]
  {10.1093/mnras/stt283}, \href
  {http://adsabs.harvard.edu/abs/2013MNRAS.431.1638H} {431, 1638}

\bibitem[\protect\citeauthoryear{{Hlavacek-Larrondo}
  et~al.,}{{Hlavacek-Larrondo} et~al.}{2015}]{hlav15}
{Hlavacek-Larrondo} J.,  et~al., 2015, \mn@doi [\apj]
  {10.1088/0004-637X/805/1/35}, \href
  {http://adsabs.harvard.edu/abs/2015ApJ...805...35H} {805, 35}

\bibitem[\protect\citeauthoryear{{Ineson}, {Croston}, {Hardcastle}  \&
  {Mingo}}{{Ineson} et~al.}{2017}]{ines17}
{Ineson} J.,  {Croston} J.~H.,  {Hardcastle} M.~J.,   {Mingo} B.,  2017,
  \mn@doi [\mnras] {10.1093/mnras/stx189}, \href
  {http://adsabs.harvard.edu/abs/2017MNRAS.467.1586I} {467, 1586}

\bibitem[\protect\citeauthoryear{{Intema}}{{Intema}}{2014}]{inte14}
{Intema} H.~T.,  2014, {SPAM: Source Peeling and Atmospheric Modeling},
  Astrophysics Source Code Library (\mn@eprint {ascl} {1408.006})

\bibitem[\protect\citeauthoryear{{Intema}, {van der Tol}, {Cotton}, {Cohen},
  {van Bemmel}  \& {R{\"o}ttgering}}{{Intema} et~al.}{2009}]{inte09}
{Intema} H.~T.,  {van der Tol} S.,  {Cotton} W.~D.,  {Cohen} A.~S.,  {van
  Bemmel} I.~M.,   {R{\"o}ttgering} H.~J.~A.,  2009, \mn@doi [\aap]
  {10.1051/0004-6361/200811094}, \href
  {http://adsabs.harvard.edu/abs/2009A%26A...501.1185I} {501, 1185}

\bibitem[\protect\citeauthoryear{{Kaiser} \& {Alexander}}{{Kaiser} \&
  {Alexander}}{1997}]{kais97}
{Kaiser} C.~R.,  {Alexander} P.,  1997, \mnras, \href
  {http://adsabs.harvard.edu/abs/1997MNRAS.286..215K} {286, 215}

\bibitem[\protect\citeauthoryear{{Kale}, {Venturi}, {Giacintucci}, {Dallacasa},
  {Cassano}, {Brunetti}, {Macario}  \& {Athreya}}{{Kale} et~al.}{2013}]{kale13}
{Kale} R.,  {Venturi} T.,  {Giacintucci} S.,  {Dallacasa} D.,  {Cassano} R.,
  {Brunetti} G.,  {Macario} G.,   {Athreya} R.,  2013, \mn@doi [\aap]
  {10.1051/0004-6361/201321515}, \href
  {http://adsabs.harvard.edu/abs/2013A%26A...557A..99K} {557, A99}

\bibitem[\protect\citeauthoryear{{Kaviraj}, {Shabala}, {Deller}  \&
  {Middelberg}}{{Kaviraj} et~al.}{2015}]{kavi15}
{Kaviraj} S.,  {Shabala} S.~S.,  {Deller} A.~T.,   {Middelberg} E.,  2015,
  \mn@doi [\mnras] {10.1093/mnras/stv1329}, \href
  {http://adsabs.harvard.edu/abs/2015MNRAS.452..774K} {452, 774}

\bibitem[\protect\citeauthoryear{{Kirkpatrick}, {McNamara}  \&
  {Cavagnolo}}{{Kirkpatrick} et~al.}{2011}]{kirk11}
{Kirkpatrick} C.~C.,  {McNamara} B.~R.,   {Cavagnolo} K.~W.,  2011, \mn@doi
  [\apjl] {10.1088/2041-8205/731/2/L23}, \href
  {http://adsabs.harvard.edu/abs/2011ApJ...731L..23K} {731, L23}

\bibitem[\protect\citeauthoryear{{Koekemoer}, {O'Dea}, {Sarazin}, {McNamara},
  {Donahue}, {Voit}, {Baum}  \& {Gallimore}}{{Koekemoer} et~al.}{1999}]{koek99}
{Koekemoer} A.~M.,  {O'Dea} C.~P.,  {Sarazin} C.~L.,  {McNamara} B.~R.,
  {Donahue} M.,  {Voit} G.~M.,  {Baum} S.~A.,   {Gallimore} J.~F.,  1999,
  \mn@doi [\apj] {10.1086/307911}, \href
  {http://adsabs.harvard.edu/abs/1999ApJ...525..621K} {525, 621}

\bibitem[\protect\citeauthoryear{{Kokotanekov} et~al.,}{{Kokotanekov}
  et~al.}{2017}]{koko17}
{Kokotanekov} G.,  et~al., 2017, preprint, \href
  {http://adsabs.harvard.edu/abs/2017arXiv170600225K} {} (\mn@eprint {arXiv}
  {1706.00225})

\bibitem[\protect\citeauthoryear{{Kormendy} \& {Ho}}{{Kormendy} \&
  {Ho}}{2013}]{korm13}
{Kormendy} J.,  {Ho} L.~C.,  2013, \mn@doi [\araa]
  {10.1146/annurev-astro-082708-101811}, \href
  {http://adsabs.harvard.edu/abs/2013ARA%26A..51..511K} {51, 511}

\bibitem[\protect\citeauthoryear{{Ledlow} \& {Owen}}{{Ledlow} \&
  {Owen}}{1996}]{ledl96}
{Ledlow} M.~J.,  {Owen} F.~N.,  1996, \mn@doi [\aj] {10.1086/117985}, \href
  {http://adsabs.harvard.edu/abs/1996AJ....112....9L} {112, 9}

\bibitem[\protect\citeauthoryear{{Lehmer} et~al.,}{{Lehmer}
  et~al.}{2007}]{lehm07}
{Lehmer} B.~D.,  et~al., 2007, \mn@doi [\apj] {10.1086/511297}, \href
  {http://adsabs.harvard.edu/abs/2007ApJ...657..681L} {657, 681}

\bibitem[\protect\citeauthoryear{{Li}, {Bryan}, {Ruszkowski}, {Voit}, {O'Shea}
  \& {Donahue}}{{Li} et~al.}{2015}]{li15}
{Li} Y.,  {Bryan} G.~L.,  {Ruszkowski} M.,  {Voit} G.~M.,  {O'Shea} B.~W.,
  {Donahue} M.,  2015, \mn@doi [\apj] {10.1088/0004-637X/811/2/73}, \href
  {http://adsabs.harvard.edu/abs/2015ApJ...811...73L} {811, 73}

\bibitem[\protect\citeauthoryear{{Lin}, {McDonald}, {Benson}  \&
  {Miller}}{{Lin} et~al.}{2015}]{lin15}
{Lin} H.~W.,  {McDonald} M.,  {Benson} B.,   {Miller} E.,  2015, \mn@doi [\apj]
  {10.1088/0004-637X/802/1/34}, \href
  {http://adsabs.harvard.edu/abs/2015ApJ...802...34L} {802, 34}

\bibitem[\protect\citeauthoryear{{Lindner} et~al.,}{{Lindner}
  et~al.}{2014}]{lind14}
{Lindner} R.~R.,  et~al., 2014, \mn@doi [\apj] {10.1088/0004-637X/786/1/49},
  \href {http://adsabs.harvard.edu/abs/2014ApJ...786...49L} {786, 49}

\bibitem[\protect\citeauthoryear{{Luo} \& {Sadler}}{{Luo} \&
  {Sadler}}{2010}]{luo10}
{Luo} Q.,  {Sadler} E.~M.,  2010, \mn@doi [\apj] {10.1088/0004-637X/713/1/398},
  \href {http://adsabs.harvard.edu/abs/2010ApJ...713..398L} {713, 398}

\bibitem[\protect\citeauthoryear{{Ma}, {McNamara}  \& {Nulsen}}{{Ma}
  et~al.}{2013}]{ma13}
{Ma} C.-J.,  {McNamara} B.~R.,   {Nulsen} P.~E.~J.,  2013, \mn@doi [\apj]
  {10.1088/0004-637X/763/1/63}, \href
  {http://adsabs.harvard.edu/abs/2013ApJ...763...63M} {763, 63}

\bibitem[\protect\citeauthoryear{{Mandelbaum}, {Li}, {Kauffmann}  \&
  {White}}{{Mandelbaum} et~al.}{2009}]{mand09}
{Mandelbaum} R.,  {Li} C.,  {Kauffmann} G.,   {White} S.~D.~M.,  2009, \mn@doi
  [\mnras] {10.1111/j.1365-2966.2008.14235.x}, \href
  {http://adsabs.harvard.edu/abs/2009MNRAS.393..377M} {393, 377}

\bibitem[\protect\citeauthoryear{{Marriage} et~al.,}{{Marriage}
  et~al.}{2011}]{marr11}
{Marriage} T.~A.,  et~al., 2011, \mn@doi [\apj] {10.1088/0004-637X/737/2/61},
  \href {http://adsabs.harvard.edu/abs/2011ApJ...737...61M} {737, 61}

\bibitem[\protect\citeauthoryear{{Mauch} \& {Sadler}}{{Mauch} \&
  {Sadler}}{2007}]{mauc07}
{Mauch} T.,  {Sadler} E.~M.,  2007, \mn@doi [\mnras]
  {10.1111/j.1365-2966.2006.11353.x}, \href
  {http://adsabs.harvard.edu/abs/2007MNRAS.375..931M} {375, 931}

\bibitem[\protect\citeauthoryear{{Mauch}, {Murphy}, {Buttery}, {Curran},
  {Hunstead}, {Piestrzynski}, {Robertson}  \& {Sadler}}{{Mauch}
  et~al.}{2003}]{mauc03}
{Mauch} T.,  {Murphy} T.,  {Buttery} H.~J.,  {Curran} J.,  {Hunstead} R.~W.,
  {Piestrzynski} B.,  {Robertson} J.~G.,   {Sadler} E.~M.,  2003, \mn@doi
  [\mnras] {10.1046/j.1365-8711.2003.06605.x}, \href
  {http://adsabs.harvard.edu/abs/2003MNRAS.342.1117M} {342, 1117}

\bibitem[\protect\citeauthoryear{{McCourt}, {Sharma}, {Quataert}  \&
  {Parrish}}{{McCourt} et~al.}{2012}]{mcco12}
{McCourt} M.,  {Sharma} P.,  {Quataert} E.,   {Parrish} I.~J.,  2012, \mn@doi
  [\mnras] {10.1111/j.1365-2966.2011.19972.x}, \href
  {http://adsabs.harvard.edu/abs/2012MNRAS.419.3319M} {419, 3319}

\bibitem[\protect\citeauthoryear{{McDonald}, {Veilleux}, {Rupke}  \&
  {Mushotzky}}{{McDonald} et~al.}{2010}]{mcdo10}
{McDonald} M.,  {Veilleux} S.,  {Rupke} D.~S.~N.,   {Mushotzky} R.,  2010,
  \mn@doi [\apj] {10.1088/0004-637X/721/2/1262}, \href
  {http://adsabs.harvard.edu/abs/2010ApJ...721.1262M} {721, 1262}

\bibitem[\protect\citeauthoryear{{McDonald}, {Veilleux}, {Rupke}, {Mushotzky}
  \& {Reynolds}}{{McDonald} et~al.}{2011}]{mcdo11}
{McDonald} M.,  {Veilleux} S.,  {Rupke} D.~S.~N.,  {Mushotzky} R.,   {Reynolds}
  C.,  2011, \mn@doi [\apj] {10.1088/0004-637X/734/2/95}, \href
  {http://adsabs.harvard.edu/abs/2011ApJ...734...95M} {734, 95}

\bibitem[\protect\citeauthoryear{{McDonald}, {Benson}, {Veilleux}, {Bautz}  \&
  {Reichardt}}{{McDonald} et~al.}{2013a}]{mcdo13b}
{McDonald} M.,  {Benson} B.,  {Veilleux} S.,  {Bautz} M.~W.,   {Reichardt}
  C.~L.,  2013a, \mn@doi [\apjl] {10.1088/2041-8205/765/2/L37}, \href
  {http://adsabs.harvard.edu/abs/2013ApJ...765L..37M} {765, L37}

\bibitem[\protect\citeauthoryear{{McDonald} et~al.,}{{McDonald}
  et~al.}{2013b}]{mcdo13c}
{McDonald} M.,  et~al., 2013b, \mn@doi [\apj] {10.1088/0004-637X/774/1/23},
  \href {http://adsabs.harvard.edu/abs/2013ApJ...774...23M} {774, 23}

\bibitem[\protect\citeauthoryear{{McDonald} et~al.,}{{McDonald}
  et~al.}{2015}]{mcdo15}
{McDonald} M.,  et~al., 2015, \mn@doi [\apj] {10.1088/0004-637X/811/2/111},
  \href {http://adsabs.harvard.edu/abs/2015ApJ...811..111M} {811, 111}

\bibitem[\protect\citeauthoryear{{McDonald} et~al.,}{{McDonald}
  et~al.}{2016}]{mcdo16}
{McDonald} M.,  et~al., 2016, \mn@doi [\apj] {10.3847/0004-637X/817/2/86},
  \href {http://adsabs.harvard.edu/abs/2016ApJ...817...86M} {817, 86}

\bibitem[\protect\citeauthoryear{{McDonald} et~al.,}{{McDonald}
  et~al.}{2017}]{mcdo17}
{McDonald} M.,  et~al., 2017, \mn@doi [\apj] {10.3847/1538-4357/aa7740}, \href
  {http://adsabs.harvard.edu/abs/2017ApJ...843...28M} {843, 28}

\bibitem[\protect\citeauthoryear{{McLeod} et~al.,}{{McLeod}
  et~al.}{2015}]{mcle15}
{McLeod} B.,  et~al., 2015, \mn@doi [\pasp] {10.1086/680687}, \href
  {http://adsabs.harvard.edu/abs/2015PASP..127..366M} {127, 366}

\bibitem[\protect\citeauthoryear{{McNamara} \& {O'Connell}}{{McNamara} \&
  {O'Connell}}{1989}]{mcna89}
{McNamara} B.~R.,  {O'Connell} R.~W.,  1989, \mn@doi [\aj] {10.1086/115275},
  \href {http://adsabs.harvard.edu/abs/1989AJ.....98.2018M} {98, 2018}

\bibitem[\protect\citeauthoryear{{McNamara} et~al.,}{{McNamara}
  et~al.}{2000}]{mcna00}
{McNamara} B.~R.,  et~al., 2000, \mn@doi [\apjl] {10.1086/312662}, \href
  {http://adsabs.harvard.edu/abs/2000ApJ...534L.135M} {534, L135}

\bibitem[\protect\citeauthoryear{{McNamara}, {Wise}  \& {Murray}}{{McNamara}
  et~al.}{2004}]{mcna04}
{McNamara} B.~R.,  {Wise} M.~W.,   {Murray} S.~S.,  2004, \mn@doi [\apj]
  {10.1086/380114}, \href {http://adsabs.harvard.edu/abs/2004ApJ...601..173M}
  {601, 173}

\bibitem[\protect\citeauthoryear{{McNamara}, {Nulsen}, {Wise}, {Rafferty},
  {Carilli}, {Sarazin}  \& {Blanton}}{{McNamara} et~al.}{2005}]{mcna05}
{McNamara} B.~R.,  {Nulsen} P.~E.~J.,  {Wise} M.~W.,  {Rafferty} D.~A.,
  {Carilli} C.,  {Sarazin} C.~L.,   {Blanton} E.~L.,  2005, \mn@doi [\nat]
  {10.1038/nature03202}, \href
  {http://adsabs.harvard.edu/abs/2005Natur.433...45M} {433, 45}

\bibitem[\protect\citeauthoryear{{McNamara} et~al.,}{{McNamara}
  et~al.}{2014}]{mcna14}
{McNamara} B.~R.,  et~al., 2014, \mn@doi [\apj] {10.1088/0004-637X/785/1/44},
  \href {http://adsabs.harvard.edu/abs/2014ApJ...785...44M} {785, 44}

\bibitem[\protect\citeauthoryear{{McNamara}, {Russell}, {Nulsen}, {Hogan},
  {Fabian}, {Pulido}  \& {Edge}}{{McNamara} et~al.}{2016}]{mcna16}
{McNamara} B.~R.,  {Russell} H.~R.,  {Nulsen} P.~E.~J.,  {Hogan} M.~T.,
  {Fabian} A.~C.,  {Pulido} F.,   {Edge} A.~C.,  2016, preprint, \href
  {http://adsabs.harvard.edu/abs/2016arXiv160404629M} {} (\mn@eprint {arXiv}
  {1604.04629})

\bibitem[\protect\citeauthoryear{{Mernier} et~al.,}{{Mernier}
  et~al.}{2017}]{mern17}
{Mernier} F.,  et~al., 2017, \mn@doi [\aap] {10.1051/0004-6361/201630075},
  \href {http://adsabs.harvard.edu/abs/2017A%26A...603A..80M} {603, A80}

\bibitem[\protect\citeauthoryear{{Mingo}, {Hardcastle}, {Croston}, {Dicken},
  {Evans}, {Morganti}  \& {Tadhunter}}{{Mingo} et~al.}{2014}]{ming14}
{Mingo} B.,  {Hardcastle} M.~J.,  {Croston} J.~H.,  {Dicken} D.,  {Evans}
  D.~A.,  {Morganti} R.,   {Tadhunter} C.,  2014, \mn@doi [\mnras]
  {10.1093/mnras/stu263}, \href
  {http://adsabs.harvard.edu/abs/2014MNRAS.440..269M} {440, 269}

\bibitem[\protect\citeauthoryear{{Mittal}, {Hudson}, {Reiprich}  \&
  {Clarke}}{{Mittal} et~al.}{2009}]{mitt09}
{Mittal} R.,  {Hudson} D.~S.,  {Reiprich} T.~H.,   {Clarke} T.,  2009, \mn@doi
  [\aap] {10.1051/0004-6361/200810836}, \href
  {http://adsabs.harvard.edu/abs/2009A%26A...501..835M} {501, 835}

\bibitem[\protect\citeauthoryear{{Mittal}, {Whelan}  \& {Combes}}{{Mittal}
  et~al.}{2015}]{mitt15}
{Mittal} R.,  {Whelan} J.~T.,   {Combes} F.,  2015, \mn@doi [\mnras]
  {10.1093/mnras/stv754}, \href
  {http://adsabs.harvard.edu/abs/2015MNRAS.450.2564M} {450, 2564}

\bibitem[\protect\citeauthoryear{{Motl}, {Hallman}, {Burns}  \&
  {Norman}}{{Motl} et~al.}{2005}]{motl05}
{Motl} P.~M.,  {Hallman} E.~J.,  {Burns} J.~O.,   {Norman} M.~L.,  2005,
  \mn@doi [\apjl] {10.1086/430144}, \href
  {http://adsabs.harvard.edu/abs/2005ApJ...623L..63M} {623, L63}

\bibitem[\protect\citeauthoryear{{Nulsen}, {Jones}, {Forman}, {Churazov},
  {McNamara}, {David}  \& {Murray}}{{Nulsen} et~al.}{2009}]{nuls09}
{Nulsen} P.,  {Jones} C.,  {Forman} W.,  {Churazov} E.,  {McNamara} B.,
  {David} L.,   {Murray} S.,  2009, in {S.~Heinz \& E.~Wilcots} ed.,  American
  Institute of Physics Conference Series Vol. 1201, American Institute of
  Physics Conference Series. pp 198--201 (\mn@eprint {arXiv} {0909.1809}),
  \mn@doi{10.1063/1.3293033}

\bibitem[\protect\citeauthoryear{{O'Dea} et~al.,}{{O'Dea}
  et~al.}{2008}]{odea08}
{O'Dea} C.~P.,  et~al., 2008, \mn@doi [\apj] {10.1086/588212}, \href
  {http://adsabs.harvard.edu/abs/2008ApJ...681.1035O} {681, 1035}

\bibitem[\protect\citeauthoryear{{O'Sullivan}, {Giacintucci}, {David}, {Gitti},
  {Vrtilek}, {Raychaudhury}  \& {Ponman}}{{O'Sullivan} et~al.}{2011}]{osul11}
{O'Sullivan} E.,  {Giacintucci} S.,  {David} L.~P.,  {Gitti} M.,  {Vrtilek}
  J.~M.,  {Raychaudhury} S.,   {Ponman} T.~J.,  2011, \mn@doi [\apj]
  {10.1088/0004-637X/735/1/11}, \href
  {http://adsabs.harvard.edu/abs/2011ApJ...735...11O} {735, 11}

\bibitem[\protect\citeauthoryear{{Oonk}, {Hatch}, {Jaffe}, {Bremer}  \& {van
  Weeren}}{{Oonk} et~al.}{2011}]{oonk11}
{Oonk} J.~B.~R.,  {Hatch} N.~A.,  {Jaffe} W.,  {Bremer} M.~N.,   {van Weeren}
  R.~J.,  2011, \mn@doi [\mnras] {10.1111/j.1365-2966.2011.18551.x}, \href
  {http://adsabs.harvard.edu/abs/2011MNRAS.414.2309O} {414, 2309}

\bibitem[\protect\citeauthoryear{{Owen} \& {White}}{{Owen} \&
  {White}}{1991}]{owen91}
{Owen} F.~N.,  {White} R.~A.,  1991, \mnras, \href
  {http://adsabs.harvard.edu/abs/1991MNRAS.249..164O} {249, 164}

\bibitem[\protect\citeauthoryear{{Panagoulia}, {Fabian}  \&
  {Sanders}}{{Panagoulia} et~al.}{2014a}]{pana14a}
{Panagoulia} E.~K.,  {Fabian} A.~C.,   {Sanders} J.~S.,  2014a, \mn@doi
  [\mnras] {10.1093/mnras/stt2349}, \href
  {http://adsabs.harvard.edu/abs/2014MNRAS.438.2341P} {438, 2341}

\bibitem[\protect\citeauthoryear{{Panagoulia}, {Fabian}, {Sanders}  \&
  {Hlavacek-Larrondo}}{{Panagoulia} et~al.}{2014b}]{pana14b}
{Panagoulia} E.~K.,  {Fabian} A.~C.,  {Sanders} J.~S.,   {Hlavacek-Larrondo}
  J.,  2014b, \mn@doi [\mnras] {10.1093/mnras/stu1499}, \href
  {http://adsabs.harvard.edu/abs/2014MNRAS.444.1236P} {444, 1236}

\bibitem[\protect\citeauthoryear{{Peterson} et~al.,}{{Peterson}
  et~al.}{2001}]{pete01}
{Peterson} J.~R.,  et~al., 2001, \mn@doi [\aap] {10.1051/0004-6361:20000021},
  \href {http://adsabs.harvard.edu/abs/2001A%26A...365L.104P} {365, L104}

\bibitem[\protect\citeauthoryear{{Pfrommer}}{{Pfrommer}}{2013}]{pfro13}
{Pfrommer} C.,  2013, \mn@doi [\apj] {10.1088/0004-637X/779/1/10}, \href
  {http://adsabs.harvard.edu/abs/2013ApJ...779...10P} {779, 10}

\bibitem[\protect\citeauthoryear{{Pizzolato} \& {Soker}}{{Pizzolato} \&
  {Soker}}{2005}]{pizz05}
{Pizzolato} F.,  {Soker} N.,  2005, \mn@doi [\apj] {10.1086/444344}, \href
  {http://adsabs.harvard.edu/abs/2005ApJ...632..821P} {632, 821}

\bibitem[\protect\citeauthoryear{{Pizzolato} \& {Soker}}{{Pizzolato} \&
  {Soker}}{2010}]{pizz10}
{Pizzolato} F.,  {Soker} N.,  2010, \mn@doi [\mnras]
  {10.1111/j.1365-2966.2010.17156.x}, \href
  {http://adsabs.harvard.edu/abs/2010MNRAS.408..961P} {408, 961}

\bibitem[\protect\citeauthoryear{{Planck Collaboration} et~al.,}{{Planck
  Collaboration} et~al.}{2014}]{plan13}
{Planck Collaboration} et~al., 2014, \mn@doi [\aap]
  {10.1051/0004-6361/201321523}, \href
  {http://adsabs.harvard.edu/abs/2014A%26A...571A..29P} {571, 29}

\bibitem[\protect\citeauthoryear{{Pracy} et~al.,}{{Pracy}
  et~al.}{2016}]{prac16}
{Pracy} M.~B.,  et~al., 2016, \mn@doi [\mnras] {10.1093/mnras/stw910}, \href
  {http://adsabs.harvard.edu/abs/2016MNRAS.460....2P} {460, 2}

\bibitem[\protect\citeauthoryear{{Prasad}, {Sharma}  \& {Babul}}{{Prasad}
  et~al.}{2015}]{pras15}
{Prasad} D.,  {Sharma} P.,   {Babul} A.,  2015, \mn@doi [\apj]
  {10.1088/0004-637X/811/2/108}, \href
  {http://adsabs.harvard.edu/abs/2015ApJ...811..108P} {811, 108}

\bibitem[\protect\citeauthoryear{{Pratt}, {Croston}, {Arnaud}  \&
  {B{\"o}hringer}}{{Pratt} et~al.}{2009}]{prat09}
{Pratt} G.~W.,  {Croston} J.~H.,  {Arnaud} M.,   {B{\"o}hringer} H.,  2009,
  \mn@doi [\aap] {10.1051/0004-6361/200810994}, \href
  {http://adsabs.harvard.edu/abs/2009A%26A...498..361P} {498, 361}

\bibitem[\protect\citeauthoryear{{Rafferty}, {McNamara}, {Nulsen}  \&
  {Wise}}{{Rafferty} et~al.}{2006}]{raff06}
{Rafferty} D.~A.,  {McNamara} B.~R.,  {Nulsen} P.~E.~J.,   {Wise} M.~W.,  2006,
  \mn@doi [\apj] {10.1086/507672}, \href
  {http://adsabs.harvard.edu/abs/2006ApJ...652..216R} {652, 216}

\bibitem[\protect\citeauthoryear{{Rafferty}, {McNamara}  \&
  {Nulsen}}{{Rafferty} et~al.}{2008}]{raff08}
{Rafferty} D.~A.,  {McNamara} B.~R.,   {Nulsen} P.~E.~J.,  2008, \mn@doi [\apj]
  {10.1086/591240}, \href {http://adsabs.harvard.edu/abs/2008ApJ...687..899R}
  {687, 899}

\bibitem[\protect\citeauthoryear{{Ramos Almeida} et~al.,}{{Ramos Almeida}
  et~al.}{2012}]{ramo12}
{Ramos Almeida} C.,  et~al., 2012, \mn@doi [\mnras]
  {10.1111/j.1365-2966.2011.19731.x}, \href
  {http://adsabs.harvard.edu/abs/2012MNRAS.419..687R} {419, 687}

\bibitem[\protect\citeauthoryear{{Reichardt} et~al.,}{{Reichardt}
  et~al.}{2013}]{reic13}
{Reichardt} C.~L.,  et~al., 2013, \mn@doi [\apj] {10.1088/0004-637X/763/2/127},
  \href {http://adsabs.harvard.edu/abs/2013ApJ...763..127R} {763, 127}

\bibitem[\protect\citeauthoryear{{Reiprich} \& {B{\"o}hringer}}{{Reiprich} \&
  {B{\"o}hringer}}{2002}]{reip02}
{Reiprich} T.~H.,  {B{\"o}hringer} H.,  2002, \mn@doi [\apj] {10.1086/338753},
  \href {http://adsabs.harvard.edu/abs/2002ApJ...567..716R} {567, 716}

\bibitem[\protect\citeauthoryear{{Ruel} et~al.,}{{Ruel} et~al.}{2014}]{ruel14}
{Ruel} J.,  et~al., 2014, \mn@doi [\apj] {10.1088/0004-637X/792/1/45}, \href
  {http://adsabs.harvard.edu/abs/2014ApJ...792...45R} {792, 45}

\bibitem[\protect\citeauthoryear{{Russell}, {McNamara}, {Edge}, {Hogan}, {Main}
   \& {Vantyghem}}{{Russell} et~al.}{2013}]{russ13}
{Russell} H.~R.,  {McNamara} B.~R.,  {Edge} A.~C.,  {Hogan} M.~T.,  {Main}
  R.~A.,   {Vantyghem} A.~N.,  2013, \mn@doi [\mnras] {10.1093/mnras/stt490},
  \href {http://adsabs.harvard.edu/abs/2013MNRAS.432..530R} {432, 530}

\bibitem[\protect\citeauthoryear{{Russell} et~al.,}{{Russell}
  et~al.}{2016}]{russ16}
{Russell} H.~R.,  et~al., 2016, preprint, \href
  {http://adsabs.harvard.edu/abs/2016arXiv161100017R} {} (\mn@eprint {arXiv}
  {1611.00017})

\bibitem[\protect\citeauthoryear{{Ruszkowski}, {Yang}  \&
  {Reynolds}}{{Ruszkowski} et~al.}{2017}]{rusz17}
{Ruszkowski} M.,  {Yang} H.-Y.~K.,   {Reynolds} C.~S.,  2017, preprint, \href
  {http://adsabs.harvard.edu/abs/2017arXiv170107441R} {} (\mn@eprint {arXiv}
  {1701.07441})

\bibitem[\protect\citeauthoryear{{Sadler} et~al.,}{{Sadler}
  et~al.}{2002}]{sadl02}
{Sadler} E.~M.,  et~al., 2002, \mn@doi [\mnras]
  {10.1046/j.1365-8711.2002.04998.x}, \href
  {http://adsabs.harvard.edu/abs/2002MNRAS.329..227S} {329, 227}

\bibitem[\protect\citeauthoryear{{Sadler} et~al.,}{{Sadler}
  et~al.}{2007}]{sadl07}
{Sadler} E.~M.,  et~al., 2007, \mn@doi [\mnras]
  {10.1111/j.1365-2966.2007.12231.x}, \href
  {http://adsabs.harvard.edu/abs/2007MNRAS.381..211S} {381, 211}

\bibitem[\protect\citeauthoryear{{Sadler}, {Ekers}, {Mahony}, {Mauch}  \&
  {Murphy}}{{Sadler} et~al.}{2014}]{sadl14}
{Sadler} E.~M.,  {Ekers} R.~D.,  {Mahony} E.~K.,  {Mauch} T.,   {Murphy} T.,
  2014, \mn@doi [\mnras] {10.1093/mnras/stt2239}, \href
  {http://adsabs.harvard.edu/abs/2014MNRAS.438..796S} {438, 796}

\bibitem[\protect\citeauthoryear{{Samuele}, {McNamara}, {Vikhlinin}  \&
  {Mullis}}{{Samuele} et~al.}{2011}]{samu11}
{Samuele} R.,  {McNamara} B.~R.,  {Vikhlinin} A.,   {Mullis} C.~R.,  2011,
  \mn@doi [\apj] {10.1088/0004-637X/731/1/31}, \href
  {http://adsabs.harvard.edu/abs/2011ApJ...731...31S} {731, 31}

\bibitem[\protect\citeauthoryear{{Santos}, {Tozzi}, {Rosati}  \&
  {B{\"o}hringer}}{{Santos} et~al.}{2010}]{sant10}
{Santos} J.~S.,  {Tozzi} P.,  {Rosati} P.,   {B{\"o}hringer} H.,  2010, \mn@doi
  [\aap] {10.1051/0004-6361/201015208}, \href
  {http://adsabs.harvard.edu/abs/2010A%26A...521A..64S} {521, A64}

\bibitem[\protect\citeauthoryear{{Shabala}, {Deller}, {Kaviraj}, {Middelberg},
  {Turner}, {Ting}, {Allison}  \& {Davis}}{{Shabala} et~al.}{2017}]{shab17}
{Shabala} S.~S.,  {Deller} A.,  {Kaviraj} S.,  {Middelberg} E.,  {Turner}
  R.~J.,  {Ting} Y.~S.,  {Allison} J.~R.,   {Davis} T.~A.,  2017, \mn@doi
  [\mnras] {10.1093/mnras/stw2536}, \href
  {http://adsabs.harvard.edu/abs/2017MNRAS.464.4706S} {464, 4706}

\bibitem[\protect\citeauthoryear{{Sharma}, {McCourt}, {Quataert}  \&
  {Parrish}}{{Sharma} et~al.}{2012}]{shar12}
{Sharma} P.,  {McCourt} M.,  {Quataert} E.,   {Parrish} I.~J.,  2012, \mn@doi
  [\mnras] {10.1111/j.1365-2966.2011.20246.x}, \href
  {http://adsabs.harvard.edu/abs/2012MNRAS.420.3174S} {420, 3174}

\bibitem[\protect\citeauthoryear{{Sif{\'o}n} et~al.,}{{Sif{\'o}n}
  et~al.}{2013}]{sifo13}
{Sif{\'o}n} C.,  et~al., 2013, \mn@doi [\apj] {10.1088/0004-637X/772/1/25},
  \href {http://adsabs.harvard.edu/abs/2013ApJ...772...25S} {772, 25}

\bibitem[\protect\citeauthoryear{{Silk} \& {Rees}}{{Silk} \&
  {Rees}}{1998}]{silk98}
{Silk} J.,  {Rees} M.~J.,  1998, \aap, \href
  {http://adsabs.harvard.edu/abs/1998A%26A...331L...1S} {331, L1}

\bibitem[\protect\citeauthoryear{{Simionescu}, {Werner}, {Finoguenov},
  {B{\"o}hringer}  \& {Br{\"u}ggen}}{{Simionescu} et~al.}{2008}]{simi08}
{Simionescu} A.,  {Werner} N.,  {Finoguenov} A.,  {B{\"o}hringer} H.,
  {Br{\"u}ggen} M.,  2008, \mn@doi [\aap] {10.1051/0004-6361:20078749}, \href
  {http://adsabs.harvard.edu/abs/2008A%26A...482...97S} {482, 97}

\bibitem[\protect\citeauthoryear{{Simpson}, {Westoby}, {Arumugam}, {Ivison},
  {Hartley}  \& {Almaini}}{{Simpson} et~al.}{2013}]{simp13}
{Simpson} C.,  {Westoby} P.,  {Arumugam} V.,  {Ivison} R.,  {Hartley} W.,
  {Almaini} O.,  2013, \mn@doi [\mnras] {10.1093/mnras/stt940}, \href
  {http://adsabs.harvard.edu/abs/2013MNRAS.433.2647S} {433, 2647}

\bibitem[\protect\citeauthoryear{{Smith}, {Brickhouse}, {Liedahl}  \&
  {Raymond}}{{Smith} et~al.}{2001}]{smit01}
{Smith} R.~K.,  {Brickhouse} N.~S.,  {Liedahl} D.~A.,   {Raymond} J.~C.,  2001,
  \mn@doi [\apjl] {10.1086/322992}, \href
  {http://adsabs.harvard.edu/abs/2001ApJ...556L..91S} {556, L91}

\bibitem[\protect\citeauthoryear{{Smol{\v c}i{\'c}} et~al.,}{{Smol{\v c}i{\'c}}
  et~al.}{2009}]{smol09}
{Smol{\v c}i{\'c}} V.,  et~al., 2009, \mn@doi [\apj]
  {10.1088/0004-637X/696/1/24}, \href
  {http://adsabs.harvard.edu/abs/2009ApJ...696...24S} {696, 24}

\bibitem[\protect\citeauthoryear{{Soker}}{{Soker}}{2006}]{soke06}
{Soker} N.,  2006, \mn@doi [New Astronomy] {10.1016/j.newast.2006.05.003},
  \href {http://adsabs.harvard.edu/abs/2006NewA...12...38S} {12, 38}

\bibitem[\protect\citeauthoryear{{Sommer}, {Basu}, {Pacaud}, {Bertoldi}  \&
  {Andernach}}{{Sommer} et~al.}{2011}]{somm11}
{Sommer} M.~W.,  {Basu} K.,  {Pacaud} F.,  {Bertoldi} F.,   {Andernach} H.,
  2011, \mn@doi [\aap] {10.1051/0004-6361/201016150}, \href
  {http://adsabs.harvard.edu/abs/2011A%26A...529A.124S} {529, A124}

\bibitem[\protect\citeauthoryear{{Song} et~al.,}{{Song} et~al.}{2012}]{song12}
{Song} J.,  et~al., 2012, \mn@doi [\apj] {10.1088/0004-637X/761/1/22}, \href
  {http://adsabs.harvard.edu/abs/2012ApJ...761...22S} {761, 22}

\bibitem[\protect\citeauthoryear{{Sun}}{{Sun}}{2009}]{sun09}
{Sun} M.,  2009, \mn@doi [\apj] {10.1088/0004-637X/704/2/1586}, \href
  {http://adsabs.harvard.edu/abs/2009ApJ...704.1586S} {704, 1586}

\bibitem[\protect\citeauthoryear{{Sun}, {Jones}, {Forman}, {Vikhlinin},
  {Donahue}  \& {Voit}}{{Sun} et~al.}{2007}]{sun07}
{Sun} M.,  {Jones} C.,  {Forman} W.,  {Vikhlinin} A.,  {Donahue} M.,   {Voit}
  M.,  2007, \mn@doi [\apj] {10.1086/510895}, \href
  {http://adsabs.harvard.edu/abs/2007ApJ...657..197S} {657, 197}

\bibitem[\protect\citeauthoryear{{Tang} \& {Churazov}}{{Tang} \&
  {Churazov}}{2017}]{tang17}
{Tang} X.,  {Churazov} E.,  2017, \mn@doi [\mnras] {10.1093/mnras/stx590},
  \href {http://adsabs.harvard.edu/abs/2017MNRAS.468.3516T} {468, 3516}

\bibitem[\protect\citeauthoryear{{Tremblay} et~al.,}{{Tremblay}
  et~al.}{2015}]{trem15}
{Tremblay} G.~R.,  et~al., 2015, \mn@doi [\mnras] {10.1093/mnras/stv1151},
  \href {http://adsabs.harvard.edu/abs/2015MNRAS.451.3768T} {451, 3768}

\bibitem[\protect\citeauthoryear{{Tremblay} et~al.,}{{Tremblay}
  et~al.}{2016}]{trem16}
{Tremblay} G.~R.,  et~al., 2016, \mn@doi [\nat] {10.1038/nature17969}, \href
  {http://adsabs.harvard.edu/abs/2016Natur.534..218T} {534, 218}

\bibitem[\protect\citeauthoryear{{Turner} \& {Shabala}}{{Turner} \&
  {Shabala}}{2015}]{turn15}
{Turner} R.~J.,  {Shabala} S.~S.,  2015, \mn@doi [\apj]
  {10.1088/0004-637X/806/1/59}, \href
  {http://adsabs.harvard.edu/abs/2015ApJ...806...59T} {806, 59}

\bibitem[\protect\citeauthoryear{{Valentini} \& {Brighenti}}{{Valentini} \&
  {Brighenti}}{2015}]{vale15}
{Valentini} M.,  {Brighenti} F.,  2015, \mn@doi [\mnras]
  {10.1093/mnras/stv090}, \href
  {http://adsabs.harvard.edu/abs/2015MNRAS.448.1979V} {448, 1979}

\bibitem[\protect\citeauthoryear{{Vantyghem} et~al.,}{{Vantyghem}
  et~al.}{2016}]{vant16}
{Vantyghem} A.~N.,  et~al., 2016, \mn@doi [\apj] {10.3847/0004-637X/832/2/148},
  \href {http://adsabs.harvard.edu/abs/2016ApJ...832..148V} {832, 148}

\bibitem[\protect\citeauthoryear{{Vikhlinin}, {Burenin}, {Forman}, {Jones},
  {Hornstrup}, {Murray}  \& {Quintana}}{{Vikhlinin} et~al.}{2007}]{vikh07}
{Vikhlinin} A.,  {Burenin} R.,  {Forman} W.~R.,  {Jones} C.,  {Hornstrup} A.,
  {Murray} S.~S.,   {Quintana} H.,  2007, in {B{\"o}hringer} H.,  {Pratt}
  G.~W.,  {Finoguenov} A.,   {Schuecker} P.,  eds, Heating versus Cooling in
  Galaxies and Clusters of Galaxies. p.~48 (\mn@eprint {}
  {arXiv:astro-ph/0611438}), \mn@doi{10.1007/978-3-540-73484-0_9}

\bibitem[\protect\citeauthoryear{{Voit} \& {Donahue}}{{Voit} \&
  {Donahue}}{2015}]{voit15}
{Voit} G.~M.,  {Donahue} M.,  2015, \mn@doi [\apjl]
  {10.1088/2041-8205/799/1/L1}, \href
  {http://adsabs.harvard.edu/abs/2015ApJ...799L...1V} {799, L1}

\bibitem[\protect\citeauthoryear{{Voit}, {Cavagnolo}, {Donahue}, {Rafferty},
  {McNamara}  \& {Nulsen}}{{Voit} et~al.}{2008}]{voit08}
{Voit} G.~M.,  {Cavagnolo} K.~W.,  {Donahue} M.,  {Rafferty} D.~A.,  {McNamara}
  B.~R.,   {Nulsen} P.~E.~J.,  2008, \mn@doi [\apjl] {10.1086/590344}, \href
  {http://adsabs.harvard.edu/abs/2008ApJ...681L...5V} {681, L5}

\bibitem[\protect\citeauthoryear{{Voit}, {Donahue}, {Bryan}  \&
  {McDonald}}{{Voit} et~al.}{2015}]{voit15b}
{Voit} G.~M.,  {Donahue} M.,  {Bryan} G.~L.,   {McDonald} M.,  2015, \mn@doi
  [\nat] {10.1038/nature14167}, \href
  {http://adsabs.harvard.edu/abs/2015Natur.519..203V} {519, 203}

\bibitem[\protect\citeauthoryear{{Voit}, {Meece}, {Li}, {O'Shea}, {Bryan}  \&
  {Donahue}}{{Voit} et~al.}{2016}]{voit16}
{Voit} G.~M.,  {Meece} G.,  {Li} Y.,  {O'Shea} B.~W.,  {Bryan} G.~L.,
  {Donahue} M.,  2016, preprint, \href
  {http://adsabs.harvard.edu/abs/2016arXiv160702212V} {} (\mn@eprint {arXiv}
  {1607.02212})

\bibitem[\protect\citeauthoryear{{Wagh}, {Sharma}  \& {McCourt}}{{Wagh}
  et~al.}{2014}]{wagh14}
{Wagh} B.,  {Sharma} P.,   {McCourt} M.,  2014, \mn@doi [\mnras]
  {10.1093/mnras/stu138}, \href
  {http://adsabs.harvard.edu/abs/2014MNRAS.439.2822W} {439, 2822}

\bibitem[\protect\citeauthoryear{{Werner} et~al.,}{{Werner}
  et~al.}{2010}]{wern10}
{Werner} N.,  et~al., 2010, \mn@doi [\mnras]
  {10.1111/j.1365-2966.2010.16755.x}, \href
  {http://adsabs.harvard.edu/abs/2010MNRAS.407.2063W} {407, 2063}

\bibitem[\protect\citeauthoryear{{Werner} et~al.,}{{Werner}
  et~al.}{2011}]{wern11}
{Werner} N.,  et~al., 2011, \mn@doi [\mnras]
  {10.1111/j.1365-2966.2011.18957.x}, \href
  {http://adsabs.harvard.edu/abs/2011MNRAS.415.3369W} {415, 3369}

\bibitem[\protect\citeauthoryear{{Werner} et~al.,}{{Werner}
  et~al.}{2014}]{wern14}
{Werner} N.,  et~al., 2014, \mn@doi [\mnras] {10.1093/mnras/stu006}, \href
  {http://adsabs.harvard.edu/abs/2014MNRAS.439.2291W} {439, 2291}

\bibitem[\protect\citeauthoryear{{Willott}, {Rawlings}, {Blundell}  \&
  {Lacy}}{{Willott} et~al.}{1999}]{will99}
{Willott} C.~J.,  {Rawlings} S.,  {Blundell} K.~M.,   {Lacy} M.,  1999, \mn@doi
  [\mnras] {10.1046/j.1365-8711.1999.02907.x}, \href
  {http://adsabs.harvard.edu/abs/1999MNRAS.309.1017W} {309, 1017}

\bibitem[\protect\citeauthoryear{{Willott}, {Rawlings}, {Blundell}, {Lacy}  \&
  {Eales}}{{Willott} et~al.}{2001}]{will01}
{Willott} C.~J.,  {Rawlings} S.,  {Blundell} K.~M.,  {Lacy} M.,   {Eales}
  S.~A.,  2001, \mn@doi [\mnras] {10.1046/j.1365-8711.2001.04101.x}, \href
  {http://adsabs.harvard.edu/abs/2001MNRAS.322..536W} {322, 536}

\bibitem[\protect\citeauthoryear{{Wylezalek} et~al.,}{{Wylezalek}
  et~al.}{2013}]{wyle13}
{Wylezalek} D.,  et~al., 2013, \mn@doi [\apj] {10.1088/0004-637X/769/1/79},
  \href {http://adsabs.harvard.edu/abs/2013ApJ...769...79W} {769, 79}

\bibitem[\protect\citeauthoryear{{Yang} \& {Reynolds}}{{Yang} \&
  {Reynolds}}{2016}]{yang16}
{Yang} H.-Y.~K.,  {Reynolds} C.~S.,  2016, preprint, \href
  {http://adsabs.harvard.edu/abs/2016arXiv160501725Y} {} (\mn@eprint {arXiv}
  {1605.01725})

\bibitem[\protect\citeauthoryear{{Yates}, {Miller}  \& {Peacock}}{{Yates}
  et~al.}{1989}]{yate89}
{Yates} M.~G.,  {Miller} L.,   {Peacock} J.~A.,  1989, \mnras, \href
  {http://adsabs.harvard.edu/abs/1989MNRAS.240..129Y} {240, 129}

\bibitem[\protect\citeauthoryear{{Yee} \& {Green}}{{Yee} \&
  {Green}}{1987}]{yee87}
{Yee} H.~K.~C.,  {Green} R.~F.,  1987, \mn@doi [\apj] {10.1086/165430}, \href
  {http://adsabs.harvard.edu/abs/1987ApJ...319...28Y} {319, 28}

\bibitem[\protect\citeauthoryear{{Yuan}, {Han}  \& {Wen}}{{Yuan}
  et~al.}{2016}]{yuan16b}
{Yuan} Z.~S.,  {Han} J.~L.,   {Wen} Z.~L.,  2016, \mn@doi [\mnras]
  {10.1093/mnras/stw1125}, \href
  {http://adsabs.harvard.edu/abs/2016MNRAS.460.3669Y} {460, 3669}

\bibitem[\protect\citeauthoryear{{Zhuravleva} et~al.,}{{Zhuravleva}
  et~al.}{2014}]{zhur14}
{Zhuravleva} I.,  et~al., 2014, \mn@doi [\nat] {10.1038/nature13830}, \href
  {http://adsabs.harvard.edu/abs/2014Natur.515...85Z} {515, 85}

\bibitem[\protect\citeauthoryear{{Zhuravleva} et~al.,}{{Zhuravleva}
  et~al.}{2016}]{zhur16}
{Zhuravleva} I.,  et~al., 2016, \mn@doi [\mnras] {10.1093/mnras/stw520}, \href
  {http://adsabs.harvard.edu/abs/2016MNRAS.458.2902Z} {458, 2902}

\bibitem[\protect\citeauthoryear{{van Weeren} et~al.,}{{van Weeren}
  et~al.}{2014}]{vanw14}
{van Weeren} R.~J.,  et~al., 2014, \mn@doi [\apjl]
  {10.1088/2041-8205/786/2/L17}, \href
  {http://adsabs.harvard.edu/abs/2014ApJ...786L..17V} {786, L17}

\makeatother
\end{thebibliography}

% Don't change these lines
\bsp	% typesetting comment
\label{lastpage}
\end{document}